\documentclass[traditabstract]{aa}
\usepackage{epsfig}    
\usepackage{lscape}
\usepackage{natbib}
\bibpunct{(}{)}{;}{a}{}{,}    
\usepackage{graphics}
\usepackage{graphicx,graphics}
\usepackage{color} 
\usepackage{times}
\usepackage{amsmath}
\usepackage{amssymb}
\begin{document}

\title {Relations between abundance characteristics and rotation velocity  
       for star-forming MaNGA galaxies}

\author {
         L.~S.~Pilyugin\inst{\ref{MAO},\ref{ARI}} \and 
         E.~K.~Grebel\inst{\ref{ARI}} \and
         I.~A.~Zinchenko\inst{\ref{MAO},\ref{ARI}} \and
         Y.~A.~Nefedyev\inst{\ref{KGU}} \and
         J.~M.~V\'{i}lchez\inst{\ref{IAA}}
         }
\institute{Main Astronomical Observatory, National Academy of Sciences of Ukraine, 27 Akademika Zabolotnoho St, 03680, Kiev, Ukraine \label{MAO} \and
Astronomisches Rechen-Institut, Zentrum f\"{u}r Astronomie der Universit\"{a}t Heidelberg, M\"{o}nchhofstr.\ 12--14, 69120 Heidelberg, Germany \label{ARI} \and
Kazan Federal University, 18 Kremlyovskaya St., 420008, Kazan, Russian Federation \label{KGU} \and
Instituto de Astrof\'{\i}sica de Andaluc\'{\i}a, CSIC, Apdo 3004, 18080 Granada, Spain \label{IAA} }

\abstract{We derive rotation curves, surface brightness profiles, and
oxygen abundance distributions for 147 late-type galaxies using the
publicly available spectroscopy obtained by the MaNGA survey.  Changes
of the central oxygen abundance (O/H)$_{0}$, the abundance at the
optical radius (O/H)$_{R_{25}}$, and the abundance gradient with
rotation velocity $V_{rot}$ are examined for galaxies with rotation
velocities from 90 km/s to 350 km/s. We found that each
relation shows a break at $V_{rot}^{*} \sim200$~km/s. The
central (O/H)$_{0}$ abundance increases with rising $V_{rot}$ and the
slope of the (O/H)$_{0}$ --  $V_{rot}$ relation is steeper for
galaxies with $V_{rot} \la V_{rot}^{*}$. The mean scatter of the
central abundances around this relation is 0.053 dex.  The relation
between the abundance at the optical radius of a galaxy and its
rotation velocity is similar; the mean scatter in abundances around
this relation is 0.081 dex.  The radial abundance gradient expressed
in dex/kpc flattens with the increase of the rotation velocity.
The slope of the relation is very low for galaxies with $V_{rot} \ga V_{rot}^{*}$.
The abundance gradient expressed in dex/$R_{25}$ is
rougly constant for galaxies with $V_{rot} \la  V_{rot}^{*}$, flattens
towards $V_{rot}^{*}$, and then again is roughly constant for galaxies
with $V_{rot} \ga V_{rot}^{*}$.  The change of the gradient expressed
in terms of dex/$h_{d}$ (where $h_{d}$ is the disc scale length), 
in terms of dex/$R_{e,d}$ (where $R_{e,d}$ is the disc effective radius), 
and in terms of dex/$R_{e,g}$ (where $R_{e,g}$ is the galaxy effective radius) 
with rotation velocity is similar to that for gradient in
dex/$R_{25}$.  The relations between abundance characteristics
and other basic parameters (stellar mass, luminosity, and radius) are
also considered.
}


\titlerunning{Relation between abundance and rotation velocity}
\authorrunning{Pilyugin et al.}
\maketitle

\section{Introduction}

The formation and evolution of galaxies in the Lambda Cold Dark Matter
($\Lambda$CDM) universe is mainly governed by the host dark matter
halo. The dark matter halo makes a dominant contribution to the
dynamical mass of a galaxy. Hence the rotation velocity of a disc
galaxy, which is a tracer of the dynamical mass is the most
fundamental parameter of that disc galaxy. The correlations between
the rotation velocity and other basic parameters (stellar mass, size,
luminosity) and, consequently, the correlations between each pair of
the basic parameters are predicted by simulations
\citep[e.g.,][]{Mo1998,Mao1998,Dutton2011,Dutton2012}.  Empirical
scaling relations between the rotation velocity and the luminosity or
stellar mass, i.e., the luminous or stellar mass Tully--Fisher (TF)
relation \citep{Tully1977,Reyes2011,McGaugh2015,Straatman2017},
between the rotation velocity and the baryonic mass, i.e., the
baryonic TF relation
\citep{Walker1999,Zaritsky2014,Lelli2016,Bradford2016,Ubler2017}, and
between the rotation velocity and the size
\citep{Tully1977,Karachentsev2013,Bohm2016,Schulz2017} for disc
galaxies at the present epoch and at different redshifts are
established in the above quoted and many other papers.  To reduce the
scatter in the relation, a second parameter beyond the rotation
velocity is sometimes added, i.e., a correlation between TF residuals
and an additional parameter is sought.

Is the chemical evolution of galaxies also governed mainly by the
dynamical mass (or the host dark matter halo)?  On the one hand, it is
established that the characteristic oxygen abundance of a galaxy
correlates with its rotation velocity, stellar mass, luminosity, and
size \citep[][among many others]{Lequeux1979,Zaritsky1994, Garnett2002,
Grebel2003, Pilyugin2004, Tremonti2004, Erb2006, Cowie2008, Maiolino2008,
Guseva2009, Thuan2010, Andrews2013, Zahid2013, Maier2014, Steidel2014,
Pilyugin2014b, Izotov2015,Sanchez2017,Barrera2017}.
On the other hand, the relation of the radial oxygen abundance distribution
(gradient) with rotation velocity and other basic parameters of a
galaxy is still not well explored.  The conclusions on the correlation
between the metallicity gradient and basic galaxy parameters reached
in recent studies are somewhat controversial.

The radial oxygen abundance distributions across the galaxies measured
by the Calar Alto Legacy Integral Field Area (CALIFA) survey
\citep{Sanchez2012a, Husemann2013, GarciaBenito2015} were investigated
by \citet{Sanchez2012b,Sanchez2014} and \citet{SanchezMenguiano2016}.
These authors found that all galaxies without clear evidence of an
interaction present a common gradient in the oxygen abundance
expressed in terms of dex/$R_{\rm e,d}$, where $R_{\rm e,d}$ is
the disc's effective radius.  The slope is independent of morphology,
the extistence of a bar, absolute magnitude, or mass.  The
distribution of slopes is statistically compatible with a random
Gaussian distribution around the mean value.  This conclusion was
confirmed by the results based on Multi-Unit Spectroscopic Explorer
(MUSE) integral field spectrograph observations of spiral galaxies
\citep{SanchezMenguiano2018}.  

\citet{Ho2015} determined metallicity gradients in 49 local field
star-forming galaxies.  They found that when the metallicity gradients
are expressed in dex/$R_{25}$ ($R_{25}$ is the $B$-band isophotal
radius at a surface brightness of 25 mag/sec$^{2}$) then there is no
correlation between the metallicity gradient and the stellar mass and
luminosity. When the metallicity gradients are expressed in dex/kpc
then galaxies with lower mass and luminosity, on average,
have steeper metallicity gradients.

\citet{Belfiore2017} determined the oxygen abundance gradients in a
sample of 550 nearby galaxies in the stellar mass range of
10$^{9}$~M$_{\odot}$ --  10$^{11.5}$~M$_{\odot}$ with spectroscopic
data from the Mapping Nearby Galaxies at Apache Point Observatory
(SDSS-IV MaNGA) survey \citep{Bundy2015}. They found that the gradient
in terms of dex/$R_{\rm e,d}$ steepens with stellar mass until
$\sim 10^{10.5}$~M$_{\odot}$ and remains roughly constant for higher
masses. The gradient in terms of dex/$R_{\rm e,g}$ (where $R_{\rm e,g}$
is the galaxy effective radius) steepens with stellar mass until
$\sim 10^{10.5}$~M$_{\odot}$ and then flattens slightly for higher
masses. The gradient in the terms of dex/kpc steepens with
stellar mass until $\sim 10^{10.5}$~M$_{\odot}$ and then becomes
flatter again.  

Thus, differing conclusions on the relation between the metallicity
gradient and basic parameters were reached in the studies quoted
above.

Our current investigation is motivated by the following deliberations.
The oxygen abundances in the quoted papers were estimated through 1-D
calibrations.  We have demonstrated in a previous study that the 1-D
N2 calibration produces either a reliable or a wrong abundance
depending on whether the excitation and the N/O abundances ratio in
the target region is close to or differs from those parameters in the
calibrating points \citep[][hereafter Paper I]{Pilyugin2018}.
\citet{Ho2015} also noted that the radial change of oxygen abundances
derived with two different 1-D calibrations can differ significantly
when the ionization parameters change systematically with radius.  We
also demonstrated that the 3-D $R$ calibration from
\citet{Pilyugin2016} produces quite reliable abundances in the MaNGA
galaxies.

Here we derive rotation curves, photometric profiles, and
$R$-calibration-based abundances for 147 MaNGA galaxies.  We use the
obtained rotation velocities, abundances, and photometric parameters
coupled with published stellar masses to examine the correlations
between chemical properties (oxygen abundances at the centres and at
optical radii, and radial gradients) with the rotation velocity and
other basic parameters (stellar mass, luminosity, radius) of our
target galaxies.

The paper is organized in the following way. The data are described in
Section 2. In Section 3 the relations between the characteristics of
our galaxies are examined. Section 4 includes the discussion of our
results, and Section 5 contains a brief summary.

Throughout the paper, we will use the rotation velocity $V_{rot}$ in
the units of km/s, the photometric ($M_{ph}$) and spectroscopic
($M_{sp}$) stellar masses of a galaxy in the units of $M_{\odot}$, and
the luminosity $L_{B}$ in the units of $L_{\odot}$.   
A cosmological model with $H_{0}$ = 73 km/s/Mpc and
$\Omega_{m}$ = 0.27 and $\Omega_{\Lambda}$ = 0.73 is adopted.

\section{Data}

\subsection{Preliminary remarks}

The spectroscopic data from the SDSS-IV MaNGA survey provide the
possibility $i)$ to measure the observed velocity fields and derive 
rotation curves for a large sample of galaxies, $ii)$ to measure
surface brightnesses and derive optical radii and luminosities, and
$iii)$ to measure emission line fluxes and derive abundance maps.  

The publicly available data obtained in the framework of the MaNGA
survey are the basis of our current study.  We selected our initial
sample of star-forming galaxies (600 galaxies) by visual examination of the
images of the SDSS-IV  MaNGA DR13 galaxies. 
The spectrum of each spaxel from the MaNGA datacubes was
reduced in the manner described in \citet{Zinchenko2016}. For each
spectrum, the fluxes of the 
[O\,{\sc ii}]$\lambda$3727+$\lambda$3729, 
H$\beta$,  
[O\,{\sc iii}]$\lambda$4959, 
[O\,{\sc iii}]$\lambda$5007,
[N\,{\sc ii}]$\lambda$6548,
H$\alpha$,  
[N\,{\sc ii}]$\lambda$6584, 
[S\,{\sc ii}]$\lambda$6717, 
[S\,{\sc ii}]$\lambda$6731 lines were measured. The determinations of
the photometrical properties of the MaNGA galaxies is described in Paper I. 
The determinations of the abundances and the rotation curves
is described below.  

The SDSS data base offers values of the stellar masses of galaxies
determined in different ways.  We have chosen the photometric
$M_{ph}$ and spectroscopic $M_{sp}$ masses of the SDSS and BOSS galaxies
(BOSS stands for the Baryon Oscillation Spectroscopic Survey in
SDSS-III, see \citet{Dawson13}).  The photometric masses, $M_{ph}$, are the
best-fit stellar masses from database table {\sc
stellarMassStarformingPort} obtained by the Portsmouth method, which
fits stellar evolution models to the SDSS photometry
\citep{Maraston2009,Maraston2013}.  The spectroscopic masses, $M_{sp}$, are the median (50th
percentile of the probability distribution function, PDF) of the
logarithmic stellar masses from table {\sc stellarMassPCAWiscBC03}
determined by the Wisconsin method \citep{Chen2012} with the stellar
population synthesis models from \citet{Bruzual2003}.
The NSA (NASA-Sloan Atlas) data base also offers values of photometric stellar
masses of the MaNGA galaxies, $M_{ma}$.
The NSA stellar masses are derived using a \citet{Chabrier2003} initial mass function
with \citet{Bruzual2003} simple stellar population models. 
Stellar masses $M_{ma}$ are taken from
the extended NSA targeting catalogue {\sc  http://nsatlas.org/}.

\subsection{Rotation curve and representative value of the rotation velocity}

\subsubsection{Approach}

\begin{figure}
\resizebox{1.00\hsize}{!}{\includegraphics[angle=000]{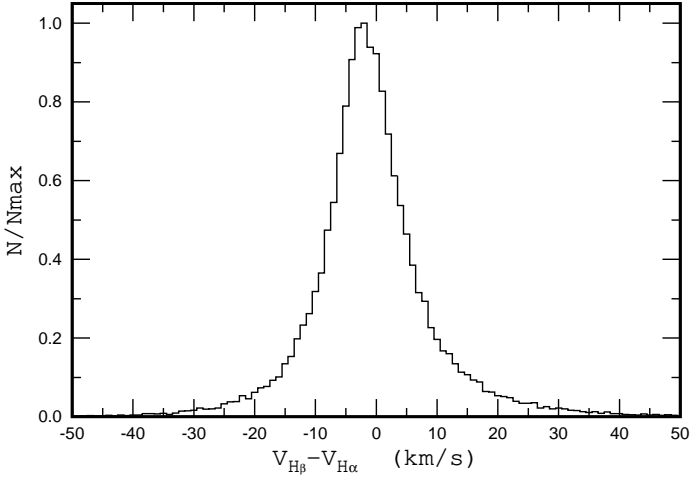}}
\caption{
  The normalized histogram of the differences $V_{\rm H\beta}$--
$V_{\rm H\alpha}$ between the line-of-sight velocities measured for
H$\beta$ and H$\alpha$ emission lines in 46,350 regions (spaxels) in
MaNGA galaxies \citep{Pilyugin2018}.
}
\label{figure:gist-dvhab}
\end{figure}

\begin{figure*}
\resizebox{1.00\hsize}{!}{\includegraphics[angle=000]{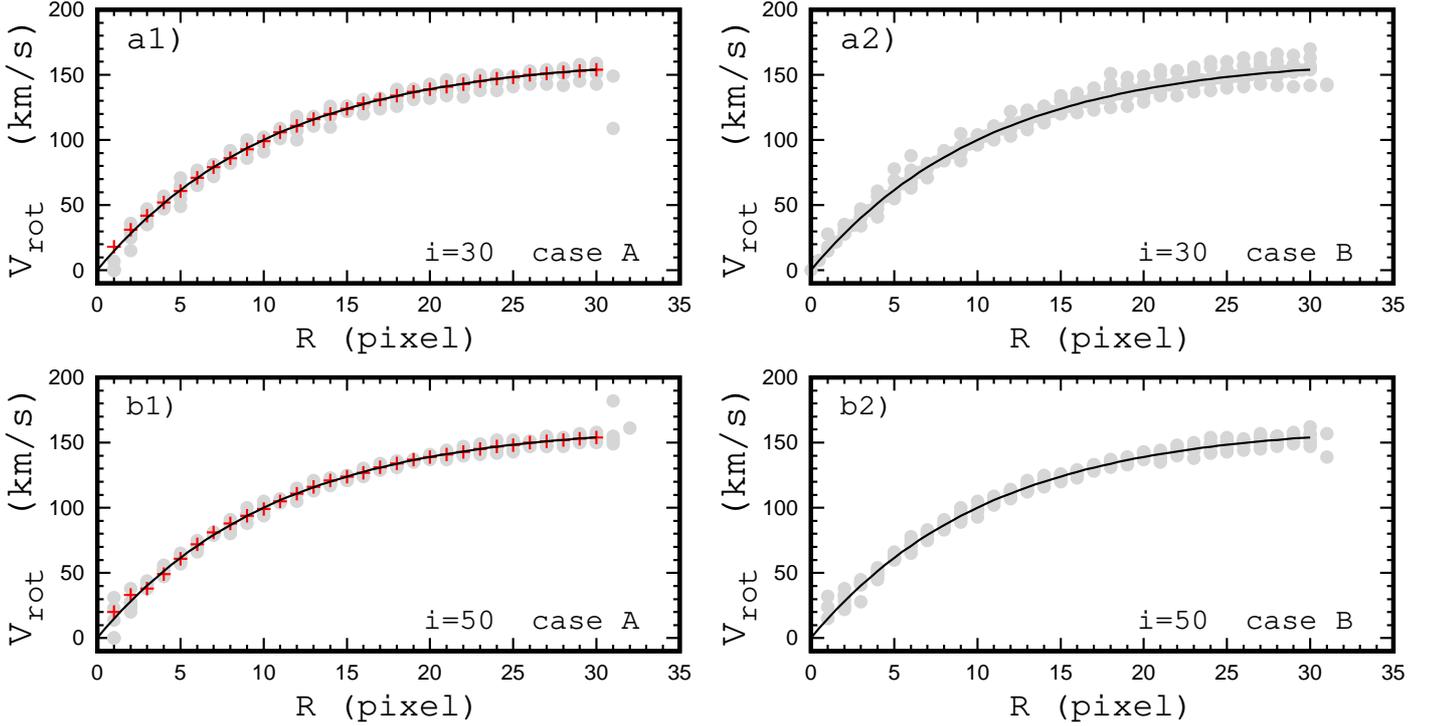}}
\caption{
  {\em Panel} $a1$ shows the rotation curves for the model of a galaxy
with an inclination angle $i = 30\degr$ obtained for the case $A$.
The solid line is the model rotation curve.  The rotation curve derived from
the modelled map of the line-of-sight velocities without errors is
shown by plus signs.  Six rotation curves derived from the modelled
maps of the line-of-sight velocities with random errors are indicated
by points.  {\em Panel} $a2$ shows the same as {\em panel} $a1$ but
for the case $B$.  {\em Panels} $b1$ and $b2$ show the same as {\em
panels} $a1$ and $a2$ but for an inclination angle of $i = 50\degr$.
}
\label{figure:rc-model}
\end{figure*}

\begin{figure}
\resizebox{1.00\hsize}{!}{\includegraphics[angle=000]{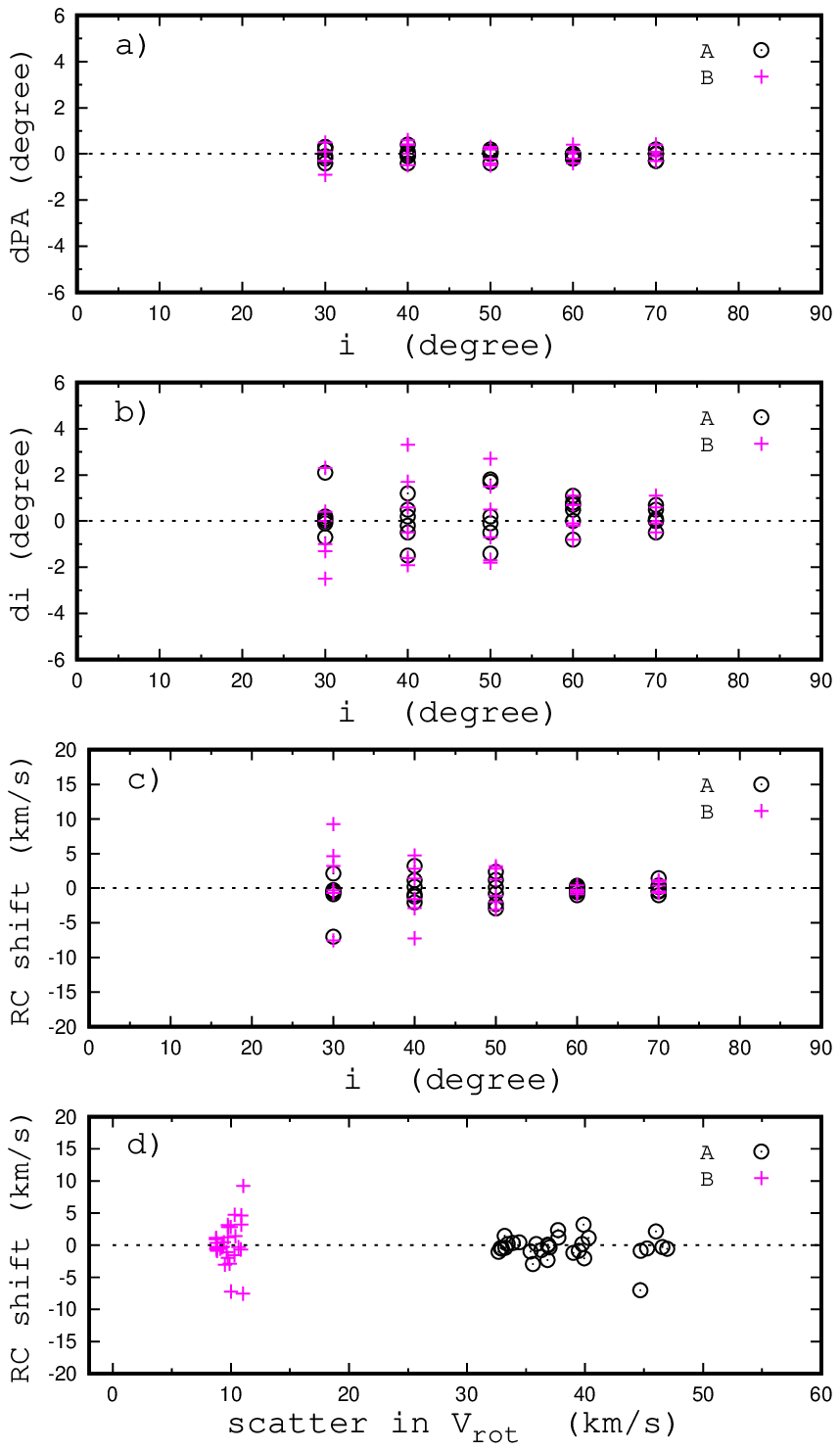}}
\caption{
  {\em Panel} $a$ shows the difference between the obtained and the
true position angle of the major axis $d(PA)$ as a function of
inclination angle for our set of models.  {\em Panel} $b$ shows the
difference between the obtained and the true inclination angle $di$ as
a function of inclination angle.  {\em Panel} $c$ shows the mean shift
of the obtained rotation curve relative to the true one as a function of
the inclination angle.  {\em Panel} $d$ shows the mean shift of the
obtained rotation curve as a function of the value of the scatter of
the spaxel rotation velocities around the obtained rotation curve.
The cases $A$ and $B$ are shown by circles and plus signs,
respectively, in each panel.
}
\label{figure:i-shift-model}
\end{figure}

The measured wavelength of the emission lines provides the velocity of
each region (spaxel). The emission line profile is fitted by a
Gaussian, and the position of the centre of the Gaussian is adopted as
the wavelength of the emission line. We measured the wavelengths of
the H$\alpha$ and H$\beta$ emission lines. We analysed the velocity
map obtained from the H$\alpha$ emission line measurements in our
current study because this stronger line can usually be measured with
higher precision than the weaker H$\beta$ line. Thus one can expect
that the error in the measured wavelength of the H$\alpha$ emission
line $e(\lambda_{0})$ is lower than that for the H$\beta$ emission
line.  Moreover, the error in the measured wavelength of the H$\alpha$
line results in a lower error of the velocity
$e(\lambda_{0})$/$\lambda_{0}$ than the error in the measured
wavelength of the H$\beta$ line.  Therefore, we obtained the position
angle of the major axis of the galaxy, the galaxy's inclination angle,
and the rotation velocity from the analysis of the H$\alpha$ velocity
field.

The determination of the rotation velocity from the observed velocity
field is performed in the standard way
\citep[e.g.,][]{Warner1973,Begeman1989,deBlok2008,Oh2018}.  The
parameters that define the observed velocity field of a galaxy with a
symmetrically rotating disc are the following: \\
-- The pixel coordinates ($x_{0}$ and $y_{0}$) of the rotation centre 
of the galaxy. \\
-- The velocity of the centre of the galaxy with respect to the Sun, 
the so-called systemic velocity $V_{sys}$. \\ 
-- The circular velocity $V_{rot}$ at the distance $R$ from the
galaxy's centre. \\ 
-- The position angle $PA$ of the major axis. \\
-- The angle $i$ between the normal to the plane of the galaxy and the 
line of sight, the inclination angle. \\ 
The observed line-of-sight velocities $V_{los}$ recorded on a set of 
pixel coordinates ($x,y$) are related to the kinematic parameters by:
\begin{equation} 
V_{los}(x,y) = V_{sys} + V_{rot}\,\cos(\theta) \, \sin(i)
\label{equation:vxy} 
\end{equation} 
where
\begin{equation}
\cos(\theta)   =   \frac{-(x-x_{0})\, \sin(PA) + (y-y_{0}) \, \cos(PA)}{R} 
\label{equation:costeta}
\end{equation}
\begin{equation}
\sin(\theta)   =   \frac{-(x-x_{0})\, \cos(PA) - (y-y_{0}) \, \sin(PA)}{R \cos(i)} 
\label{equation:sinteta}
\end{equation}
where $R$
\begin{eqnarray}
       \begin{array}{lll}
 R & = & [ \{ - (x-x_0) \sin(PA) + (y-y_0) \cos(PA) \}^{2}  \\  
 & + &  \{((x-x_0) \cos(PA) + (y-y_0) \sin(PA))/\cos i\} ^{2} ]^{1/2}
\end{array}
\label{equation:rdisc}
\end{eqnarray}
is the radius of a ring in the plane of the galaxy 
\citep{Warner1973,Begeman1989,deBlok2008,Oh2018}. 

The deprojected galaxy plane is divided into rings with a width of
1 in pixel units.  The rotation velocity is assumed to be the same
for all the points within the ring.  The position angle of the major
axis and the galaxy inclination angle are assumed to be the same for
all the rings, i.e., constant within the disc.  The parameters $x_{0}$,
$y_{0}$, $V_{sys}$, $PA$, $i$, and the rotation curve  $V_{rot}(R)$
are derived through the best fit to the observed velocity field
$V_{los}(x,y)$, i.e., we require that the deviation $\sigma_{V_{los}}$
given by
\begin{equation}
\sigma_{V_{los}} = \sqrt{ [\sum\limits_{j=1}^n (V_{los,j}^{cal} - V_{los,j}^{obs})^{2}]/n}  
\label{equation:sigmavlos}
\end{equation}
is minimized. Here the $V_{los,j}^{obs}$ is the measured line-of-sight 
velocity of the $j$-th spaxel and $V_{los,j}^{cal}$ is the velocity 
computed through Eq.~(\ref{equation:vxy}) for the corresponding pixel 
coordinates $x$ and $y$. 

We use two different approaches to derive rotation curves for each
galaxy.  In the first approach, the values of $x_{0}$, $y_{0}$,
$V_{sys}$, $PA$, $i$, and the rotation curve  $V_{rot}(R)$ are derived
using all the spaxels with measured $V_{los}$. We will refer to this
method as the case $A$.

In the second approach, we use an iterative procedure to determine the
rotation curve.  At each step, data points with large deviations from
the rotation curve obtained in the previous step are rejected, and new
values of the parameters $x_{0}$, $y_{0}$, $V_{sys}$, $PA$, $i$, and
of the rotation curve  $V_{rot}(R)$ are derived.  In the first step,
the values of $x_{0}$, $y_{0}$, $V_{sys}$, $PA$, $i$, and the
rotation curve determined via case $A$ are adopted as values of the
previous step. The chosen value of $d(V_{rot})_{max}$ is the adopted
value of the criterion of the reliability of the spaxel rotation
velocity and is used to select data points with reliable rotation
velocity $V_{rot}$, i.e., spaxels where the deviation of the rotation
velocity from the rotation curve exceeds $d(V_{rot})_{max}$ are
rejected.  A ring with a width of 1 pixel usually contains several
dozens data points; however, the number of data points in the rings
near the centre and at the very periphery of the galaxy can be small.
If the number of data points in the ring is lower than 3 then this
ring is excluded from consideration.  The iteration is stopped when
the absolute values of the difference of $x_{0}$ (and $y_{0}$)
obtained in successive steps is less than 0.1 pixels, the difference
of $PA$ (and $i$) is smaller than 0.1$\degr$, and the rotation curves
agree within 1 km/s (at each radius).  The iteration converges
after up to 10 steps.  We will refer to this approach as the case $B$.

\subsubsection{Estimation of the uncertainty of the rotation curve
  caused by errors in the velocity measurements}

\begin{figure*}
\resizebox{1.00\hsize}{!}{\includegraphics[angle=000]{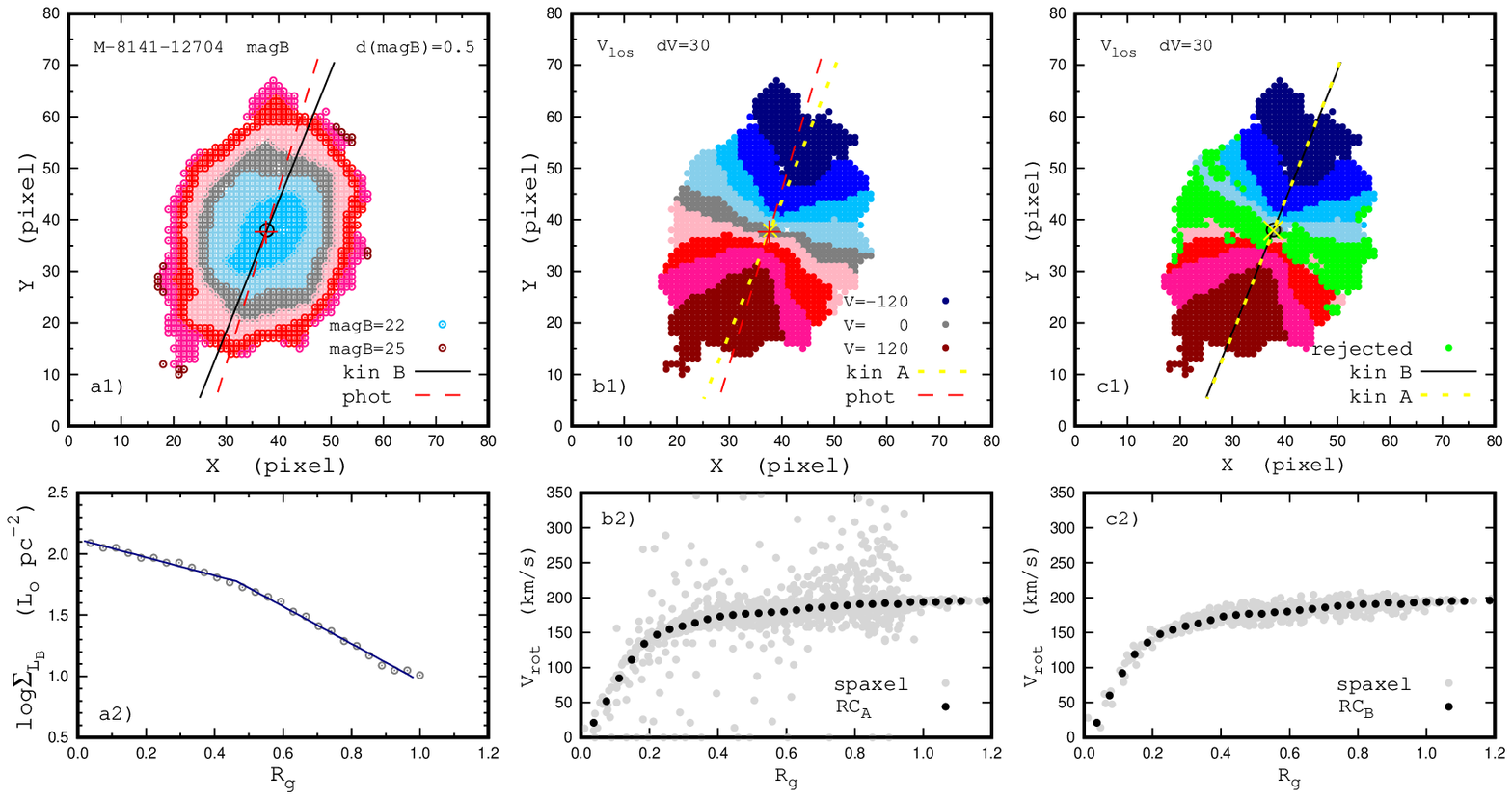}}
\caption{
  {\em Panel} $a1$ shows the surface brightness distribution across
the image of the galaxy M-8141-12704 in sky coordinates (pixels).  The
value of the surface brightness is colour-coded with a stepsize of
d(magB) = 0.5.  The plus sign marks our derived photometric centre of
the galaxy. The dashed line is the major photometric axis of the
galaxy.  The circle denotes the inferred kinematic centre of the
galaxy. The solid line indicates the major kinematic axis of the
galaxy for the case $B$.  {\em Panel} $a2$ shows the derived surface
brightness profile (points, where the surface brightness is given in
units of $L_{\sun}$/pc$^{2}$ as a function of the fractional optical
radius $R_{g}$ = $R/R_{25}$) and the fit within the isophotal radius
with a S\'{e}rsic profile for the bulge and a broken exponential for
the disc (line). The surface brightness profile was obtained with
kinematic (case $B$) geometric parameters of the galaxy.  {\em Panel}
$b1$ shows the observed velocity field in pixel coordinates.  The
value of the observed  velocity is colour-coded with a step size of 30
km/s.  The positions of the photometric centre and the major
axis come from {\em panel} $a1$.  The cross marks the kinematic centre
of the galaxy.  The dotted line shows the major kinematic axis of the
galaxy for the case $A$.  {\em Panel} $b2$ shows the rotation curve
derived for case $A$.  The grey points stand for the obtained rotation
velocities of the individual spaxels; the dark points are the mean
values of the rotation velocities in rings of a width of 1 pixel.
{\em Panel} $c1$ shows the observed velocity field in pixel
coordinates as in {\em panel} $b1$ but the spaxels rejected in the
determination of the rotation curve  for the case $B$ are shown by the
green colour.  The circle indicates the kinematic centre of the
galaxy, and the solid line is the major kinematic axis of the galaxy
for case $B$.  The cross is the kinematic centre and the dotted line
is the major kinematic axis of the galaxy for the case $A$.  {\em
Panel} $c2$ shows the rotation curve derived for case $B$.  The grey
points are the measured rotation velocities for individual spaxels;
the dark points indicate the mean values of the rotation velocities in
rings of a width of 1 pixel.
}
\label{figure:rc-ab}
\end{figure*}

\begin{figure*}
\resizebox{1.00\hsize}{!}{\includegraphics[angle=000]{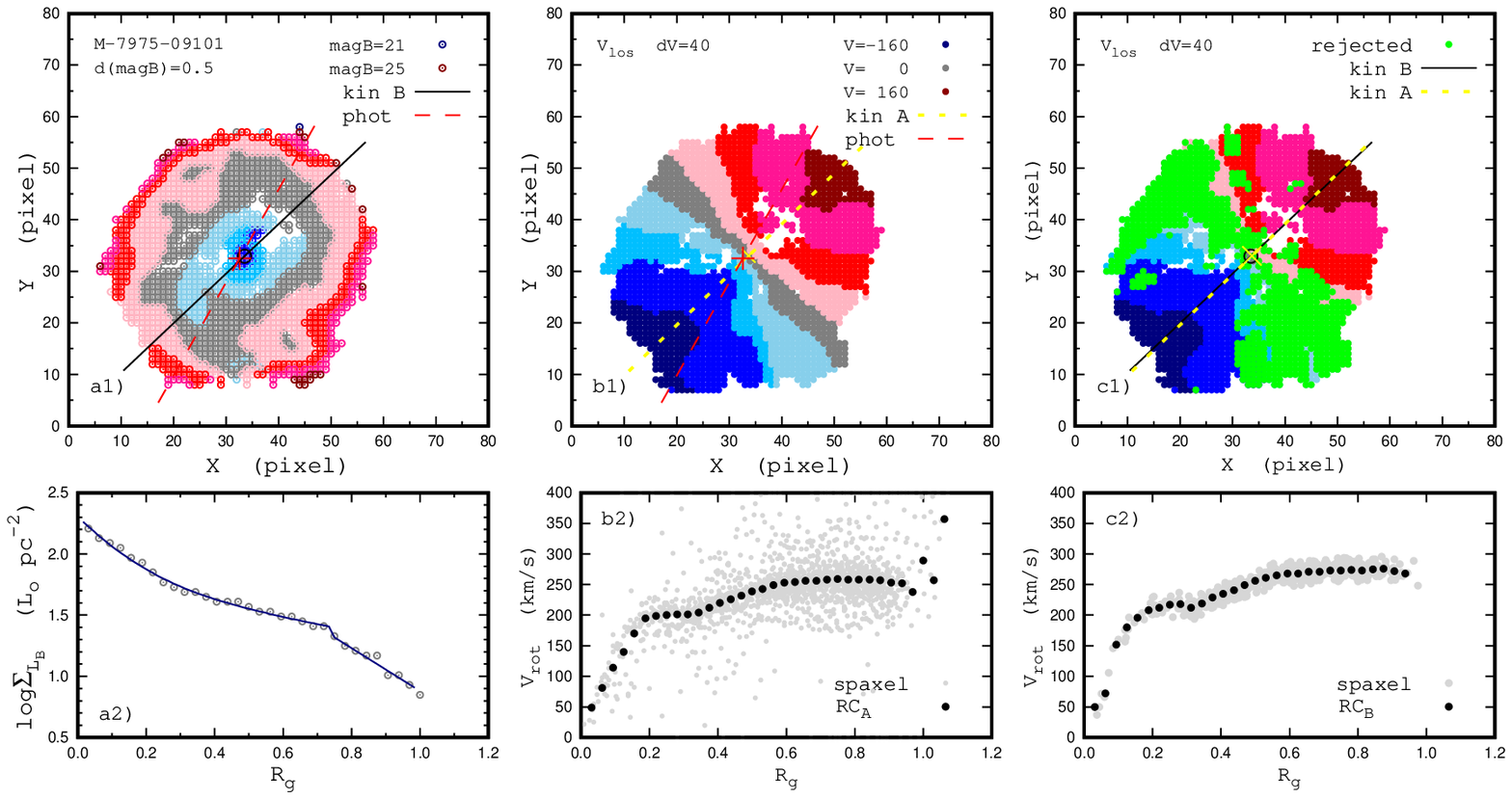}}
\caption{
  The same as Fig.~\ref{figure:rc-ab} but for the galaxy M-7975-09101.
}
\label{figure:rc-ab-2}
\end{figure*}

The errors in the measured line-of-sight velocities $V_{los}$ of the
spaxels can distort the derived rotation curve. We expect that the
validity of the obtained rotation curve may depend on the value of the
errors in the line-of-sight velocities and galaxy inclination.

The difference between the values of the spaxel velocity obtained from
the measured wavelengths of the H$\beta$ and H$\alpha$ lines
$V_{los,{\rm H\beta}}$-- $V_{los,{\rm H\alpha}}$ can be considered as
some kind of estimation of the error in the spaxel velocity
measurements in the MaNGA spectra. Fig.~\ref{figure:gist-dvhab} shows
the normalized histogram of the differences $V_{los,{\rm H\beta}}$--
$V_{los,{\rm H\alpha}}$  between the measured  H$\beta$ and H$\alpha$
velocities in 46,350  regions (spaxels) in the MaNGA galaxies from
\citet{Pilyugin2018}. The mean value of the differences is $\sim -0.8$
km/s.  The mean value of the absolute differences is $\sim 7$
km/s.  Thus we assume that the mean value of the errors in the
spaxel velocities in the MaNGA galaxies is near 7 km/s.

To examine the influence of the errors in the measured line-of-sight
spaxel velocities $V_{los}$ on the obtained rotation curve we
construct  a set of toy models. We fixed the rotation curve and the
position angle of the major axis (using $PA$ = 45$\degr$). The fields
of line-of-sight velocities $V_{los}$ were constructed for five values
of the inclination angle $i = 30\degr$, $40\degr$,  $50\degr$,
$60\degr$, and $70\degr$ using Eq.~(\ref{equation:vxy}) to
Eq.~(\ref{equation:rdisc}).  Then we introduced errors in the
line-of-sight velocity $V_{los}$ of each spaxel.  These errors were
randomly chosen from a set of errors that follow a Gaussian
distribution with $\sigma = 7$. 

In case $A$, the rotation curve RC$_{A}$ is determined from each
modelled field of line-of-sight velocities using all the data points.
In case $B$, the rotation curve RC$_{B}$ is determined from each
modelled set of line-of-sight velocities using the iterative procedure
described above.  If the deviation of the spaxel rotation velocity
from the rotation curve of the previous iteration exceeds a fixed
value then this spaxel is rejected.  It should be noted that the error
in the rotation velocity of the point (spaxel) $d(V_{rot})$ depends
not only on the error in the line-of-sight velocity $d(V_{los})$ but
also on the galaxy inclination $i$ and the position of the region
(spaxel) in the galaxy (angle $\theta$). Those errors are related by
the expression  $d(V_{rot}) = d(V_{los})/(\cos\theta \sin i)$ (see
Eq.~(\ref{equation:vxy})), i.e., the error in the $V_{rot}$ exceeds
the error in the $V_{los}$ by a factor of $1/(\cos \theta\sin i)$.  We
adopt the value of the criterion of the reliability of the spaxel
rotation velocity $d(V_{rot})_{max} = 21$ km/s, which exceeds
the mean random error in the line-of-sight velocity $V_{los}$
measurements by a factor of 3. 

We have considered 30 models containing six sets of random errors of
line-of-sight velocities $V_{los}$ for each of the five values of the
inclination angle. The models without errors in the line-of-sight
velocity $V_{los}$ measurements are also considered for comparison.

{\em Panel} $a1$ of Fig.~\ref{figure:rc-model} shows the rotation
curves for the model of a galaxy with an inclination angle $i =
30\degr$ obtained for case $A$.  The model rotation curve is shown by
the solid line, the rotation curve derived from the modelled map of
the line-of-sight velocities without errors by plus signs, and the six
rotation curves derived from the modelled map of the line-of-sight
velocities with random errors by points.  {\em Panel} $a2$ of
Fig.~\ref{figure:rc-model} shows the same as {\em panel} $a1$ but for
the case $B$ of the rotation curve determination.  {\em Panels} $b1$
and $b2$ show the same as {\em panels} $a1$ and $a2$ but for an
inclination angle of $i = 50\degr$.

The quality of each obtained rotation curve can be specified by the two
following characteristics. The first characteristic is the scatter
$\sigma_{V_{rot}}$ in the rotation velocities of individual spaxels
relative to the obtained rotation curve, which is given by the
expression 
\begin{equation}
\sigma_{V_{rot}} = \sqrt{ [\sum\limits_{j=1}^n (V_{rot,j}^{spaxel} - V_{rot,j}^{RC})^{2}]/n}  
\label{equation:sigmavrot}
\end{equation}
where $V_{rot,j}^{spaxel}$ is the rotation velocity obtained through 
Eq.~(\ref{equation:vxy}) to Eq.~(\ref{equation:rdisc}) for the $j$-th 
spaxel of the constructed field of line-of-sight velocities and 
$V_{rot,j}^{RC}$ is the rotation velocity given by the inferred 
rotation curve at the corresponding galactocentric distance.
The second characteristic is the mean shift of the obtained rotation 
curve in comparison to the true one, i.e. the mean value of the deviations 
$d(RC)$ of the derived rotation curve from the true one (used in the 
model construction), which is estimated through the expression 
\begin{equation}
d(RC) = \sqrt{ [\sum\limits_{j=1}^n (V_{rot,R_j}^{RC} - V_{rot,R_j}^{mod})^{2}]/n}  
\label{equation:dRC}
\end{equation}
where $V_{rot,R_j}^{RC}$ is the obtained rotation velocity at the 
radius $R_{j}$ and $V_{rot,R_j}^{mod}$ is the rotation velocity 
given by the true rotation curve at the corresponding galactocentric 
distance.

{\em Panel} $a$ of Fig.~\ref{figure:i-shift-model} shows the
difference between the obtained and true position angle of the major
axis as a function of inclination angle for our set of models.  {\em
Panel} $b$ shows the difference between the inferred and true
inclination angle as a function of inclination angle.  {\em Panel} $c$
shows the mean shift of the derived rotation curve relative to the
true one, Eq.~(\ref{equation:dRC}), as a function of inclination
angle.  {\em Panel} $d$ shows the mean shift of the obtained rotation
curve in comparison to the true one as a function of the value of the
scatter in the spaxel rotation velocities around the inferred rotation
curve, Eq.~(\ref{equation:sigmavrot}).  The cases $A$ and $B$ are
shown in each panel by the circles and the plus signs, respectively. 

Examination of Fig.~\ref{figure:rc-model} and Fig.~\ref{figure:i-shift-model}
results in the following conclusions. \\
-- The rotation curve derived for the case $A$ may involve data points
with large deviations.  \\
-- The mean shift of the obtained rotation curve relative to the true 
one does not exceed $\sim 5$ km/s for galaxies with an
inclination of $i \ga 40\degr$ and can reach up to $\sim 10$
km/s for galaxies with $i \sim 30\degr$ if the line-of-sight 
velocity measurements involve random errors with $\sigma_{V_{los}} = 7$
km/s. This is the mean value of the velocity measurement 
errors estimated from the comparison between the line-of-sight 
velocities measured for H$\beta$ and H$\alpha$ emission lines in 
46,350 regions (spaxels) in MaNGA galaxies as mentioned earlier. \\
--  There is no one-to-one correspondence between $d(RC)$ and 
$\sigma_{V_{rot}}$, i.e., the value of the scatter $\sigma_{V_{rot}}$ 
in the rotation velocities of individual spaxels around the obtained 
rotation curve is not a reliable indicator of the deviation $d(RC)$ 
of the inferred rotation curve from the true one.

Thus, a value of the $d(V_{rot})_{max} = 21$ km/s is adopted in
the determinations of the rotation curves of the target MaNGA
galaxies.  One can expect that the deviation $d(RC)$ of the obtained
rotation curve from the true rotation curve caused by errors in the
velocity measurements does not exceed 10 km/s for galaxies with
inclination $i \ga 30\degr$.

\subsubsection{Rotation curves of target MaNGA galaxies}

\begin{figure}
\resizebox{1.00\hsize}{!}{\includegraphics[angle=000]{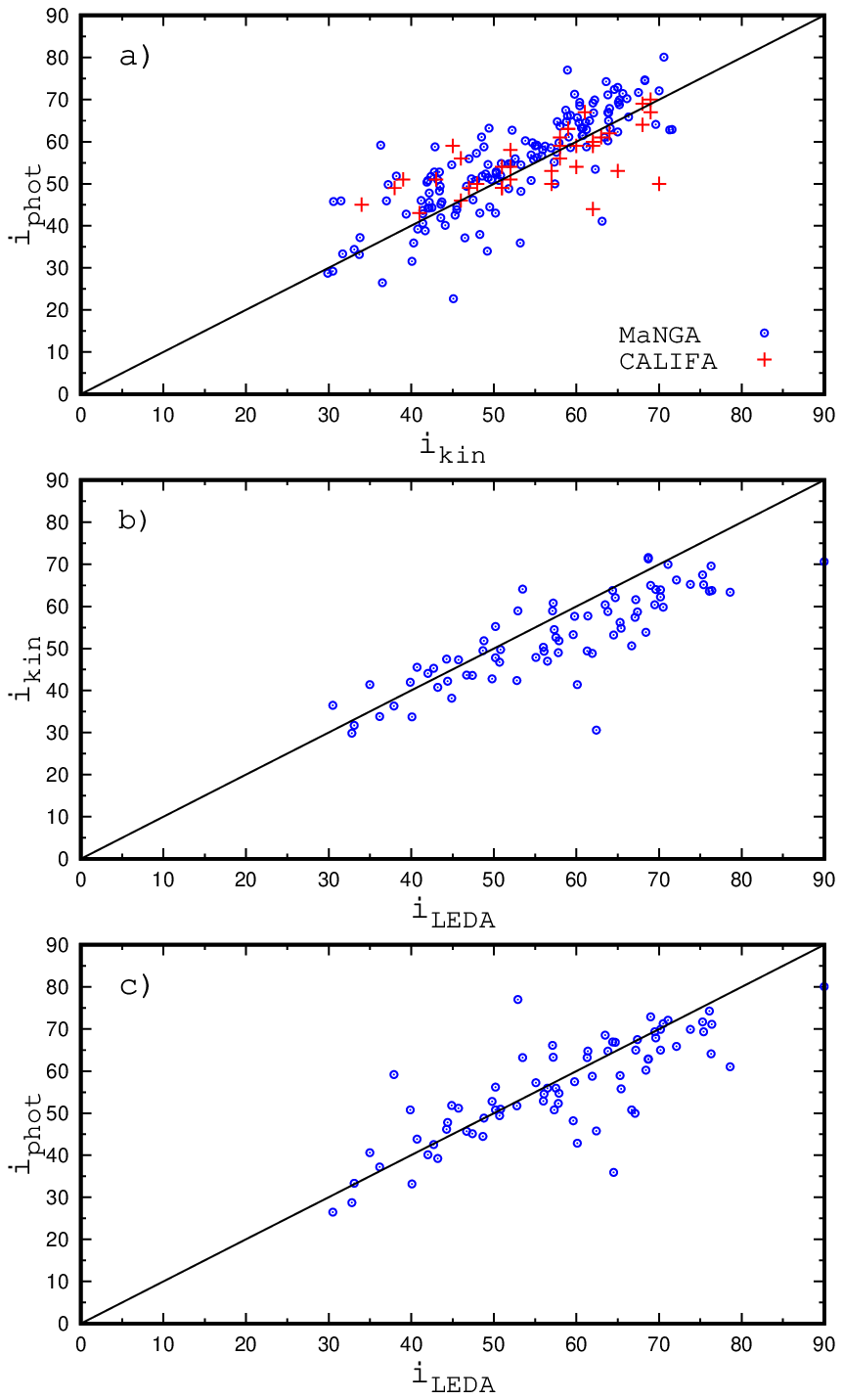}}
\caption{
  {\em Panel} $a$ shows the comparison between the values of the
inclination angle obtained from the analysis of the velocity fields,
$i_{kin}$, and from the analysis of the photometric maps, $i_{phot}$,
for our sample of MaNGA galaxies. The points stand for individual
galaxies.  The line indicates unity.  The plus signs indicate data for
CALIFA galaxies from \citet{Holmes2015}.  {\em Panel} $b$ shows the
comparison between the kinematic inclination angles derived here and
the inclination angles from the HyperLeda database.  {\em Panel} $c$
shows the comparison between the photometric inclination angles
derived here and the inclination angles from the HyperLeda database.
}
\label{figure:i-i}
\end{figure}

\begin{figure}
\resizebox{1.00\hsize}{!}{\includegraphics[angle=000]{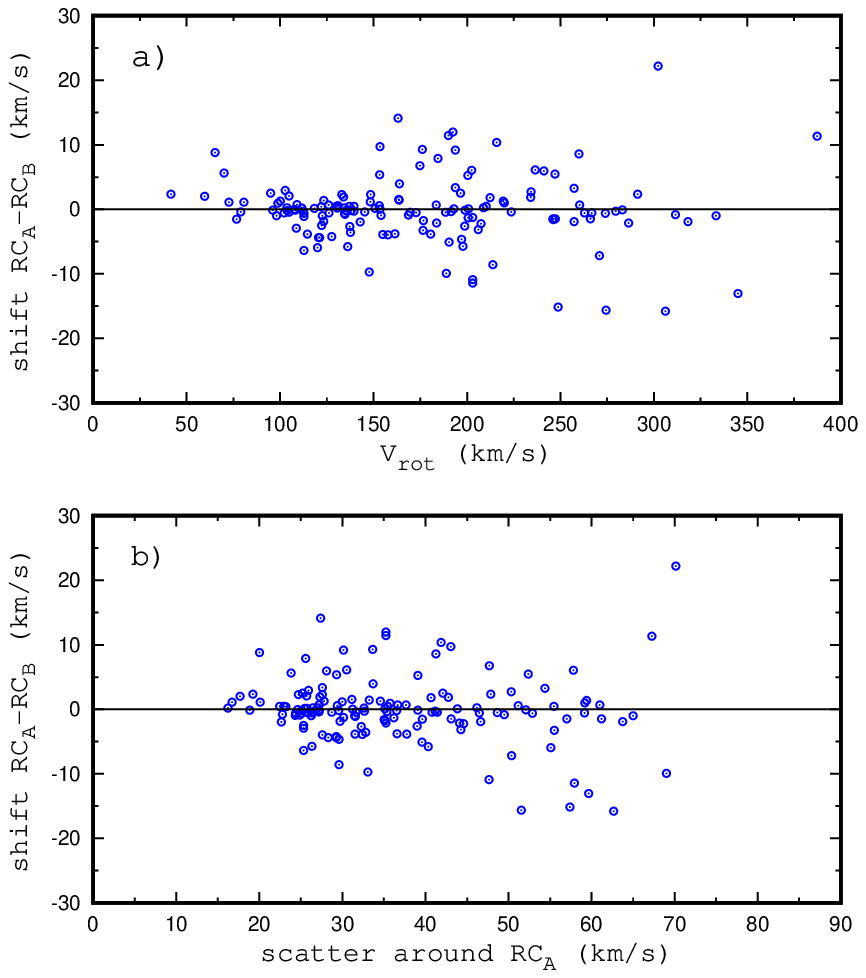}}
\caption{
The shift of the initial rotation curve (RC$_{A}$) relative to the
final rotation curve (RC$_{B}$) as a function of rotation velocity
$V_{rot}$ ({\em panel} $a$) and as a function of the scatter in the
spaxel velocities around the initial rotation curve ({\em panel} $b$)
for our sample of the MaNGA galaxies.
}
\label{figure:v-shiftab}
\end{figure}

We derive surface brightness profiles, rotation curves for cases $A$
and $B$, and abundance distributions for our target MaNGA galaxies.
Fig.~\ref{figure:rc-ab} shows the galaxy M-8141-12704, which is one of
those galaxies with very good agreement between the rotation curves
RC$_{A}$ and RC$_{B}$. The shift of RC$_{A}$ relative to RC$_{B}$ is
only $-0.5$ km/s.  Fig.~\ref{figure:rc-ab-2} shows the galaxy
M-7975-09101, which is an example of a galaxy with an appreciable
difference between RC$_{A}$ and RC$_{B}$; here the shift between
RC$_{A}$ and RC$_{B}$ is $-15.6$  km/s.
Fig.~\ref{figure:rc-ab} and Fig.~\ref{figure:rc-ab-2} show the surface
brightness distribution across the image of the galaxy in pixel
coordinates ({\em panel} $a1$), the derived surface brightness profile
(points) and the fit within the isophotal radius with a S\'{e}rsic
profile for the bulge and a broken exponential for the disc (solid
line) ({\em panel} $a2$), the observed velocity $V_{los}$ field in
pixel coordinates ({\em panel} $b1$), the rotation curve derived for
case $A$ ({\em panel} $b2$), the spaxels rejected in the determination
of the rotation curve for case $B$ ({\em panel} $c1$), and the
rotation curve derived for case $B$ ({\em panel} $c2$).  A prominent
feature of both Fig.~\ref{figure:rc-ab}  and Fig.~\ref{figure:rc-ab-2}
is that the spaxels rejected in the determination of the rotation
curve for the case $B$ are usually located near the minor kinematic
axis (the rotation axis) of the galaxies since even a small error in
the line-of-sight velocity $V_{los}$ measurement results in a large
error in the rotation velocity  $V_{rot}$.  

We selected the final sample of the MaNGA galaxies to be analyzed by
visual inspection of the derived surface brightness profile, rotation
curve, and abundance distribution for each galaxy. We used the
following criteria: \\
-- Spaxels with measured emission lines and surface brightness need
to be well distributed across the galactic disks, covering more than 
$\sim 0.8 R_{25}$. 
Those conditions provide the possibility to estimate reliable 
intersect values of the rotation velocity, the surface brightness, 
and the oxygen abundance both at the centre and at the optical radius 
since the extrapolation is relatively small (if any).
It should be noted that only the spaxel spectra where all the
used lines are measured with S/N $>$ 3 were considered.  Therefore the
spaxels with the reliable measured spectra can cover less than
$\sim 0.8 R_{25}$ even if the spaxel spectra beyond 0.8 $R_{25}$
are available.
151 galaxies were rejected according to this criterion. \\
-- The rotation curve should be more or less smooth. If the
obtained rotation velocities show a large irregular variation
at some galactocentric distances then that galaxy was excluded from 
consideration.
The presence of irregular variations in the rotation velocities
implies that either the quality of the spectra is not good enough
or the rotation is distorted. This prevents an estimation of a
reliable rotation curve. 
The geometric angles were poorly determined in a number of galaxies,
i.e., the kinematic angles differ significantly from the photometric angles.
\citet{Barrera2014,Barrera2015} have found that disagreement between
the kinematic and photometric position angles can be
caused by interactions. 
A galaxy was also excluded from consideration if the values of the
inclination angle obtained from the analysis of the velocity field and
from the analysis of the photometric map differed by more than 12$\degr$. 
In several cases, one can see clearly that the spiral arms
(or bright star formation regions) influence the derived values of
the photometric angles. Such galaxies were not rejected even if the
disagreement between the kinematic and photometric angles is large.
The rotation curves or/and kinematic angles were poorly determined
in 157 galaxies. Those galaxies are excluded from further consideration. \\
-- Galaxies with an inclination angle less than $\sim 30 \degr$ were 
rejected since even a small error in the inclination angle can result 
in a large error in the rotation velocity for nearly face-on galaxies.
Such small values of the inclination angle were obtained in 84
  galaxies. \\
-- Galaxies with an inclination angle larger than $\sim 70 \degr$ were 
also rejected (61 galaxies). The fit of the H$\alpha$ velocity field in galaxies 
with a large ratio of the major to minor axis can produce unrealistic 
values of the inclination angle. On the one hand, this may be caused 
by the problem of the determination of the rotation curves in galaxies 
with large inclination angles from the H$\alpha$ velocity field as was 
noted and discussed by \citet{Epinat2008}. 
On the other hand, a seemingly high inclination angle may instead
imply that some of those galaxies are not galaxies with thin discs.
Since our sample includes low-mass galaxies (the stellar masses of 
our target galaxies lie in the range from $\sim 10^{9}{\rm M}_{\odot}$ 
to  $\sim 10^{10.5}{\rm M}_{\odot}$) some of them may be irregular 
galaxies. The intrinsic (3-dimensional) shapes of irregular galaxies 
have been subject for discussion for a long time
\citep[][among others]{Hubble1926,HodgeHitchcock1966,vandenBergh1988,Roychowdhury2013,Johnson2017}. 
It is usually assumed that the more massive irregular galalaxies may 
be considered (at least as the first-order approximation) as a disc. 
\citet{Roychowdhury2013} found that the intrinsic shapes of irregular 
galaxies change systematically with luminosity, with fainter galaxies
being thicker. The most luminous irregular galaxies have thin discs, 
and these discs tend to be slightly elliptical (axial ratio $\sim 0.8$). 

Our final sample of galaxies selected using the above criteria
comprises 147 MaNGA galaxies out of 600 considered galaxies.
The selected galaxies are listed in Table \ref{table:sample}.

{\em Panel} $a$ of Fig.~\ref{figure:i-i} shows the comparison between
the values of the inclination angle obtained from the analysis of the
velocity fields, $i_{kin}$, and from the analysis of the pohometric
maps, $i_{phot}$. The points stand for the individual galaxies. The
line indicates equal values. For comparison, the data for the CALIFA
galaxies from \citet{Holmes2015} are presented by plus signs.  Our
$i_{kin}$ -- $i_{phot}$ diagram for the MaNGA galaxies is similar to
the diagram for the CALIFA galaxies from \citet{Holmes2015}.  {\em Panel}
$b$ of Fig.~\ref{figure:i-i} shows the comparison between the
kinematic inclination angles derived here and the inclination angles
from the HyperLeda\footnote{http://leda.univ-lyon1.fr/} database
\citep{Paturel2003,Makarov2014}.  {\em Panel} $c$  shows the
comparison between the photometric inclination angles derived here and
the inclination angles from the HyperLeda data base.  Examination of
Fig.~\ref{figure:i-i} shows that the HyperLeda inclination angles are
in better agreement with our photometric inclination angles than with
the kinematic inclination angles.  The credibility of out kinematic
inclination angles will be tested below.

Fig.~\ref{figure:v-shiftab} shows the shift of the initial rotation
curve (RC$_{A}$) relative to the final rotation curve (RC$_{B}$) as a
function of rotation velocity $V_{rot}$ ({\em panel} $a$) and as a
function of the scatter in the spaxel velocities around the initial
rotation curve ({\em panel} $b$) for our sample of the MaNGA galaxies.
Close examination of  Fig.~\ref{figure:v-shiftab} shows that the
absolute value of the shift RC$_{A}$ -- RC$_{B}$ is lower that 5
km/s in the majority of our galaxies (112 galaxies out of 147)
and exceeds 10 km/s only in a few galaxies (12 out of 147).

\begin{figure}
\resizebox{1.00\hsize}{!}{\includegraphics[angle=000]{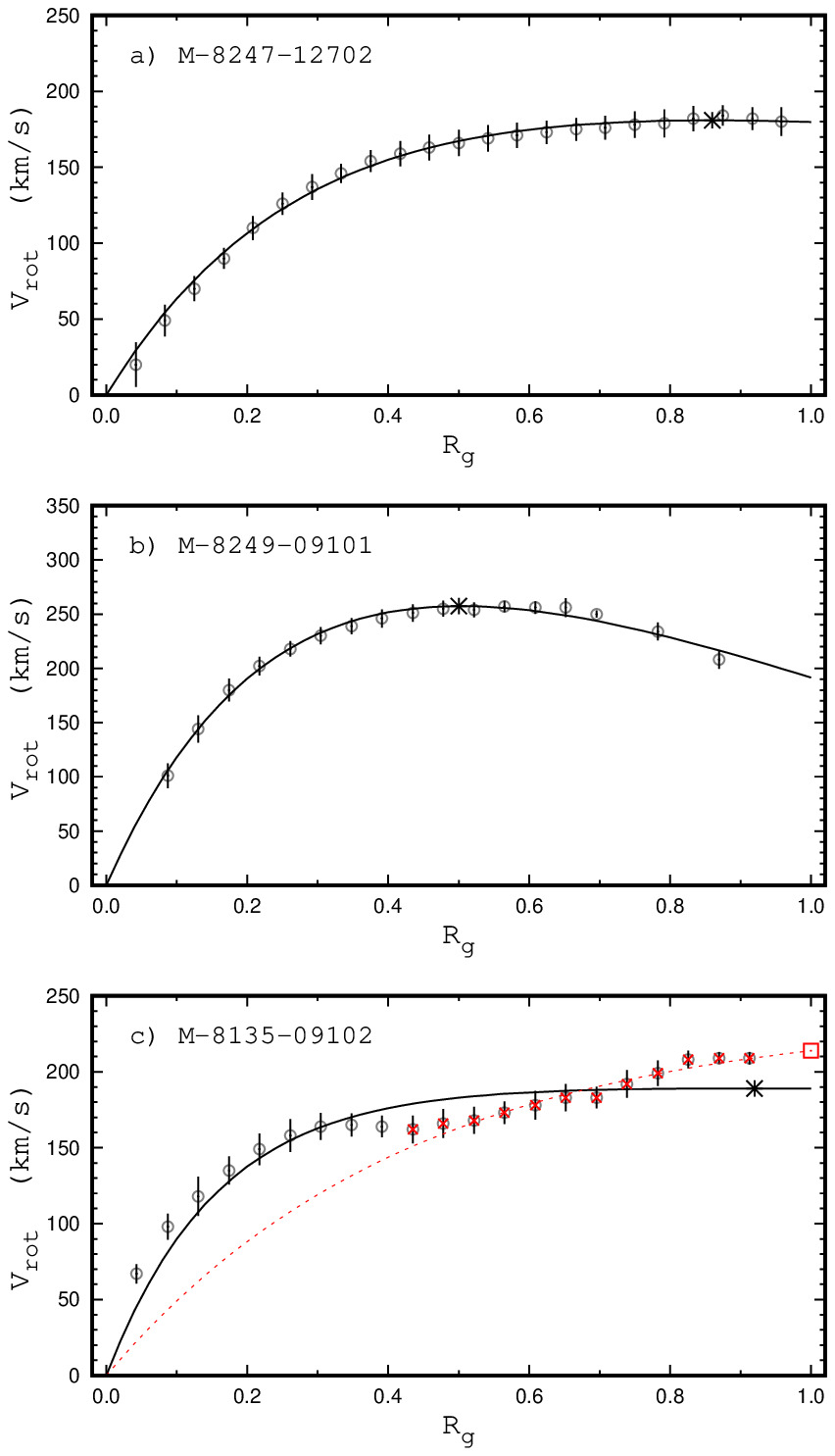}}
\caption{
  The obtained rotation curve and the fit with a Polyex curve for
three MaNGA galaxies are shown. $R_{g}$ is the fractional radius
(normalized to the optical radius $R_{25}$).  In each panel,
the obtained rotation curve is represented by grey circles.  The bars
indicate the mean scatter of the spaxel rotation velocities in rings
of 1 pixel width. The solid line is the fit with a Polyex curve.  The
asterisk is the maximum value of the rotation velocity in the rotation
curve fit within the optical radius.  In {\em panel} $c$, the Polyex
curve fit to the outer part of the rotation curve (points marked with
red crosses) is presented by a dashed line and the corresponding
maximum value of the rotation velocity is indicated by a square.
}
\label{figure:rc-fit}
\end{figure}

\subsubsection{The  representative value of the rotation velocity}

We carried out a least-squares fit of the rotation curve using an
empirical ``Polyex'' curve \citep{Giovanelli2002,Spekkens2005}:
\begin{equation}
V_{pe}(r)   =  V_{o} (1 - e^{-r/r_{pe}})(1+\beta \, r/r_{pe}) 
\label{equation:vpe}
\end{equation}
where $V_{0}$ sets the amplitude of the fit, $r_{pe}$ is a scale
length that governs the inner rotation curve slope, and $\beta$
determines the outer rotation curve slope.  The obtained rotation
curves for three MaNGA galaxies are shown by grey circles in
Fig.~\ref{figure:rc-fit}. The bars denote the mean scatter of the
spaxel rotation velocities in rings of 1 pixel width. The solid
line is the fit with the Polyex curve.  The asterisk indicates the
maximum value of the rotation velocity in the rotation curve fit
within the optical radius.  Examination of Fig.~\ref{figure:rc-fit}
shows that the Polyex curve is a good approximation for the rotation
curves with simple shapes ({\em panels} $a$ and $b$).  But the Polyex
curve is not a satisfactory approximation for complex rotation curves,
e.g., rotation curves with a hump ({\em panel} $c$).  In this case,
the rotation curve fit results in a wrong maximum value of the
rotation velocity  within the optical radius.  To obtain a correct
maximum value of the rotation velocity within the optical radius in
such cases, we did not fit the whole rotation curve with a Polyex
curve but only the outer part of the rotation curve. The points of the
rotation curve used in the construction of this fit are marked with
crosses in {\em panel} $c$ of  Fig.~\ref{figure:rc-fit}.  The obtained
fit is presented by the dashed line, and the maximum value of the rotation
velocity within the optical radius is shown by the square. 

The selection of the representative value of the rotation velocity on
the rotation curve for the baryonic Tully-Fisher relation is discussed
by \citet{Stark2009} and \citet{Lelli2016}. They argued that the
rotation velocity along the flat part of the rotation curve,
$V_{flat}$, is the preferable representative value.

We obtain the $V_{flat}$ value using a simple algorithm similar to
that of \citet{Lelli2016}. Starting from the centre, $R =0$, we obtain
the value of $V_{mean}$ for different galactocentric distances with
a step size of $0.01 R_{25}$ 
\begin{equation}
V_{mean}   =  0.5(V_{R} + V_{R+h_{d}}) 
\label{equation:vmean}
\end{equation}
where $h_{d}$ is the disc's exponential scale length. At each step we 
check the condition 
\begin{equation}
\frac{|V_{R+h_{d}} - V_{R}|}{V_{mean}}  < 0.05 
\label{equation:dv}
\end{equation}
We adopt $V_{flat} = V_{mean}$ when this condition is satisfied. The 
exponential scale length $h_{d}$ is estimated using a single 
exponential profile for the surface brightness distribution of the 
disc. The surface brightnesses at the optical radius, 
$S_{L_{B},R_{25}}$, and at the centre, $S_{L_{B},0}$, are linked by 
the expression
\begin{equation}
S_{L_{B},R_{25}} = S_{L_{B},0} \exp(-R_{25}/h_{d}) .
\label{equation:hd}
\end{equation}
The value of the central surface brightness of the disc is obtained
from bulge-disc decomposition of the surface brightness profile. The
surface brightness at the optical radius is the same for all galaxies
by definition. Eq.~(\ref{equation:hd}) is used to estimate the value
of $h_{d}$ for each galaxy.  The obtained values of $h_{d}$ for the
target galaxies are listed in Table \ref{table:sample}.  It should be
noted that this definition of the disc scale length $h_{d}$ may be
debatable since the surface brightness profiles of the discs of many
galaxies are broken, see, e.g., {\em panel} $a2$ of
Fig.~\ref{figure:rc-ab}. 

Using the criterium of Eq.~(\ref{equation:dv}), we obtained the
rotation velocity along the flat part of the rotation curve $V_{flat}$
for 81 out of 147 MaNGA galaxies of our sample.  We used the maximum
rotation velocity within the optical radius as the representative
value of the rotation velocity.  The influence of the choice of the
representative value of $V_{rot}$ on the result will be discussed
below.

\setcounter{table}{0}
\begin{table*}
\caption[]{\label{table:sample}
Properties of our sample of MaNGA galaxies.  The columns show
the name (the MaNGA number),
the galaxy distance $d$ in Mpc,
the position angle of the major axis $PA$, 
galaxy inclination angle $i$,
the maximum value of the rotation velocity within the optical radius
$V_{rot}$ in km/s,
the spectroscopic stellar mass $M_{sp}$ in solar masses,
the optical radius $R_{25}$ in kpc,
the disc scale length $h_{d}$ in kpc,
the luminosity $L_{B}$ in solar luminosities, 
the central oxygen abundance (O/H)$_{0}$,
the oxygen abundance at the optical radius (O/H)$_{R_{25}}$,
the number of points (spaxels) N$_{A}$ with a measured rotation 
velocity N$_{A}$, and the number of points used in derivation of 
the final rotation curve N$_{B}$.
}
\begin{center}
\begin{tabular}{cccccccccccrr} \hline \hline
Name                  &
$d$                   &
{\em PA}              &           
$i$                   &           
$V_{rot}$             &
log$M_{sp}$           &
log$L_{B}$            & 
$R_{25}$              &
$h_{d}$               &
12+log(O/H)$_{0}$     &
12+log(O/H)$_{R_{25}}$  &
N$_{A}$               &
N$_{B}$               \\
                     &
[Mpc]                &
[$\degr$]            &
[$\degr$]            &
[km/s]              &
[$M_{\odot}$]        &
[$L_{\odot}$]        &
[kpc]                &
[kpc]                &
                     &
                     &
                     &
                     \\  \hline
 7443 06101 &  133.8 &  111.5 &   39.4 &  122.6 &    9.23 &    9.17 &    5.19 &    2.15 &   8.436 &   8.129 &   333 &   261 \\
 7443 06103 &   82.8 &  200.5 &   58.0 &  112.4 &    9.56 &    9.62 &    5.82 &    1.79 &   8.411 &   8.320 &  1139 &   876 \\
 7443 12705 &  271.1 &   40.5 &   63.9 &  212.4 &   10.94 &   10.29 &   19.71 &    9.55 &   8.626 &   8.410 &  1214 &   864 \\
 7495 12703 &  130.7 &   15.1 &   63.9 &  200.7 &   10.60 &   10.26 &   16.16 &    6.49 &   8.674 &   8.457 &  2159 &  1604 \\
 7495 12704 &  127.5 &  352.0 &   57.3 &  220.0 &   10.87 &   10.43 &   17.31 &    5.97 &   8.669 &   8.572 &  2189 &  1404 \\
 7815 09101 &  117.9 &  305.3 &   55.0 &  154.1 &   10.17 &    9.78 &    8.86 &    3.66 &   8.603 &   8.428 &  1229 &  1059 \\
 7815 09102 &  169.3 &   79.6 &   59.3 &  109.0 &    9.41 &    9.73 &    8.62 &    4.37 &   8.518 &   8.185 &   474 &   399 \\
 7815 12702 &  163.1 &  287.1 &   52.2 &  120.5 &    9.58 &    9.60 &    9.09 &    4.31 &   8.439 &   8.227 &   521 &   447 \\
 7815 12704 &  178.1 &   75.3 &   66.1 &  219.4 &   11.07 &   10.21 &   15.11 &    5.53 &   8.586 &   8.597 &   745 &   594 \\
 7957 06104 &  111.2 &  261.4 &   60.8 &  138.1 &    9.76 &    9.99 &    8.36 &    2.31 &   8.467 &   8.294 &  1261 &   887 \\
 7957 09102 &  111.4 &  238.6 &   46.7 &  163.5 &   10.17 &   10.11 &   10.53 &    3.18 &   8.579 &   8.360 &  1898 &  1418 \\
 7957 12705 &  307.5 &  210.2 &   42.5 &  247.1 &   11.10 &   10.49 &   22.36 &    8.38 &   8.681 &   8.483 &   905 &   618 \\
 7962 12702 &  204.5 &  208.6 &   59.3 &  169.7 &   10.36 &   10.22 &   19.33 &    8.76 &   8.592 &   8.166 &  1532 &  1140 \\
 7975 09101 &  261.3 &  313.9 &   33.1 &  274.4 &   10.90 &   10.51 &   20.27 &    9.83 &   8.673 &   8.424 &  1949 &  1226 \\
 7977 06103 &  143.5 &   78.0 &   60.6 &  111.7 &    9.54 &    9.31 &    6.26 &    3.09 &   8.528 &   8.216 &   453 &   380 \\
 7991 09101 &  120.6 &   11.6 &   65.2 &  139.6 &   10.18 &    9.61 &    8.19 &    3.52 &   8.628 &   8.453 &   817 &   725 \\
 7991 12701 &  123.6 &  185.1 &   62.1 &  202.5 &    0.00 &   10.26 &   12.58 &    4.10 &   8.667 &   8.461 &  2224 &  1376 \\
 7992 09101 &   83.7 &   19.3 &   52.6 &  102.7 &    9.01 &    9.23 &    5.88 &    2.79 &   8.285 &   8.009 &   594 &   439 \\
 8082 12701 &  106.5 &   10.3 &   50.3 &  183.7 &   10.71 &   10.11 &   11.88 &    4.08 &   8.659 &   8.452 &  2652 &  2006 \\
 8083 12705 &   83.6 &  324.0 &   62.2 &  123.3 &   10.29 &   10.19 &   13.98 &    4.15 &   8.571 &   8.261 &  3048 &  1793 \\
 8131 06101 &  213.1 &  127.5 &   49.3 &  197.1 &   10.53 &   10.14 &   11.88 &    3.15 &   8.657 &   8.400 &   869 &   609 \\
 8131 12701 &   72.5 &   92.3 &   61.6 &  104.8 &    0.00 &    9.57 &    7.56 &    2.63 &   8.460 &   8.069 &  1643 &  1320 \\
 8131 12703 &  177.9 &  269.9 &   68.3 &  153.5 &   10.37 &    9.74 &   10.78 &    5.28 &   8.608 &   8.322 &   736 &   650 \\
 8132 09101 &   96.2 &   50.2 &   60.4 &  163.8 &   10.54 &    9.72 &    7.46 &    2.45 &   8.570 &   8.457 &  1068 &   810 \\
 8133 09101 &  231.9 &   84.5 &   62.3 &  234.1 &   11.06 &   10.19 &   13.49 &    5.38 &   8.718 &   8.530 &   623 &   362 \\
 8134 12701 &   88.0 &   71.3 &   41.4 &  104.8 &    9.21 &    9.51 &    8.32 &    3.45 &   8.543 &   8.225 &  1480 &  1113 \\
 8134 12703 &   88.1 &  114.7 &   42.0 &  127.6 &    9.65 &    9.42 &    7.47 &    3.46 &   8.518 &   8.283 &  1276 &  1003 \\
 8134 12704 &  332.7 &   41.7 &   42.2 &  241.1 &   11.50 &   10.54 &   27.42 &   14.57 &   8.666 &   8.458 &  1068 &   885 \\
 8134 12705 &   76.4 &  124.9 &   48.8 &  118.3 &    9.73 &    9.42 &    7.41 &    4.28 &   8.592 &   8.353 &  1296 &  1090 \\
 8135 09102 &  188.9 &   40.9 &   43.4 &  213.8 &   10.40 &    9.90 &   10.53 &    3.88 &   8.603 &   8.337 &   522 &   399 \\
 8135 12702 &  216.8 &  309.6 &   40.3 &  193.0 &   10.68 &   10.21 &   15.24 &    5.83 &   8.673 &   8.479 &  1322 &   958 \\
 8135 12703 &  210.1 &  351.4 &   55.9 &  198.9 &   10.69 &   10.08 &   14.26 &    5.29 &   8.706 &   8.432 &  1113 &   960 \\
 8137 09102 &  129.8 &  314.1 &   58.9 &  191.6 &   10.70 &   10.17 &   16.05 &    6.65 &   8.661 &   8.547 &  1678 &  1146 \\
 8138 06104 &  158.4 &   11.0 &   45.1 &   79.0 &    9.65 &    9.50 &    7.30 &    3.51 &   8.474 &   8.293 &   613 &   471 \\
 8138 09101 &  219.2 &   38.6 &   43.6 &  270.9 &   11.19 &   10.51 &   21.78 &    7.99 &   8.699 &   8.453 &  1737 &  1120 \\
 8140 12704 &  100.9 &  195.0 &   49.2 &  126.0 &    9.88 &    9.72 &   10.76 &    6.26 &   8.465 &   8.159 &  1287 &   987 \\
 8140 12705 &  179.6 &  224.1 &   46.5 &  161.4 &   10.09 &    9.90 &   11.75 &    4.74 &   8.512 &   8.192 &   899 &   727 \\
 8141 12702 &  130.6 &  237.6 &   53.8 &  168.7 &   10.67 &   10.18 &   14.56 &    3.67 &   8.586 &   8.451 &  1757 &  1463 \\
 8141 12704 &  199.1 &  158.5 &   41.4 &  188.5 &   10.63 &   10.16 &   13.03 &    4.77 &   8.673 &   8.404 &  1486 &  1210 \\
 8143 06102 &  170.1 &  202.0 &   50.2 &  100.1 &    9.73 &    9.77 &   10.31 &    3.94 &   8.465 &   8.191 &   915 &   678 \\
 8143 09102 &  182.4 &  239.6 &   44.9 &  193.8 &   10.31 &   10.04 &   12.82 &    3.97 &   8.636 &   8.421 &   990 &   849 \\
 8143 12702 &  185.3 &  178.0 &   48.5 &  197.8 &   10.24 &    9.92 &   12.13 &    4.22 &   8.595 &   8.510 &   606 &   480 \\
 8243 09101 &  179.8 &  319.2 &   54.5 &  136.1 &   10.13 &   10.19 &   11.77 &    3.18 &   8.432 &   8.252 &   743 &   523 \\
 8243 12701 &  186.4 &   80.6 &   69.6 &  283.1 &   11.64 &   10.74 &   28.47 &   12.88 &   8.637 &   8.518 &  2435 &  1446 \\
 8243 12705 &  189.1 &   78.8 &   60.4 &  145.2 &   10.07 &    9.99 &   14.67 &    7.86 &   8.532 &   8.212 &  1104 &   944 \\
 8247 12701 &   62.1 &  223.0 &   57.4 &   72.6 &    9.04 &    9.10 &    5.42 &    2.87 &   8.245 &   8.058 &  1116 &   906 \\
 8247 12702 &  181.4 &   94.9 &   42.8 &  180.6 &   10.35 &   10.03 &   10.55 &    3.83 &   8.565 &   8.429 &  1015 &   855 \\
 8247 12703 &  221.4 &  265.2 &   55.2 &  200.5 &   10.59 &   10.24 &   14.49 &    4.98 &   8.685 &   8.423 &  1294 &   967 \\
 8249 09101 &  215.5 &   93.3 &   50.3 &  257.4 &   10.84 &   10.18 &   12.01 &    6.87 &   8.641 &   8.553 &   364 &   276 \\
 8250 06102 &  105.2 &   99.0 &   37.2 &  176.2 &    9.94 &    9.66 &    7.14 &    3.53 &   8.616 &   8.489 &   989 &   762 \\
 8250 12702 &  188.9 &   34.6 &   57.7 &  148.3 &   10.19 &    9.90 &   11.45 &    4.39 &   8.627 &   8.312 &  1063 &   818 \\
 8253 09102 &  124.4 &   48.0 &   49.4 &  157.8 &   10.07 &   10.02 &   10.55 &    3.21 &   8.526 &   8.268 &  1503 &  1106 \\
 8254 06102 &  109.0 &  151.5 &   43.4 &  153.1 &    9.98 &    9.72 &    8.45 &    2.75 &   8.594 &   8.526 &   660 &   538 \\
 8254 12704 &  204.3 &   74.0 &   51.8 &  172.8 &   10.47 &   10.27 &   18.32 &    6.92 &   8.585 &   8.368 &   980 &   741 \\
 8256 12703 &  244.9 &  309.3 &   42.2 &  259.9 &   11.03 &   10.30 &   15.43 &    6.70 &   8.653 &   8.607 &   618 &   436 \\
 8257 03704 &   88.8 &   81.0 &   47.5 &   59.6 &    0.00 &    9.53 &    6.24 &    2.22 &   8.267 &   8.084 &   587 &   473 \\
 8257 06101 &  124.8 &  340.1 &   43.7 &  190.1 &   10.54 &   10.11 &   10.29 &    2.92 &   8.686 &   8.488 &  1407 &  1058 \\
 8257 06102 &   95.7 &  103.8 &   47.9 &  112.8 &    9.71 &    9.61 &    6.50 &    2.57 &   8.528 &   8.387 &   991 &   769 \\
 8257 09101 &  109.0 &   95.2 &   64.6 &  150.8 &   10.08 &    9.65 &    7.93 &    3.04 &   8.607 &   8.442 &  1028 &   862 \\
 8257 12703 &  107.5 &  113.4 &   63.8 &  208.9 &   10.52 &    9.91 &    9.64 &    3.41 &   8.620 &   8.568 &  1357 &  1103 \\
 8258 06101 &   93.3 &  109.0 &   48.3 &   70.0 &    9.16 &    9.32 &    6.33 &    2.91 &   8.276 &   8.145 &   815 &   689 \\
                    \hline
\end{tabular}\\
\end{center}
\end{table*}

\setcounter{table}{0}
\begin{table*}
\caption[]{
  Continued
}
\begin{center}
\begin{tabular}{cccccccccccrr} \hline \hline
Name                  &
$d$                   &
{\em PA}              &           
$i$                   &           
$V_{rot}$              &
log$M_{sp}$           &
log$L_{B}$            & 
$R_{25}$              &
$h_{d}$               &
12+log(O/H)$_{0}$     &
12+log(O/H)$_{R_{25}}$  &
N$_{i}$               &
N$_{f}$               \\
                     &
[Mpc]                &
[$\degr$]            &
[$\degr$]            &
[km/s]              &
[$M_{\odot}$]        &
[$L_{\odot}$]        &
[kpc]                &
[kpc]                &
                     &
                     &
                     &
                     \\  \hline
 8258 12701 &  110.3 &  297.0 &   64.1 &  130.9 &   10.05 &    9.88 &   12.57 &    8.59 &   8.569 &   8.285 &  1860 &  1311 \\
 8258 12702 &   93.9 &  117.5 &   70.0 &  103.9 &    9.54 &    9.30 &    6.37 &    2.88 &   8.530 &   8.251 &   950 &   792 \\
 8258 12703 &  276.4 &  222.5 &   68.3 &  210.3 &   11.07 &   10.36 &   22.78 &   12.56 &   8.629 &   8.418 &  1189 &   749 \\
 8259 12701 &   87.1 &  315.1 &   65.6 &   80.6 &    9.13 &    9.44 &    7.60 &    3.81 &   8.315 &   8.135 &   935 &   835 \\
 8261 12705 &  109.8 &  183.2 &   65.1 &  183.5 &    0.00 &   10.19 &   13.84 &    4.34 &   8.561 &   8.438 &  1975 &  1491 \\
 8263 06104 &  167.9 &  173.8 &   60.3 &  246.2 &   10.80 &   10.15 &   11.80 &    3.40 &   8.645 &   8.615 &   642 &   538 \\
 8263 12704 &  163.2 &  334.8 &   43.6 &  114.7 &    9.47 &    9.46 &    7.12 &    3.45 &   8.443 &   8.356 &   473 &   373 \\
 8313 12702 &  144.3 &   30.0 &   47.3 &  188.9 &   10.50 &   10.58 &   16.44 &    4.93 &   8.594 &   8.348 &  2850 &  1610 \\
 8313 12705 &  136.1 &  169.3 &   63.4 &  311.5 &   11.26 &   10.57 &   18.14 &    9.96 &   8.715 &   8.516 &  2009 &  1312 \\
 8315 12703 &  326.1 &  120.4 &   37.0 &  345.0 &   11.58 &   10.57 &   22.92 &    8.36 &   8.679 &   8.553 &   813 &   554 \\
 8317 09101 &  212.5 &  227.8 &   43.2 &  192.5 &   11.06 &   10.19 &   12.88 &    3.84 &   8.677 &   8.564 &   973 &   748 \\
 8317 12705 &  165.3 &  144.1 &   62.0 &  108.1 &    9.85 &    9.59 &    9.62 &    5.12 &   8.497 &   8.292 &   588 &   529 \\
 8318 09102 &  115.1 &   57.0 &   38.2 &  134.0 &   10.06 &    9.87 &    8.37 &    2.97 &   8.551 &   8.443 &  1405 &  1116 \\
 8318 12703 &  167.3 &   57.3 &   50.6 &  266.0 &   11.36 &   10.63 &   20.28 &    8.03 &   8.646 &   8.558 &  2293 &  1517 \\
 8318 12705 &  125.8 &  175.4 &   47.0 &  154.9 &   10.22 &   10.15 &   11.28 &    3.39 &   8.645 &   8.294 &  2286 &  1739 \\
 8319 12704 &  125.1 &  116.6 &   36.3 &  203.1 &   10.78 &   10.35 &   15.16 &    5.40 &   8.646 &   8.560 &  2280 &  1559 \\
 8320 09102 &  222.9 &  111.5 &   49.0 &  266.7 &   11.31 &   10.64 &   20.53 &    7.13 &   8.707 &   8.404 &  1832 &  1358 \\
 8320 12701 &  180.3 &   57.4 &   55.2 &  142.9 &    9.70 &    9.71 &   10.05 &    3.67 &   8.518 &   8.268 &   574 &   492 \\
 8320 12703 &  132.7 &  352.6 &   70.6 &  134.6 &   10.35 &    9.68 &    9.97 &    4.68 &   8.526 &   8.252 &  1085 &   851 \\
 8325 12701 &  182.9 &  224.5 &   50.9 &  148.4 &    9.96 &    9.78 &   11.53 &    5.15 &   8.571 &   8.361 &   892 &   716 \\
 8325 12703 &  121.9 &  100.1 &   71.3 &  262.9 &   10.82 &   10.21 &   14.18 &   10.55 &   8.674 &   8.579 &  1117 &   634 \\
 8325 12705 &  168.4 &  277.7 &   48.6 &  133.1 &    9.93 &    9.75 &   10.21 &    4.16 &   8.615 &   8.307 &   861 &   726 \\
 8326 09102 &  300.3 &   68.1 &   30.5 &  291.4 &   11.26 &   10.62 &   25.48 &    7.90 &   8.730 &   8.445 &  1537 &  1050 \\
 8326 12701 &  121.9 &  100.7 &   71.6 &  274.0 &   10.88 &   10.18 &   14.48 &   11.38 &   8.653 &   8.580 &  1312 &   785 \\
 8326 12702 &  118.6 &   85.4 &   31.7 &  174.9 &   10.06 &    9.89 &   10.06 &    3.77 &   8.633 &   8.350 &  1928 &  1377 \\
 8326 12703 &  220.2 &   35.9 &   56.9 &  163.7 &   10.35 &   10.29 &   16.01 &    4.97 &   8.597 &   8.265 &  1428 &  1134 \\
 8329 06103 &  135.3 &  207.3 &   54.8 &  203.1 &   10.57 &   10.20 &   10.49 &    3.50 &   8.638 &   8.610 &   961 &   732 \\
 8329 12701 &  150.2 &  137.1 &   33.7 &  248.8 &   11.18 &   10.55 &   18.57 &    5.31 &   8.650 &   8.594 &  2473 &  1590 \\
 8329 12702 &   61.3 &   27.3 &   58.9 &   41.7 &    9.05 &    9.05 &    5.94 &    4.91 &   8.332 &   8.055 &   805 &   632 \\
 8329 12703 &  172.5 &  295.4 &   59.8 &  153.2 &    9.69 &    9.81 &   11.29 &    4.12 &   8.518 &   8.268 &   575 &   478 \\
 8329 12704 &   72.9 &  300.0 &   56.2 &  122.3 &    9.62 &    9.88 &    8.84 &    3.37 &   8.474 &   8.165 &  2412 &  1958 \\
 8330 06102 &  207.2 &  349.9 &   29.9 &  193.8 &   10.35 &   10.12 &   10.05 &    2.87 &   8.671 &   8.504 &   651 &   483 \\
 8330 12701 &  224.0 &  270.3 &   42.4 &  203.0 &   10.67 &   10.37 &   18.46 &    6.90 &   8.697 &   8.313 &  1658 &  1183 \\
 8330 12703 &  117.9 &   64.7 &   44.1 &  190.4 &   10.50 &   10.10 &   11.15 &    4.62 &   8.660 &   8.457 &  1880 &  1428 \\
 8332 12701 &  123.3 &   45.0 &   50.8 &  120.0 &    9.82 &   10.07 &   12.25 &    4.81 &   8.498 &   8.257 &  2317 &  1615 \\
 8332 12702 &  387.3 &  306.3 &   54.5 &  333.3 &   11.46 &   10.65 &   26.29 &   10.08 &   8.715 &   8.554 &   528 &   293 \\
 8332 12703 &  376.6 &  158.9 &   41.7 &  306.2 &   11.50 &   10.82 &   34.69 &   12.45 &   8.645 &   8.544 &   828 &   527 \\
 8332 12705 &  143.2 &  145.4 &   41.4 &  286.5 &   11.13 &   10.48 &   15.62 &    3.46 &   8.690 &   8.537 &  1077 &   837 \\
 8335 06102 &   86.8 &   83.3 &   49.7 &  108.6 &    9.55 &    9.59 &    6.73 &    1.98 &   8.422 &   8.198 &  1077 &   767 \\
 8335 12701 &  266.3 &   82.5 &   65.0 &  247.2 &   10.93 &   10.33 &   18.72 &    8.04 &   8.647 &   8.498 &   795 &   510 \\
 8341 09101 &  216.3 &    9.2 &   45.3 &  207.5 &   10.75 &   10.34 &   15.21 &    4.14 &   8.614 &   8.320 &  1180 &   872 \\
 8341 09102 &  107.4 &   18.4 &   54.7 &  137.2 &   10.01 &    9.68 &    7.55 &    2.74 &   8.577 &   8.435 &  1044 &   886 \\
 8439 06101 &  111.9 &  133.5 &   57.6 &  130.7 &    9.71 &    9.76 &    8.14 &    2.59 &   8.570 &   8.307 &  1378 &  1101 \\
 8439 09101 &  159.0 &  272.4 &   61.2 &  103.6 &    9.67 &    9.36 &    6.17 &    2.85 &   8.492 &   8.271 &   317 &   277 \\
 8439 12701 &   71.4 &   24.9 &   63.8 &   98.8 &    9.38 &    9.21 &    6.92 &    3.95 &   8.431 &   8.182 &  1329 &  1108 \\
 8439 12702 &  114.0 &   31.4 &   53.2 &  257.4 &   10.94 &   10.30 &   15.47 &    3.35 &   8.683 &   8.590 &  1247 &   780 \\
 8440 12703 &  110.8 &  219.1 &   58.8 &  130.1 &    9.79 &    9.63 &    8.06 &    3.24 &   8.514 &   8.221 &  1354 &  1140 \\
 8447 12703 &  150.8 &  198.6 &   60.1 &   96.1 &    9.40 &    8.99 &    5.48 &    4.25 &   8.477 &   8.199 &   241 &   201 \\
 8448 06104 &   95.5 &   79.2 &   51.8 &  137.6 &    9.27 &    9.79 &    8.57 &    2.53 &   8.395 &   8.266 &  1073 &   821 \\
 8451 12703 &  160.4 &  182.0 &   42.2 &  147.7 &    9.53 &    9.88 &   10.89 &    4.85 &   8.412 &   8.215 &   759 &   548 \\
 8452 06101 &  104.6 &   76.0 &   58.7 &  122.1 &    0.00 &    9.57 &    7.35 &    2.84 &   8.496 &   8.340 &   713 &   622 \\
 8452 06104 &  179.7 &   28.9 &   41.9 &  184.4 &   10.08 &    9.70 &    7.41 &    2.40 &   8.663 &   8.444 &   444 &   377 \\
 8452 12703 &  253.9 &   64.4 &   30.6 &  318.2 &   11.59 &   10.68 &   24.62 &   11.61 &   8.619 &   8.603 &  1167 &   826 \\
 8453 12701 &  107.3 &  284.6 &   33.8 &  163.2 &   10.32 &   10.02 &   11.70 &    5.49 &   8.635 &   8.533 &  1010 &   842 \\
 8453 12702 &  160.7 &  166.0 &   60.7 &  134.5 &    9.94 &    9.57 &    9.74 &    6.11 &   8.535 &   8.313 &   739 &   630 \\
 8453 12703 &  263.0 &  149.0 &   61.1 &  245.9 &   11.04 &   10.39 &   21.04 &    8.66 &   8.651 &   8.489 &  1021 &   775 \\
 8459 12702 &   70.3 &  155.3 &   42.0 &   65.2 &    8.12 &    9.36 &    6.65 &    3.14 &   8.300 &   8.159 &  1516 &  1279 \\
 8459 12703 &  109.0 &  311.6 &   65.2 &  139.6 &    9.99 &    9.62 &    8.19 &    3.32 &   8.546 &   8.360 &  1223 &   997 \\
 8459 12705 &   71.9 &  338.3 &   36.5 &   76.8 &    9.52 &    9.42 &    7.15 &    3.57 &   8.522 &   8.263 &  1975 &  1544 \\
 8464 09102 &  104.4 &  113.9 &   45.5 &  135.2 &   10.12 &    9.92 &    9.62 &    4.56 &   8.617 &   8.375 &  1775 &  1342 \\
 8464 12701 &  103.1 &   89.7 &   60.8 &  200.8 &   10.31 &   10.03 &   15.24 &    6.68 &   8.639 &   8.440 &  1891 &  1447 \\
 8464 12704 &  207.7 &  148.7 &   43.5 &  176.7 &   10.30 &   10.27 &   12.59 &    4.10 &   8.595 &   8.319 &  1101 &   911 \\
 8466 09101 &  106.8 &   87.3 &   63.8 &  134.7 &    9.80 &    9.57 &    8.03 &    3.70 &   8.562 &   8.274 &  1067 &   909 \\
 8466 09102 &   77.8 &   27.5 &   67.5 &  102.2 &    9.12 &    9.51 &    5.85 &    1.89 &   8.282 &   8.068 &  1219 &  1022 \\
 8466 12702 &  122.3 &  319.2 &   53.3 &  176.5 &   10.53 &   10.12 &   12.75 &    4.47 &   8.712 &   8.329 &  2534 &  1802 \\
                    \hline
\end{tabular}\\
\end{center}
\end{table*}

\setcounter{table}{0}
\begin{table*}
\caption[]{
  Continued
}
\begin{center}
\begin{tabular}{cccccccccccrr} \hline \hline
Name                  &
$d$                   &
{\em PA}              &           
$i$                   &           
$V_{rot}$             &
log$M_{sp}$           &
log$L_{B}$            & 
$R_{25}$              &
$h_{d}$               &
12+log(O/H)$_{0}$     &
12+log(O/H)$_{R_{25}}$  &
N$_{i}$               &
N$_{f}$               \\
                     &
[Mpc]                &
[$\degr$]            &
[$\degr$]            &
[km/s]              &
[$M_{\odot}$]        &
[$L_{\odot}$]        &
[kpc]                &
[kpc]                &
                     &
                     &
                     &
                     \\  \hline
 8466 12704 &  226.5 &  196.1 &   45.5 &  153.4 &   10.62 &   10.36 &   20.86 &    7.41 &   8.666 &   8.311 &  1549 &  1138 \\
 8482 12705 &  176.8 &  117.4 &   57.9 &  279.6 &   11.12 &   10.45 &   20.14 &    5.36 &   8.649 &   8.590 &  1066 &   824 \\
 8484 06103 &  107.6 &  193.6 &   49.3 &  123.3 &   10.10 &    9.76 &    8.35 &    2.77 &   8.619 &   8.398 &  1194 &   958 \\
 8484 12702 &  238.7 &  201.3 &   55.8 &  236.6 &   11.09 &   10.21 &   16.20 &    5.67 &   8.670 &   8.511 &   470 &   367 \\
 8485 06101 &  155.6 &  189.3 &   41.2 &  137.3 &    9.73 &    9.60 &    7.54 &    2.31 &   8.636 &   8.351 &   452 &   382 \\
 8485 09102 &  114.2 &  182.3 &   47.8 &  198.8 &   10.78 &   10.24 &   13.56 &    5.09 &   8.676 &   8.497 &  1871 &  1384 \\
 8485 12701 &  102.7 &   92.0 &   63.1 &   98.1 &    9.01 &    9.68 &    8.96 &    7.43 &   8.341 &   8.142 &   912 &   690 \\
 8485 12705 &  161.9 &  134.2 &   59.1 &  112.8 &    9.75 &    9.46 &    7.46 &    3.22 &   8.630 &   8.368 &   419 &   312 \\
 8548 06103 &   89.9 &   30.9 &   49.5 &   94.9 &    9.90 &    9.80 &    8.28 &    2.60 &   8.451 &   8.215 &  1519 &  1132 \\
 8549 06103 &  177.2 &  236.9 &   48.3 &  121.1 &    9.62 &    9.89 &   12.03 &    5.21 &   8.433 &   8.222 &   841 &   648 \\
 8549 12702 &  185.1 &  279.0 &   52.1 &  260.2 &   11.13 &   10.58 &   19.74 &    7.32 &   8.677 &   8.509 &  2298 &  1523 \\
 8549 12703 &  198.0 &  110.6 &   61.2 &  126.2 &   10.21 &    9.87 &   12.00 &    5.20 &   8.594 &   8.355 &  1041 &   838 \\
 8550 06103 &  108.6 &  118.7 &   40.8 &  215.9 &   10.67 &   10.09 &   10.00 &    2.11 &   8.652 &   8.522 &  1279 &   891 \\
 8551 09101 &  172.4 &   13.9 &   42.9 &  112.7 &    9.46 &    9.66 &    8.36 &    3.23 &   8.518 &   8.226 &   414 &   305 \\
 8552 12703 &  325.4 &   31.3 &   40.1 &  302.3 &   11.33 &   10.61 &   27.61 &   12.55 &   8.688 &   8.496 &   502 &   339 \\
 8554 12701 &  396.5 &   52.0 &   31.5 &  387.3 &   11.35 &   10.65 &   25.95 &   10.33 &   8.686 &   8.519 &   605 &   411 \\
 8554 12705 &  144.3 &  196.6 &   65.0 &  234.3 &   10.96 &   10.26 &   16.79 &    7.34 &   8.636 &   8.566 &  1109 &   795 \\
 8555 12705 &  136.6 &  110.8 &   53.3 &  223.7 &   10.74 &   10.20 &   12.58 &    4.06 &   8.651 &   8.589 &  1368 &  1178 \\
 8601 12705 &  129.1 &   51.8 &   66.3 &  206.1 &   10.80 &   10.29 &   16.27 &    6.41 &   8.639 &   8.566 &  1189 &   772 \\
 8612 12703 &   71.3 &  121.6 &   63.6 &  107.6 &    9.55 &    9.49 &    6.22 &    2.07 &   8.446 &   8.359 &  1321 &  1094 \\
 8618 06101 &  121.8 &  320.0 &   64.0 &  196.5 &   10.70 &   10.22 &   13.58 &    4.19 &   8.561 &   8.559 &  1265 &   831 \\
                    \hline
\end{tabular}\\
\end{center}
\end{table*}

\subsection{The stellar masses of our galaxies}

\begin{figure*}
\resizebox{1.00\hsize}{!}{\includegraphics[angle=000]{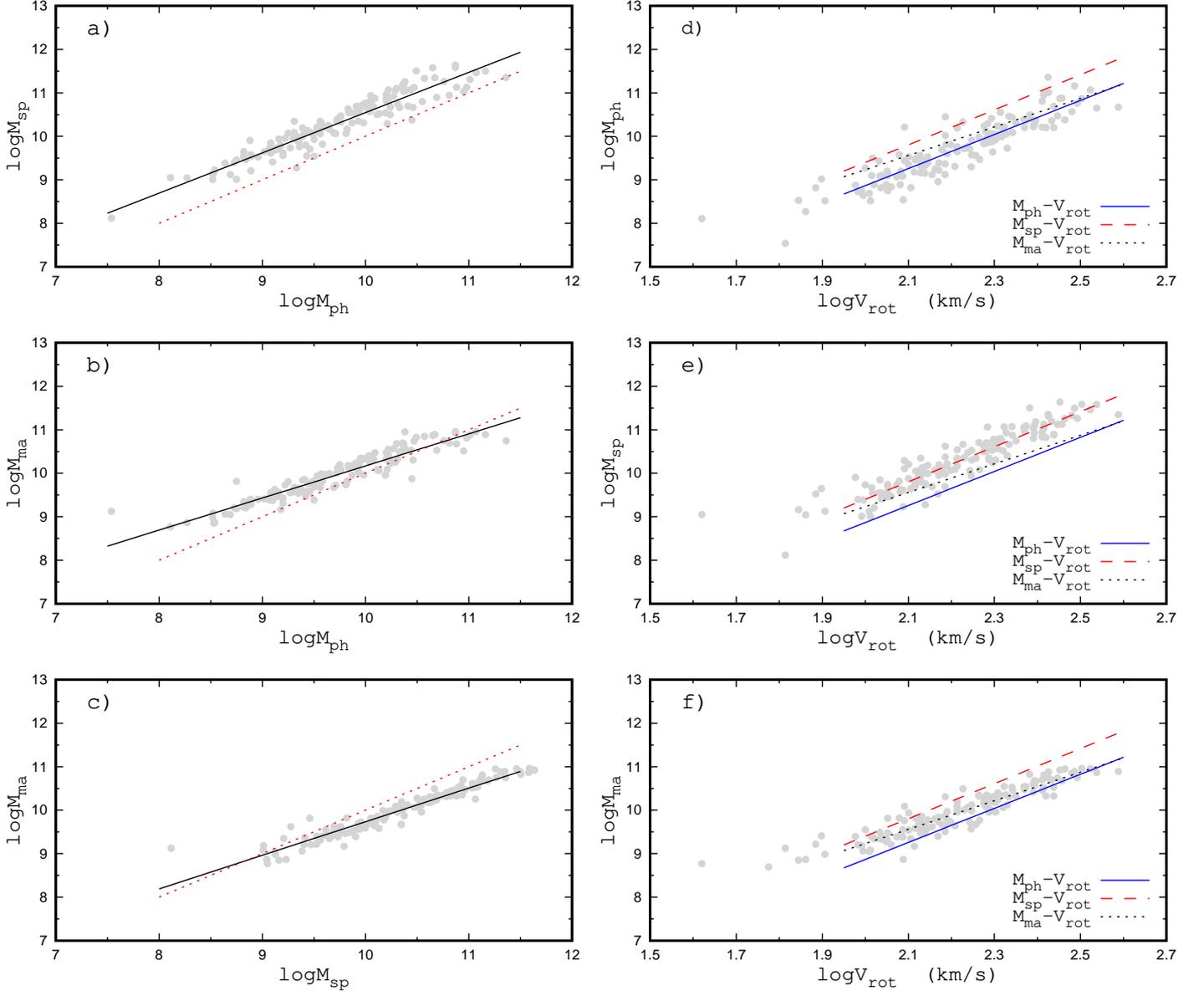}}
\caption{
{\em Panel} $a$ shows the spectroscopic stellar mass of our galaxies,
$M_{sp}$, as a function of their photometric stellar mass, $M_{ph}$.
The points represent the individual MaNGA galaxies of our sample.  The
solid line is the linear best fit to those points.  The dotted line
indicates exact correpondence in the masses.
{\em Panel} $b$ shows the MaNGA stellar mass, $M_{ma}$, as a function of
photometric stellar mass, $M_{ph}$, and {\em panel} $c$ shows the MaNGA
stellar mass, $M_{ma}$, as a function of spectroscopic stellar mass $M_{sp}$.
The notations are the same as in {\em panel} $a$. 
{\em Panels} $d$,  $e$ and $f$ show $M_{ph}$, $M_{sp}$, and $M_{ma}$ as a
function of the rotation velocity $V_{rot}$, respectively.
The points denote individual galaxies, the
solid line is the $M_{ph}$ -- $V_{rot}$ relation (best fit), the
long-dashed line is the $M_{sp}$ -- $V_{rot}$ relation, and the 
short-dashed line is the $M_{ma}$ -- $V_{rot}$ relation. 
}
\label{figure:v-mph-msp}
\end{figure*}

\begin{figure}
\resizebox{1.00\hsize}{!}{\includegraphics[angle=000]{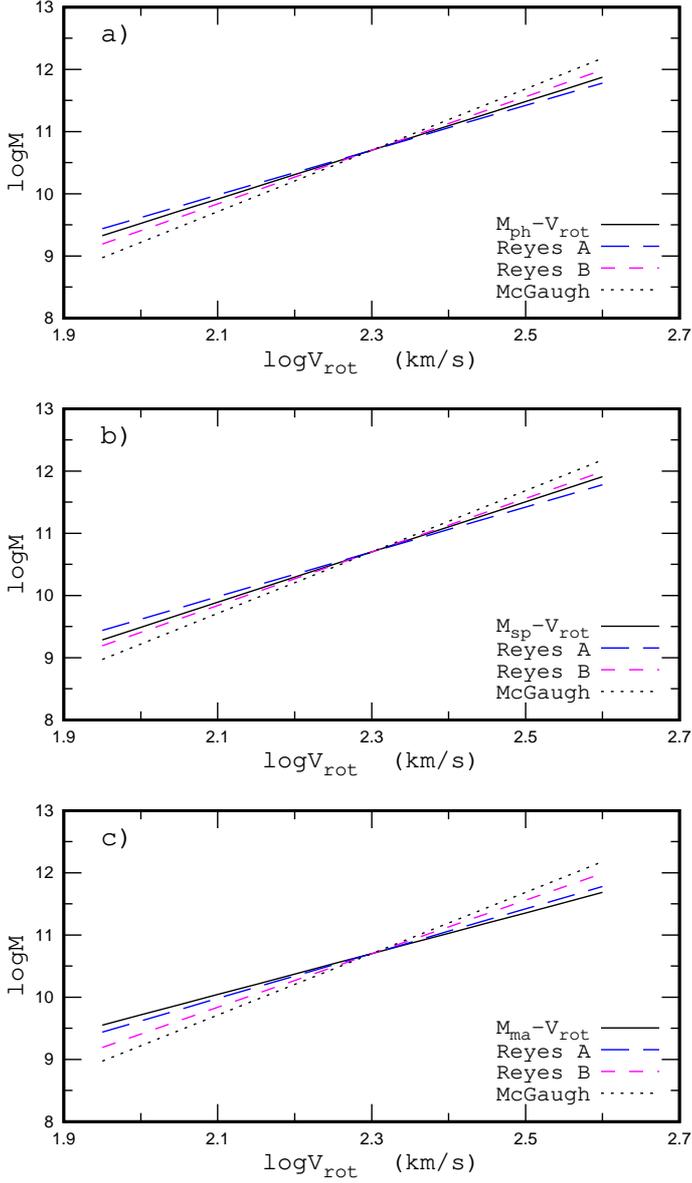}}
\caption{
{\em Panel} $a$ shows the comparison of our $M_{ph}$ -- $V_{rot}$
relation with the stellar mass Tully-Fisher relation from
\citet{McGaugh2015} and two relations (for two kinds of stellar mass
estimates) from \citet{Reyes2011}.  The meaning of each line is described
in the legend. Each relation is shifted along the M-axis in such a way
that the log$M$ = 10.7 at log$V_{rot}$ = 2.3.
{\em Panels} $b$ and $c$  show the same as {\em panel} $a$ but
for the $M_{sp}$ -- $V_{rot}$ and  $M_{ma}$ -- $V_{rot}$ relations, respectively.
}
\label{figure:tf-comparison}
\end{figure}

\begin{figure}
\resizebox{1.00\hsize}{!}{\includegraphics[angle=000]{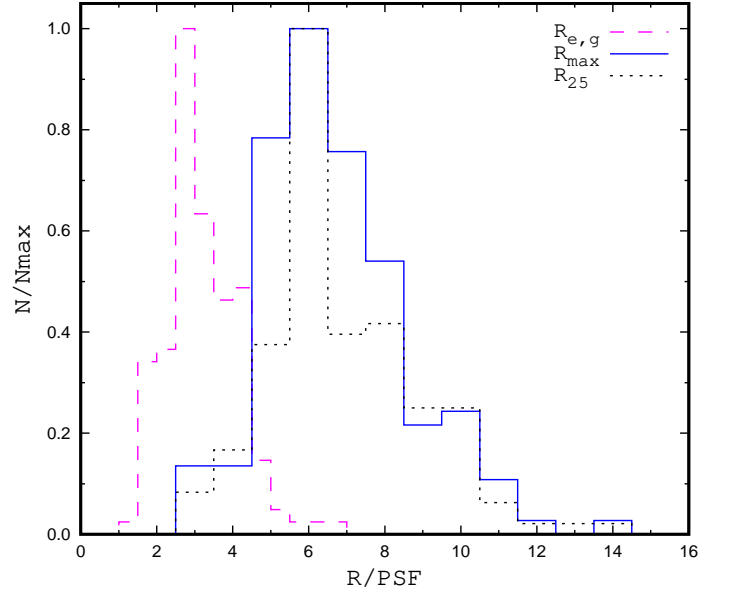}}
\caption{
  The long-dashed line shows the normalized histogram of the ratios of
  the galactic effective radius $R_{e,g}$ to the point spread
  function PSF (full width at the half maximum) for our sample
  of galaxies. The short-dashed line indicates the normalized histograms of
  the optical radius $R_{25}$ to the PSF ratios. The solid line
  denotes the histogram of the radii (expressed in units of PSF)
  up to which the measurements are available.
}
\label{figure:psf}
\end{figure}

\begin{figure}
\resizebox{1.00\hsize}{!}{\includegraphics[angle=000]{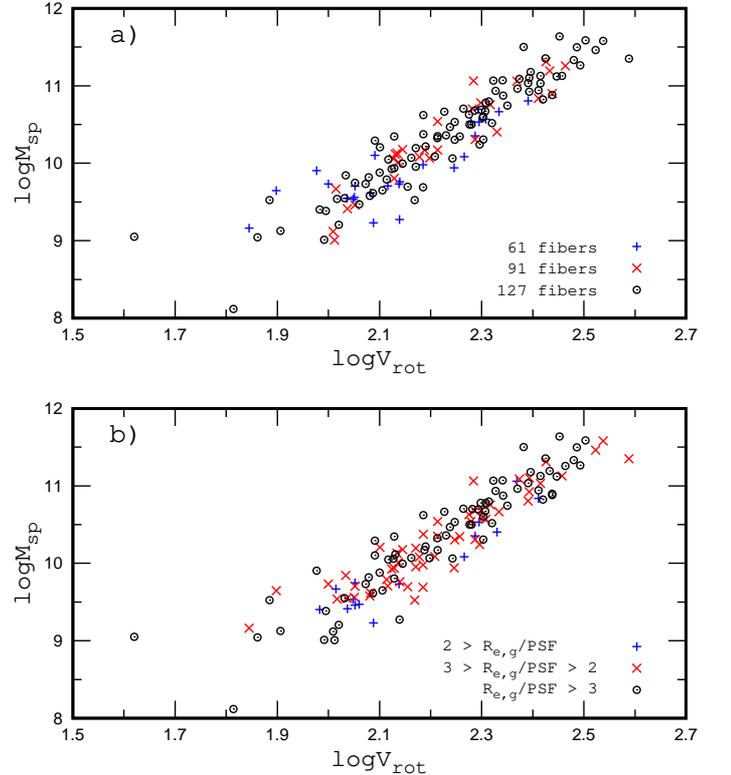}}
\caption{
{\em  Panel} $a$ shows the stellar mass Tully-Fisher relation 
for our sample of MaNGA galaxies. Galaxies measured with
different numbers of fibers are indicated by different symbols.
{\em  Panel} $b$ shows the same as {\em panel}
$a$ but the galaxies with different $R_{e,g}$/PSF (the ratio
of the effective radius of the galaxy to the full width at
half maximum of the point spread function)  are indicated by
different symbols.
}
\label{figure:tf-psf}
\end{figure}

\begin{figure}
\resizebox{1.00\hsize}{!}{\includegraphics[angle=000]{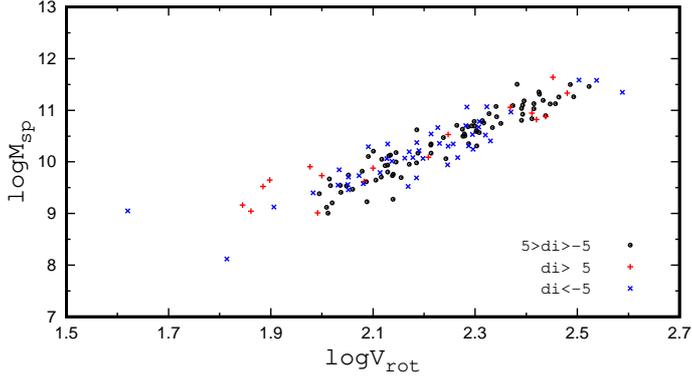}}
\caption{
The spectroscopic stellar mass of our galaxies, $M_{sp}$, as a
function of rotation velocity, $V_{rot}$. The individual MaNGA
galaxies of our sample with different values of d$i = i_{kin} -
i_{phot}$ are shown by different symbols.
}
\label{figure:v-msp-di}
\end{figure}

\begin{figure}
\resizebox{1.00\hsize}{!}{\includegraphics[angle=000]{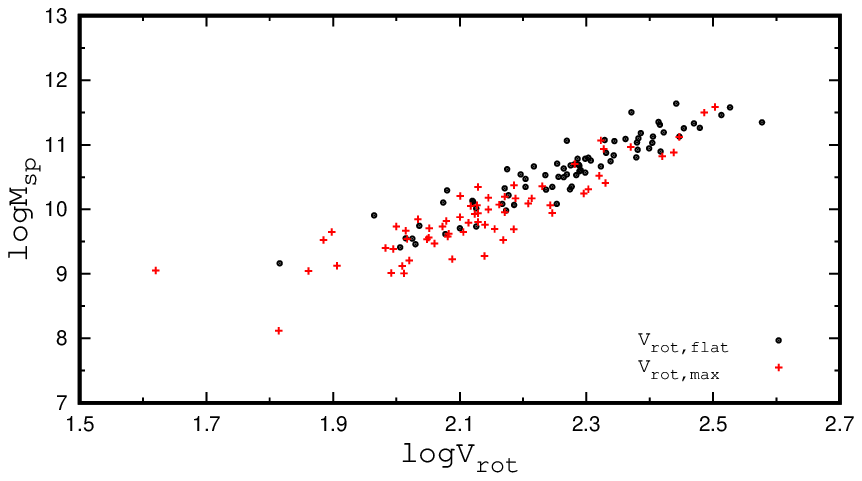}}
\caption{
The spectroscopic stellar mass of our galaxies, $M_{sp}$, as a
function of rotation velocity, $V_{rot}$. The rotation velocity along
the flat part of the rotation curve, $V_{flat}$, is shown for the
galaxy if this value was obtained (circles).  For the other galaxies,
the maximum value of the rotation velocity within optical radius is
plotted (plus signs).
}
\label{figure:v-msp-flat}
\end{figure}

The estimation of the mass of a galaxy is also not a trivial task.
The SDSS data base offers values of the stellar masses of galaxies
determined in different ways, e.g., photometric $M_{ph}$ and
spectroscopic $M_{sp}$ masses.  \citet{Kannappan2007} have
demonstrated that different stellar mass estimation methods yield
relative mass scales that can disagree by a factor $\ga 3$. The
disagreement between the values of the mass of an individual galaxy
produced by the different methods can be around an order of magnitude
at $10^{9} {\rm M}_{\odot}$.
Even in the case of the best studied galaxies (with accurate
distances based on the Cepheid period--luminosity relation or on the
brightness of the tip of the red giant branch), the values of the
stellar mass determined through the different methods can differ
by several times \citep{Ponomareva2018}.
This can result in an additional scatter
around the mass -- metallicity and mass -- metallicity gradient
diagrams and may mask possible correlations if present.

There is a tight correlation between the rotation velocity and  
the baryonic (star+gas) mass of a galaxy, the baryonic TF
relation, \citep[][among many
others]{Walker1999,Zaritsky2014,Lelli2016}.  \citet{Lelli2016} have
found that the intrinsic scatter of the baryonic Tully-Fisher relation
is $\sim 0.1$ dex for galaxies in the range from $10^{8} {\rm
M}_{\odot}$ to $\sim 10^{11} {\rm M}_{\odot}$. The residuals around
this relation show no trend with galaxy size or surface brightness.
Thus, the rotation velocity is a reliable indicator of the baryonic
(star+gas) mass of a galaxy. There is also a tight correlation between
the stellar mass of a galaxy and its rotation velocity, the 
stellar mass TF relation, \citep[][among many
others]{Tully1977,Reyes2011,McGaugh2015} since the gas mass is a
monotonic function of the rotation velocity
\citep[e.g.,][]{McGaugh2015,Zasov2017}.  Thus, the stellar mass of a
galaxy can also be estimated from its rotation velocity.
 
Panel $a$ of Fig.~\ref{figure:v-mph-msp} shows the spectroscopic
stellar mass as a function of photometric stellar mass for our sample
of galaxies. The points represent data for the individual galaxies.
The linear least-squares fit to those data 
\begin{equation}
\log M_{sp} = 0.925(\pm0.026)\log M_{ph} + 1.296(\pm0.258) 
\label{equation:msp-mph}
\end{equation}
is shown  by the solid line in panel $a$ of
Fig.~\ref{figure:v-mph-msp}. The dashed line indicates exact
correspondence in the masses. There is a significant difference
between the absolute values of the stellar masses $M_{sp}$ and $M_{ph}$
for our sample of MaNGA galaxies, i.e., there is a shift between
the zero-points of the $M_{sp}$ and $M_{ph}$ mass scales. 
\citet{Maraston2013} have compared their photometric stellar masses
with the spectral stellar masses for a large number of the SDSS galaxies.
They found a systematic offset of around 0.2 dex, with the spectroscopic masses
being larger than their photometric masses. The difference between the
values of the photometric and spectrsoscopic masses for individual galaxies
can exceed an order of magnitude (see their figure A1).
They discuss the sources of difference between the spectroscopic and 
photometric stellar masses.
Their mean value of the scatter around the $M_{sp}$ -- $M_{ph}$ relation is 0.218 dex.

Panel $b$ of Fig.~\ref{figure:v-mph-msp} shows the MaNGA
stellar mass as a function of photometric stellar mass for our sample
of galaxies. The points represent the data for the individual galaxies.
The linear least-squares fit to those data 
\begin{equation}
\log M_{ma} = 0.738(\pm0.023)\log M_{ph} + 2.785(\pm0.222) 
\label{equation:mma-mph}
\end{equation}
is shown  by the solid line in panel $b$ of
Fig.~\ref{figure:v-mph-msp}. The dashed line indicates exact
correspondence in the masses. The mean value of the scatter around
the $M_{ma}$ -- $M_{ph}$ relation is 0.186 dex.

Panel $c$ of Fig.~\ref{figure:v-mph-msp} shows the MaNGA
stellar mass as a function of spectroscopic stellar mass for our sample
of galaxies. The points represent data for the individual galaxies.
The linear least-squares fit to those data 
\begin{equation}
\log M_{ma} = 0.772(\pm0.019)\log M_{sp} + 2.017(\pm0.194) 
\label{equation:mma-msp}
\end{equation}
is shown  by the solid line in panel $c$ of
Fig.~\ref{figure:v-mph-msp}. As before, the dashed line indicates exact
correspondence in the masses. The mean value of the scatter around
the $M_{ma}$ -- $M_{sp}$ relation is 0.150 dex.
 
The mean value of the scatter around the $M_{x}$ -- $M_{y}$ relations
is between 0.150 dex and 0.218 dex.
This value can be interpreted as a mean random error of the
relative stellar mass determinations for our MaNGA sample. 
 
Panels $d$, $e$ and $f$ of Fig.~\ref{figure:v-mph-msp} show the
photometric $M_{ph}$, the spectroscopic $M_{sp}$, and the MaNGA $M_{ma}$
stellar masses as
a function of rotation velocity, respectively.  The points in each
panel denote the data for the individual galaxies. The solid line is
the linear $M_{ph}$ -- $V_{rot}$ relation for galaxies with $V_{rot} >
90$ km/s,
\begin{equation}
\log M_{ph} = 3.917(\pm0.169)\log V_{rot} + 1.034(\pm0.379) 
\label{equation:mph-vr}
\end{equation}
the long-dashed line is the linear $M_{sp}$ -- $V_{rot}$ relation for 
those galaxies, 
\begin{equation}
\log M_{sp} = 4.034(\pm0.140)\log V_{rot} + 1.332(\pm0.313) .
\label{equation:msp-vr}
\end{equation}
the short-dashed line is the linear $M_{ma}$ -- $V_{rot}$ relation for 
those galaxies, 
\begin{equation}
\log M_{ma} = 3.280(\pm0.116)\log V_{rot} + 2.674(\pm0.260) .
\label{equation:mma-vr}
\end{equation}
The mean value of the scatter around the $M_{ph}$ -- $V_{rot}$
relation is 0.283 dex,
the mean value of the scatter around the
$M_{sp}$ -- $V_{rot}$ relation is 0.233 dex, and
the mean value of the scatter around the
$M_{ma}$ -- $V_{rot}$ relation is 0.195 dex, i.e., the values of the
scatter around these $M$ -- $V_{rot}$ relations are close to the values of the scatter
around the $M_{X}$ -- $M_{Y}$ relations.  This suggests that the
random errors of the relative stellar mass determinations contribute
significantly to the scatter around the $M$ -- $V_{rot}$ 
relations for our MaNGA galaxy sample. 

The stellar mass Tully-Fisher relation has been investigated in many works. 
The comparison of the $M_{ph}$ -- $V_{rot}$ diagram obtained
here with the two stellar mass TF relations from \citet{Reyes2011}
(for two kinds of stellar mass estimates) and with the relation from
\citet{McGaugh2015} are shown in panel $a$ of Fig.~\ref{figure:tf-comparison}.
 To make a comparison clear, 
each relation is shifted along the M-axis in such way that 
the log$M$ = 10.7 at log$V_{rot}$ = 2.3.
Panels $b$ and $c$  show the same as panel $a$ but
for the $M_{sp}$ -- $V_{rot}$ and  $M_{ma}$ -- $V_{rot}$ relations, respectively.
The shifts along the M-axis are 0.029 dex and 0.060 dex for the TF relations
$A$ and $B$ from \citet{Reyes2011}, --0.029 dex for the relation from
\citet{McGaugh2015}, 0.657 dex for the   $M_{ph}$ -- $V_{rot}$ relation, 
0.090 dex for the   $M_{sp}$ -- $V_{rot}$ relation, and  0.482 dex for
the   $M_{ma}$ -- $V_{rot}$ relation.
The values of the $M_{sp}$ are in much better agreement with the stellar mass
TF relations from \citet{Reyes2011} and \citet{McGaugh2015} than the
$M_{ph}$ and $M_{ma}$ values.
The slopes of the $M_{ph}$ -- $V_{rot}$ and $M_{sp}$ -- $V_{rot}$ relations are close
to each other. The slope of the $M_{ma}$ -- $V_{rot}$ relation is slightly lower,
i.e., the $M_{ma}$ masses of the low-mass galaxies may be slightly overestimated
or/and the $M_{ma}$ masses of the massive galaxies may be slightly underestimated
(see also Fig.\ 5 in \citet{Barrera2018}).

Thus,  the $M_{sp}$ -- $V_{rot}$ diagram is in much better agreement
with the stellar mass TF relations from \citet{Reyes2011} and
\citet{McGaugh2015} than the $M_{ph}$ -- $V_{rot}$ and the $M_{ma}$ -- $V_{rot}$
diagram. Therefore, the spectroscopic stellar masses $M_{sp}$ will be used below.

The point spread function (PSF) of the MaNGA measurements is estimated
to have a full width at half maximum of 2.5 arcsec or 5 pixel
\citep{Bundy2015,Belfiore2017}.
Fig.~\ref{figure:psf} shows the normalized histograms of 
the galaxy effective radii, the optical radii, and the
radii up to which the measurements are available for
our sample of galaxies expressed in the units of the
PSF (full width at the half maximum).
The value of the effective radius of the galaxy is estimated using
the photometric profile obtained during the bulge -- disk decomposition.  

One can expect that the spatial resolution may affect the determined
rotation curve especially in the inner part of a galaxy where the change
of the rotation velocity with radius is large. We consider here
the maximum rotation velocity value, which is reached in the outer
part of a galaxy. Since the rotation curve in the outer part of a
galaxy is rather flat then we can expect that the influence of the  spatial
resolution on the obtained value of the maximum rotation velocity is small, if any. 
Depending on their angular size, the number of fibers covering the
galaxies varied during the observations with the integral field units
employed by MaNGA.
The panel $a$ of Fig.~\ref{figure:tf-psf} shows the stellar mass Tully-Fisher
relation for our sample of MaNGA galaxies. The galaxies measured with 
the different numbers of fibers are indicated by different symbols.
Inspection of the panel $a$ of Fig.~\ref{figure:tf-psf} shows that 
there is no systematic shift in the positions of the galaxies measured
with the different numbers of fibers in the stellar Tully-Fisher diagram.
The panel $b$ of Fig.~\ref{figure:tf-psf} shows the comparison of the
locations of galaxies with different ratios of galactic effective radius 
to the point spread function, $R_{e,g}$/PSF, in the
stellar mass -- rotation velocity diagram. Galaxies with different
$R_{e,g}$/PSF are indicated by different symbols.
Examination of the panel $b$ of Fig.~\ref{figure:tf-psf} shows that 
the locations of the galaxies with the different $R_{e,g}$/PSF ratios
follow the same trend.
Those two diagrams provide a strong evidence in favour that
the influence of the  spatial resolution on the obtained value of the
maximum rotation velocity in our sample of galaxies is small, if any. 

The stellar mass TF relation can be used to test the credibility of
the kinematic inclination angles determined here.  If the difference
between $i_{kin}$ and $i_{phot}$ is caused by the error in the
determination of the $i_{kin}$ value then the positions of galaxies
with positive values of d$i$ = $i_{kin}$ - $i_{phot}$ should show a
systematic shift relative to the positions of galaxies with negative
values of d$i$.  The MaNGA galaxies of our sample with different
values of d$i$ are shown by different symbols in
Fig.~\ref{figure:v-msp-di}.  There is no systematic shift of the
positions of galaxies with different values of d$i$.  This suggests
that the kinematic inclination angles derived here are quite reliable
and the difference between $i_{kin}$ and $i_{phot}$ is not caused by
the error in the determination of the $i_{kin}$ value. 

The stellar mass TF relation can also be used to test the reliability
of the choice of the representative value of the rotation velocity.
In Fig.~\ref{figure:v-msp-flat}, we plot $M_{sp}$ as a function of
rotation velocity along the flat part of the rotation curve,
$V_{rot,flat}$, for the galaxies where the $V_{rot,flat}$ value was
obtained (circles).  In Fig.~\ref{figure:v-msp-flat}, we show $M_{sp}$
as a function of maximum rotation velocity within the optical radius
$V_{rot,max}$ for the other galaxies (plus signs).  Inspection of
Fig.~\ref{figure:v-msp-flat} demonstrates that the $M_{sp}$ --
$V_{rot,flat}$ and the $M_{sp}$ -- $V_{rot,flat}$ diagrams are in
satisfactory agreement, i.e., the use of the $V_{rot,max}$ instead of
the $V_{rot,flat}$ does not change the general picture. 

This suggests that the use of the maximum rotation velocity within the
optical radius $V_{rot,max}$ as the representative value of the
rotation velocity is justified.

\section{Abundance and abundance gradient as a function of rotation
  velocity and other macroscopic characteristics of a galaxy}

\begin{figure}
\resizebox{1.00\hsize}{!}{\includegraphics[angle=000]{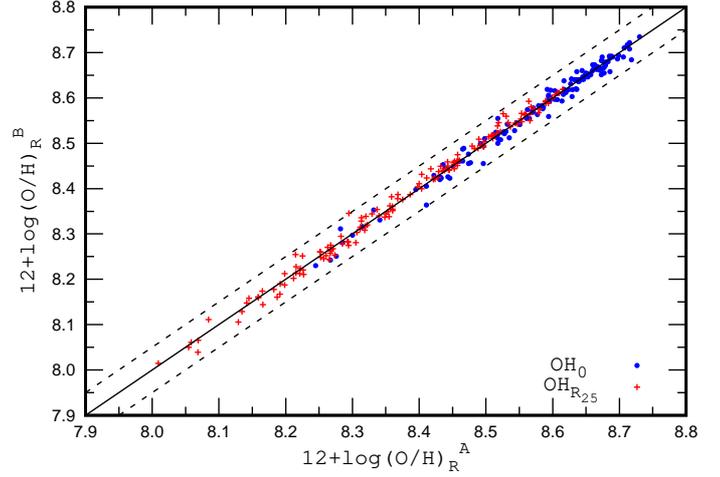}}
\caption{
  The circles show the central intersect oxygen abundance 12+log(O/H)$^{B}$
  for the radial gradient based on the points within 0.2 $<$ $R_{g}$ $<$ 0.8 
  as a function of central intersect oxygen abundance 12+log(O/H)$^{A}$
  for the radial gradient based on the points within $R_{g}$ $<$ 1. 
  The plus symbols show the same for the intersect oxygen abundance
  at the optical radius $R_{25}$.
  The solid line indicates equal values: the dashed lines show the
  $\pm$0.05 dex deviation from unity. 
  The oxygen abundances are determined through the $R$ calibration.
}
\label{figure:oh-oh-02-08}
\end{figure}

\begin{figure}
\resizebox{1.00\hsize}{!}{\includegraphics[angle=000]{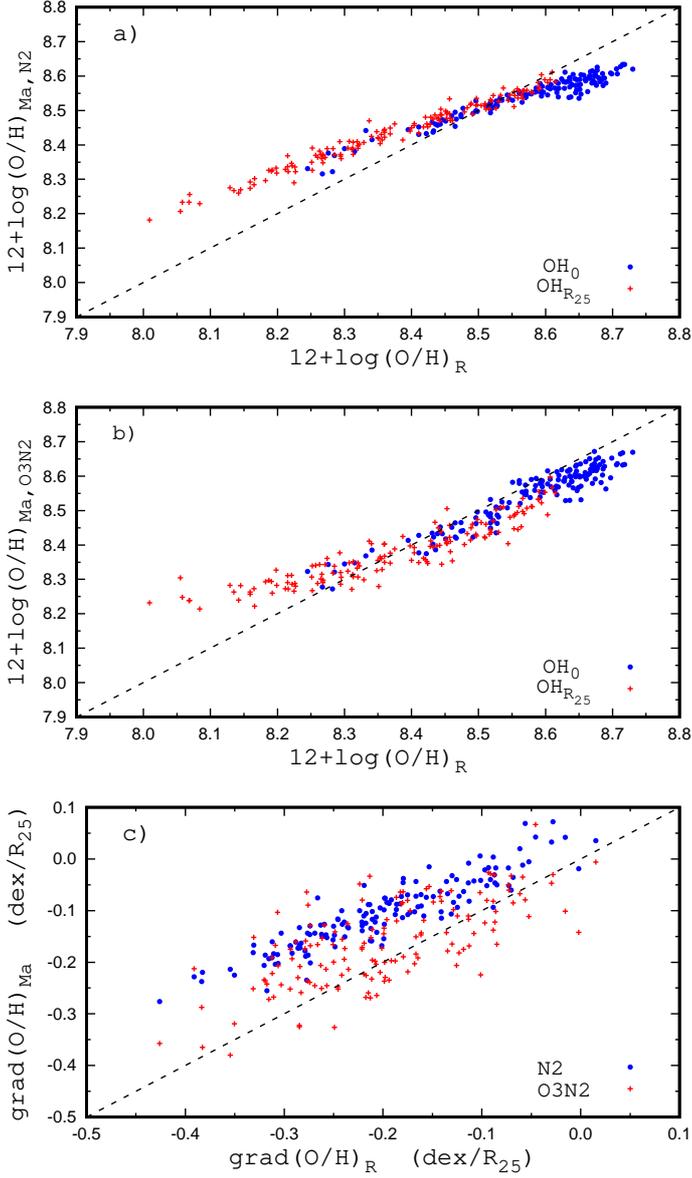}}
\caption{
  The circles in {\em Panel} $a$ show the comparison between central
  intersect oxygen abundances based on the abundances obtained through
  the $R$ calibration of \citet{Pilyugin2016} and that based on the abundances
  obtained through the $N2$ calibration of \citet{Marino2013}.
  The plus symbols shows the comparison between 
  intersect oxygen abundances at the optical radius $R_{25}$.
  The circles in {\em Panel} $b$ show the comparison between central
  intersect oxygen abundances based on the abundances obtained through
  the $R$ calibration and that based on the abundances
  obtained through the $O3N2$ calibration of \citet{Marino2013}.
  The plus symbols show the comparison between 
  intersect oxygen abundances at the optical radius $R_{25}$.
  The circles in {\em Panel} $c$ show the abundance gradient
  for the $N2$-based abundances as a function of the abundance gradient
  for $R$-based abundances.
  The plus symbols show the abundance gradient
  for the $O3N2$-based abundances as a function of abundance gradient
  for $R$-based abundances.
  The dashed line in each panel indicates equal values.
}
\label{figure:ohr-ohma}
\end{figure}

\begin{figure}
\resizebox{1.00\hsize}{!}{\includegraphics[angle=000]{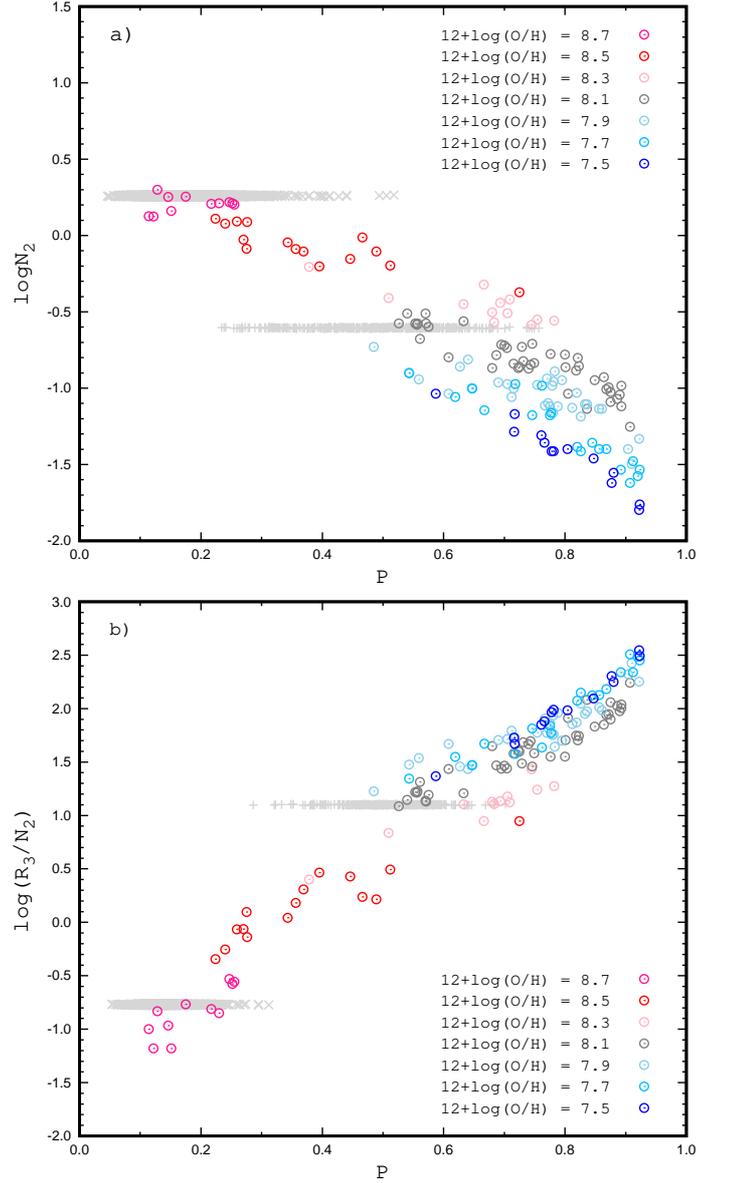}}
\caption{
  {\em Panel} $a$ shows the nitrogen line N2 intensity vs.\ excitation
  parameter $P$.
  The circles show the positions of the spectra of the calibrating objects
  of different metallicities (colour-coded) from  \citet{Pilyugin2016}.
  The plus symbols show the positions of the MaNGA spaxel spectra with
  $-0.600 > \log {\rm N}_{2} > -0.610$ (which correspond to an
  N2-based abundance of 12+log(O/H) $\sim 8.2$).
  The crosses show the positions of the MaNGA spaxel spectra with
  $0.265 > \log {\rm N}_{2} > 0.255$ (which correspond to an N2-based abundance
  of 12+log(O/H) $\sim 8.6$).
  {\em Panel} $b$ shows the R$_{3}$/N$_{2}$ line ratio vs excitation
  parameter $P$.
  The plus symbols show the positions of the MaNGA spaxel spectra with
  $1.105 > \log({\rm R}_{3}/{\rm N}_{2}) > 1.095$ (which correspond to an O3N2-based abundance
  of 12+log(O/H) $\sim 8.2$).
  The crosses show the positions of the MaNGA spaxel spectra with
  $-0.765 > \log({\rm R}_{3}/{\rm N}_{2}) > -0.775$ (which correspond to an O3N2-based abundance
  of 12+log(O/H) $\sim 8.6$).
  The notations of the calibrating objects are the same as in {\em panel} $a$.
}
\label{figure:p-n2-on}
\end{figure}

\begin{figure*}
\resizebox{1.00\hsize}{!}{\includegraphics[angle=000]{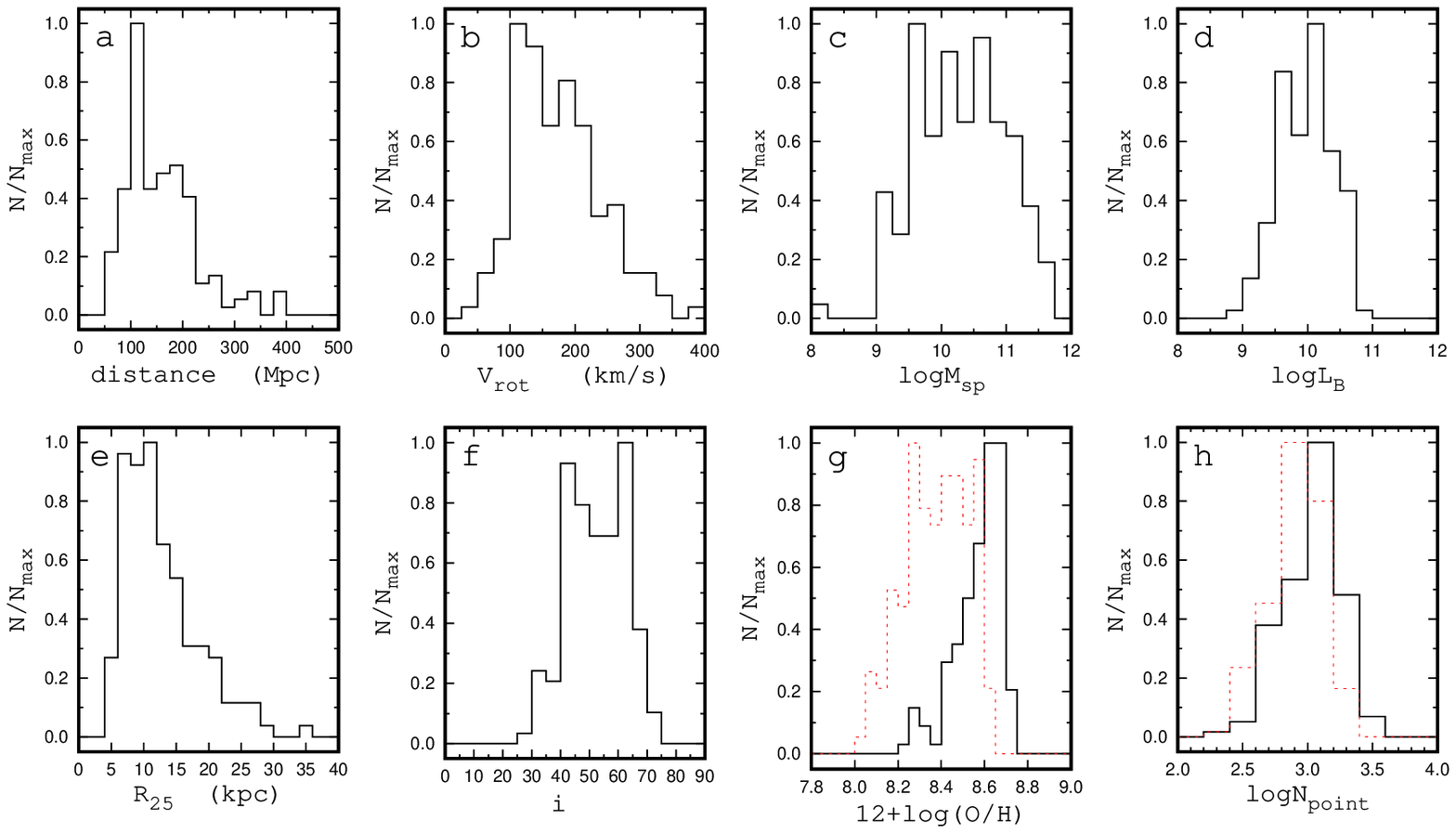}}
\caption{
  Properties of our sample of galaxies.
  The panels show the normalized histograms of the
  distances to our galaxies in Mpc ({\em panel} $a$),
  rotation velocities $V_{rot}$ in km/s ({\em panel} $b$),
  spectroscopic stellar masses $M_{sp}$ in solar units ({\em panel} $c$),
  luminosities $L_{B}$ in solar units ({\em panel} $d$),
  optical radii $R_{25}$ in kpc ({\em panel} $e$),
  inclination angles $i$ in degrees ({\em panel} $f$),
  central (intersect) oxygen abundances 12+log(O/H)$_{0}$ (solid line)
  and oxygen abundances at the optical radii 12+log(O/H)$_{R_{25}}$
  (dashed line) ({\em panel} $g$),
  number of points in the velocity map (solid line),
  and number of points used in the determination of the final rotation
  curve (dashed line) ({\em panel} $h$).
}
\label{figure:general}
\end{figure*}

\subsection{Abundance and abundance gradient}

We found in a previous study \citep{Pilyugin2018} that the
three-dimensional $R$ calibration, which uses the N$_{2}$, R$_{2}$ and R$_{3}$
lines, produces reliable abundances for regions (spaxels) with 
H\,{\sc ii} region-like spectra in the MaNGA galaxies.
One should distinguish clearly two questions: the nature of the
line-emitting gas (the sources of the ionizing radiation) and the
diagnostic of the emitting gas (the abundance determination).
The abundance determination through the strong line methods
(e.g., our and other calibrations) for objects with 
H\,{\sc ii} region-like spectra is based  on the fundamental 
assumption that if two objects have similar spectra
(similar relative intensities of the emission lines,
normalized to the H$\beta$ line) then the physical conditions and 
abundances are similar in those objects. This suggests that the
calibrations are applicable to any object with an H\,{\sc ii}
region-like spectrum independent of the sources of the ionizing
radiation.

The demarcation line between AGNs and H\,{\sc ii} regions from 
\citet{Kauffmann2003} in the standard diagnostic diagram of
the [N\,{\sc ii}]$\lambda$6584/H$\alpha$ versus the
[O\,{\sc iii}]$\lambda$5007/H$\beta$ line ratios suggested by
\citet{Baldwin1981} is a 
useful criterion to reject spectra with significantly distorted
strengths of the N$_{2}$ and R$_{3}$ lines. 
To examine the distortion of the R$_{2}$ line,  
we have compared the behaviour of the line intensities and
the abundances estimated through the $R$ calibration for   
samples of slit spectra of  H\,{\sc ii} regions in nearby galaxies,
of the fibre spectra from the SDSS, and of the spaxel spectra of the
MaNGA survey for objects located left (or below) of the Kauffmann et al.'s
demarcation line in the BPT diagram  \citep{Pilyugin2018}. 
It has been found that the mean distortion of the R$_{2}$ (and  N$_{2}$)  
is less than a factor of $\sim$1.3. This suggests that the 
Kauffmann et al.'s demarcation line in the BPT diagram is
also a relatively reliable criterium to reject the spectra with
distorted strength of the R$_{2}$ line. 
Thus, the Kauffmann et al.'s demarcation line in the BPT diagram serves to
select objects with H\,{\sc ii} region-like spectra, i.e. to reject the
spectra with distorted strengths of the  R${_3}$, N$_{2}$, and R$_{2}$ lines
used in determination of the oxygen abundance through the $R$ calibration. 
It should be noted that the Kauffmann et al.'s demarcation line can be
insufficient to distinguish the nature of the line-emitting gas although
the regions where photoionization is dominated by hot, low-mass, evolved
stars (hDIG regions) are located right (or above) of Kauffmann et al.'s
demarcation line in the BPT diagram  \citep{Lacerda2018}.
We use here Kauffmann et al.'s demarcation line to select
spaxels with H\,{\sc ii}-region-like spectra. All those
spaxels are used for the investigation of the abundance distribution across
a galaxy.
The oxygen abundances for individual spaxels are estimated through
the 3-D $R$ calibration from \citet{Pilyugin2016}.
It should be noted that the calibrating data points for our 3-D $R$ 
calibration were also selected using the  Kauffmann et al.'s demarcation
line, i.e. both the calibrating objects and the objects to which the
calibration is applied are selected using the same criterion.
For comparison, the oxygen abundances are also determined through the
widely used N2 and O3N2 indexes introduced by \citet{Pettini2004}.
The N2 and O3N2 calibration relations suggested by \citet{Marino2013}
are used.   
 
The radial oxygen abundance distribution in a spiral galaxy is
traditionally described by a straight line of the form 
\begin{equation}
{\rm (O/H)}^{*} = {\rm (O/H)}_{0}^{*} + grad \, \times \, R_{g}
\label{equation:gradient}
\end{equation}
where (O/H)$^{*}$ $\equiv$ 12 + log(O/H)($R$) is the oxygen abundance
at the fractional radius $R_{g}$ (normalized to the optical radius $R_{25}$),
(O/H)$_{0}^{*}$ $\equiv$ 12 + log(O/H)$_{0}$ is the intersect central oxygen abundance,
and $grad$ is the slope of the oxygen abundance gradient
expressed in terms of dex/$R_{25}$. 
The radial abundance gradients within the
optical radius in the discs of the majority of spiral galaxies are
reasonably well fitted by this relation although in some cases breaks
in the radial abundance gradients near the centre or near the optical
radius can occur
\citep[e.g.,][]{VilaCostas1992,Edmunds1984,Zaritsky1994,vanZee1998,Pilyugin2001,Pilyugin2003,Sanchez2012b,
Sanchez2014,Ho2015,Pilyugin2014,Pilyugin2017b,Zinchenko2016,SanchezMenguiano2016,SanchezMenguiano2018}.
To examine the influence of a possible break in the radial abundance
gradient on the central (intersect) abundance and on the abundance
at the optical radius (intersect), the value of the gradient based
on the points with galactocentric distances within 0.8 $>$ $R_{g}$ $>$ 0.2
is also obtained for each galaxy.
Fig.~\ref{figure:oh-oh-02-08} shows the comparison between the central
abundances (and the abundances at the optical radius) obtained from the gradient
for the points with galactocentric distances within 
0 $<$ $R_{g}$ $<$ 1 (labeled as case A in the Fig.~\ref{figure:oh-oh-02-08})
and the central abundances (and the abundances at the optical radius)
obtained from the gradient for the points with galactocentric distances
within 0.2 $<$ $R_{g}$ $<$ 0.8 (labeled as case B in the  Fig.~\ref{figure:oh-oh-02-08}).    
The circles show the central intersect oxygen abundances, and 
the plus symbols show the intersect oxygen abundance
at the optical radius $R_{25}$.
The solid line indicates equal values: the dashed lines show the
$\pm$0.05 dex deviation from unity. 
Inspection of Fig.~\ref{figure:oh-oh-02-08} shows that the central intersect
abundances (and the intersect abundances at the optical radius) for the cases $A$ and $B$
are in agreement, the mean difference is $\sim$0.013 dex.
This is not surprising since it was found in our previous study that the maximum
absolute difference between the abundances in a disk given by broken and
pure linear relations is less than 0.05 dex for the majority of galaxies
\citep{Pilyugin2017b}. The values of the  central intersect abundance,
the abundance at the optical radius and the radial abundance gradient 
obtained using the points with galactocentric distances within  $0 < R_{g} < 1$
will be examined below.

Panel $a$ in Fig.~\ref{figure:ohr-ohma} shows the comparison
between the abundances obtained through the $R$ calibration of \citet{Pilyugin2016}
and through the $N2$ calibration of \citet{Marino2013}.
The circles are the central intersect oxygen abundances and 
the plus symbols are the intersect oxygen abundances at the optical radius $R_{25}$.
Panel $b$ in Fig.~\ref{figure:ohr-ohma} shows the comparison
between the abundances obtained through the $R$ calibration 
and through the $O3N2$ calibration.
Again the circles are the central intersect oxygen abundances and 
the plus symbols are the intersect oxygen abundances at the optical radius $R_{25}$.
Panel $c$ in Fig.~\ref{figure:ohr-ohma} shows the comparison
between the abundance gradients for the $R$-based, the $N2$-based
and the $O3N2$-based abundances. The circles show the $N2$-based
gradients versus the $R$-based gradients. The plus signs denote the $O3N2$-based
gradients versus the $R$-based gradients.
Inspection of Fig.~\ref{figure:ohr-ohma} shows that the N2-based (and O3N2-based)
abundances differ from the $R$-based abundances, and the difference changes
with metallicity.

  Fig.~\ref{figure:p-n2-on} demonstrates clearly the origin of the
  difference between the N2-based (and O3N2-based) and the $R$-based
  abundances and the change of the difference with metallicity.
  The positions of the spectra of the calibrating objects
  of different metallicities from  \citet{Pilyugin2016} in the
  nitrogen line N2 intensity vs excitation parameter $P$ diagram
  are shown by the circles in panel $a$ of Fig.~\ref{figure:p-n2-on}. 
  The plus symbols show the positions of the MaNGA spaxel spectra with
  the nitrogen line N$_{2}$ intensity within --0.600 $>$ logN$_{2}$ $>$ --0.610. 
  The crosses show the positions of the MaNGA spaxel spectra with
  0.265 $>$ logN$_{2}$ $>$ 0.255.
  The 1-D N2 calibration is based on the assumption that there is a one-to-one
  correspondence between the oxygen abundance and the nitrogen line N2 intensity,
  i.e., (O/H)$_{N2}$ = $f$(N2). In that case all the objects with the same  nitrogen line N2
  intensity have the same (O/H)$_{N2}$ abundance. In particular, all the
  spaxels shown by the plus symbols  in the panel $a$ in Fig.~\ref{figure:p-n2-on}
  have an abundance of 12 + log(O/H)$_{N2}$ $\sim$8.2, and all the
  spaxels shown by the crosses  in the panel $a$ in Fig.~\ref{figure:p-n2-on}
  have 12 + log(O/H)$_{N2}$ $\sim$8.6.
  It is evident from the general consideration, that the nitrogen line N2 intensity
  in the spectrum of the object depends not only on its oxygen abundance but
  also on the nitrogen-to-oxygen ratio N/O and on the excitation $P$, 
  i.e. (O/H) = $f$(N2,N/O,$P$). The positions of the calibrating objects
  in the nitrogen line N2 intensity vs excitation parameter $P$ diagram
  show clearly that the objects of different metallicities can show
  a similar nitrogen line N2 intensity depending on its excitation.
  Then the N2 calibration produces a realistic oxygen abundance in the
  objects with a given  nitrogen line N2 intensity for one value of the
  excitation parameter $P^{*}$ only. The N2 calibration results in overestimated
  oxygen abundances for objects with an excitation parameter lower than $P^{*}$
  and results in underestimated oxygen abundances for objects with an excitation
  parameter higher than $P^{*}$.

  It is difficult to indicate the exact value of the $P^{*}$ for a given value
  of the  nitrogen line N2 intensity. 
  The N2 calibration relation from \citet{Marino2013}
  is based on H\,{\sc ii} regions with abundances derived through
  the direct method. Then, the value of $P^{*}$ lies within the interval of
  excitation parameters covered by the calibrating
  data points. Inspection of {\em panel} $a$ in Fig.~\ref{figure:p-n2-on} shows that
  the positions of the MaNGA spaxel spectra with $-0.600 > \log {\rm N}_{2} > -0.610$
  extend significantly towards the lower excitation parameters as compared to the
  calibrating points. The oxygen abundances in those MaNGA spaxels produced by the N2
  calibration are overestimated. 
  On contrary, the positions of the MaNGA spaxels spectra with   0.265 $>$ logN$_{2}$ $>$ 0.255 
  expand towards the higher excitation parameter as compared to the
  calibrating points. The oxygen abundances in those MaNGA spaxels produced by the N2
  calibration are underestimated. Thus, one can expect that the N2-based oxygen
  abundances should be overestimated, on average,  at low metallicities and should
  be underestimated, on average,   at high metallicities.
  Fig.~\ref{figure:ohr-ohma} shows just such behaviour of the N2-based abundances in comparison
  to the $R$-based abundances.

It should be emphasized that the variation in the N/O ratios is also
neglected in 1-D calibrations. This results in an additional uncertanty in
the oxygen abundances determined through the 1-D calibrations.

Panel $b$ of Fig.~\ref{figure:p-n2-on}  shows the R$_{3}$/N$_{2}$ line ratio
vs.\ excitation parameter $P$.
The plus symbols show the positions of the MaNGA spaxel spectra with
$1.105 > \log({\rm R}_{3}/{\rm N}_{2}) > 1.095$ (which correspond to O3N2-based abundance
of 12+log(O/H) $\sim$8.2).
The crosses show the positions of the MaNGA spaxel spectra with
$-0.765 > \log({\rm R}_{3}/{\rm N}_{2}) > -0.775$ (which correspond to O3N2-based abundance
of 12+log(O/H) $\sim$8.6).
The circles show the positions of the spectra of the calibrating objects
of different metallicities (colour-coded) from  \citet{Pilyugin2016}.
Examination of panel $b$ in Fig.~\ref{figure:p-n2-on} suggests that  
the O3N2 calibration results in overestimated
oxygen abundances for objects with an excitation parameter lower than $P^{*}$
and results in underestimated oxygen abundances for objects with an excitation
parameter higher than $P^{*}$.

Since the value of the over(under)estimation of the oxygen abundance
obtained through the 1-D (N2- and O3N2) calibration depends on the metallicity
the radial abundance gradient derived using a 1-D calibration is
not beyond question. The overestimation of the oxygen abundance 
increases with galactocentric distance (with the decrease of metallicity) and,
as a consequence, the use of the 1-D calibration for the abundance determination
results in an underestimation of the absolute value of the slope of the gradient.
Panel $c$ of Fig.~\ref{figure:ohr-ohma} shows clearly that the absolute values of the slopes
of the gradients based on the (O/H)$_{\rm N2}$  abundances
are smaller than those for the gradients based on the (O/H)$_{\rm R}$ abundances.
The absolute values of the slopes   of the gradients based on the (O/H)$_{\rm O3N2}$
abundances are also smaller, on average, than those for the gradients based on
the (O/H)$_{\rm R}$ abundances.
  
The change of the over(under)estimation of the oxygen abundance determined
through the 1-D calibration with galactocentric   distance (with metallicity) 
can also result in an artificial break in the  abundance gradient.

Thus, the (O/H)$_{\rm N2}$  and the (O/H)$_{\rm O3N3}$ abundances
are excuded from consideration.
The abundance gradients determined for the (O/H)$_{R}$ abundances
in the points with galactocentric distances within  0 $<$ $R_{g}$ $<$ 1
and corresponding central abundances and abundances at the optical
radius $R_{25}$ will be used in analysis below.

\subsection{Properties of the final sample of the galaxies}

The maximum rotation velocity within the optical radius is adopted as
the representative value of the rotation velocity of a galaxy,
$V_{rot}$. We use the rotation curve and the values of the position
angle of the major axis of the galaxy, {\em PA}, and the inclination
angle, $i$, derived for the case $B$. The surface brightness profile
(and, consequently, isophotal radius $R_{25}$ and disc scale length
$h_{d}$) for each target galaxy and the galactocentric distances of
the spaxels used in the determination of the radial abundance gradient
are also estimated with these values of the position angle of the
major axis and the inclination angle.  The oxygen abundances for
individual spaxels are estimated through the 3-D $R$ calibration from
\citet{Pilyugin2016}.  We adopt the spectroscopic stellar masses
$M_{sp}$ of the galaxies.  These properties are listed in Table
\ref{table:sample}.

Fig.~\ref{figure:general} shows the properties of our final sample of
galaxies, i.e., the normalized histograms of the distances to our
galaxies in Mpc (panel $a$), rotation velocities $V_{rot}$ in
km/s (panel $b$), spectroscopic stellar masses $M_{sp}$
in solar units (panel $c$), luminosities $L_{B}$ in solar units
(panel $d$), optical radii $R_{25}$ in kpc (panel $e$),
inclination angles $i$ in degrees (panel $f$), central
(intersect) oxygen abundances 12+log(O/H)$_{0}$ (solid line) and
oxygen abundances at the optical radii 12+log(O/H)$_{R_{25}}$ (dashed
line) (panel $g$), and number of points in the velocity map
(solid line) and number of points used in the determination of final
rotation curve (dashed line) (panel $h$).  The distances were
taken from the NASA Extragalactic Database ({\sc ned})\footnote{The
NASA/IPAC Extragalactic Database ({\sc ned}) is operated by the Jet
Propulsion Laboratory, California Institute of Technology, under
contract with the National Aeronautics and Space Administration.  {\tt
http://ned.ipac.caltech.edu/}}.  The {\sc ned} distances use flow
corrections for Virgo, the Great Attractor, and Shapley Supercluster
infall. The spectroscopic stellar masses are taken from the SDSS data
base, table {\sc stellarMassPCAWiscBC03}.  Other parameters were
derived in our current study.  Inspection of Fig.~\ref{figure:general}
shows that the selected galaxies are located at distances from $\sim
50$ to $\sim 400$ Mpc and show a large variety of physical
characteristics. The optical radii of the galaxies cover the interval
from around 5 kpc to around 35 kpc.  The velocity maps typically
contain hundreds or a few thousand data points.

\subsection{Relations between oxygen abundance and other macroscopic parameters}

\begin{figure}
\resizebox{1.00\hsize}{!}{\includegraphics[angle=000]{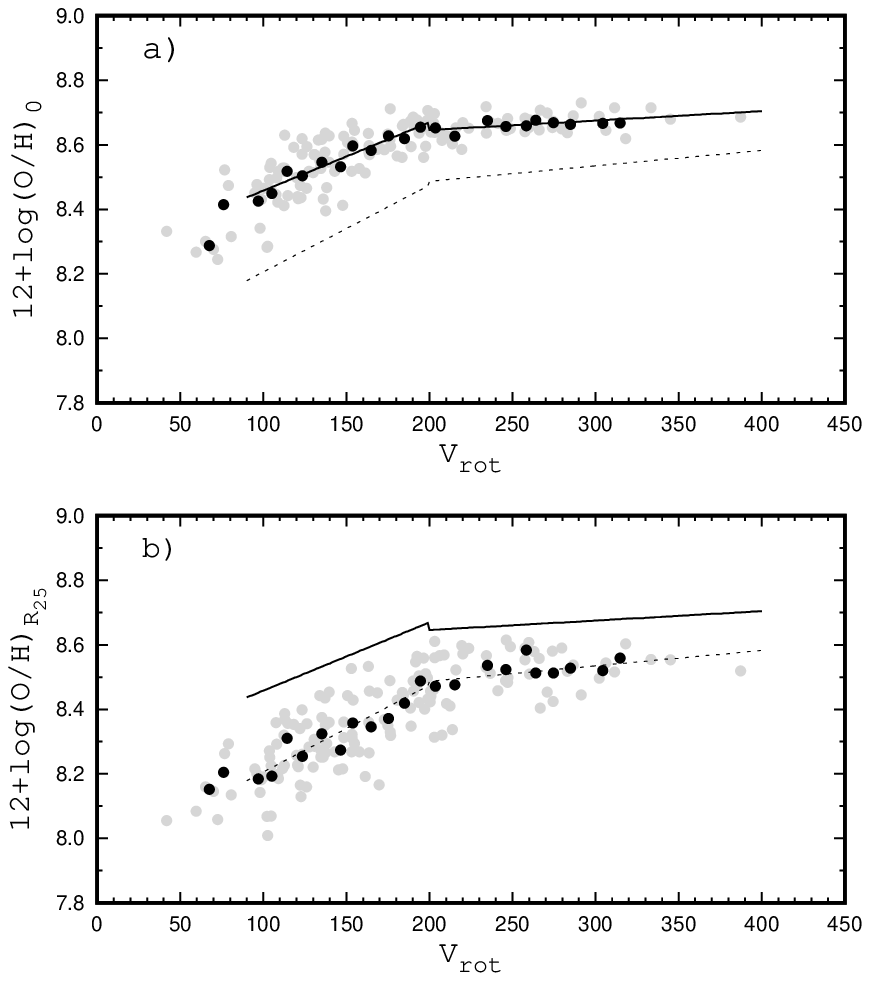}}
\caption{
The central oxygen abundance (O/H)$_{0}$ ({\em panel} $a$) and the
oxygen abundance at the optical radius (O/H)$_{R_{25}}$ ({\em panel}
$b$) as a function of the rotation velocity $V_{rot}$.  The grey
points stand for the individual galaxies; the dark points are the
binned mean values.  The solid line is the broken (O/H)$_{0}$ -- $V_{rot}$
relation, and the dashed line is the broken (O/H)$_{R_{25}}$ --
$V_{rot}$ relation. 
}
\label{figure:v-oh}
\end{figure}

\begin{figure*}
\resizebox{1.00\hsize}{!}{\includegraphics[angle=000]{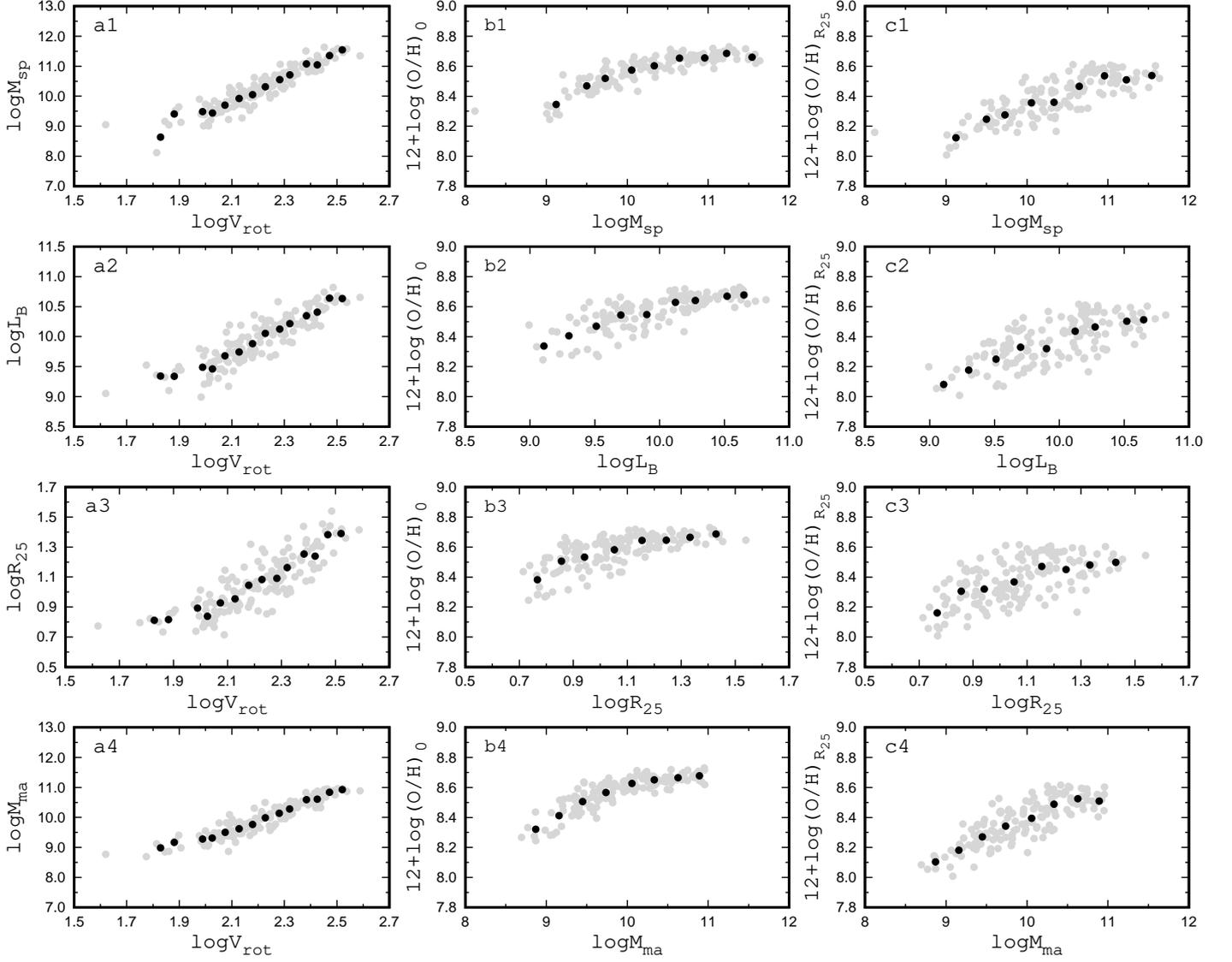}}
\caption{
The {\em left column panels} show the parameters $X$ (= $M_{sp}$,
$L_{B}$, $R_{25}$, $M_{ma}$; from top to bottom) as a function of the value of 
the rotation velocity
$V_{rot}$.  The {\em middle column panels} show the central oxygen
abundance (O/H)$_{0}$ as a function of the parameters $X$.  The {\em
right column panels} show the oxygen abundance at the optical radius
(O/H)$_{R_{25}}$ as a function of the parameters $X$.  The grey points
at each panel represent the data for individual MaNGA galaxies from our
sample; the dark points are the mean values in bins.
}
\label{figure:v-x-oh}
\end{figure*}

\begin{figure*}
\resizebox{1.00\hsize}{!}{\includegraphics[angle=000]{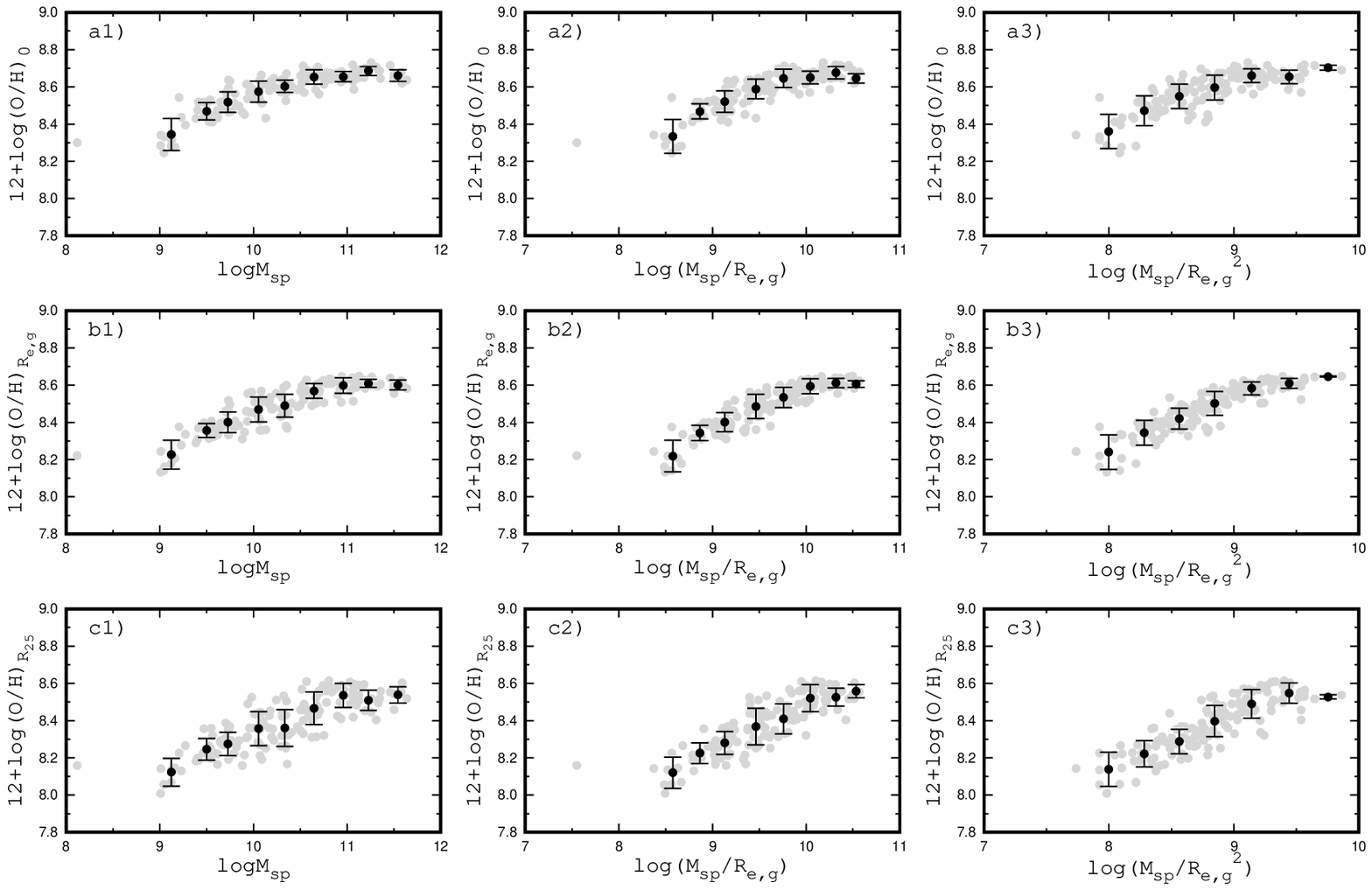}}
\caption{The central oxygen abundance ({\em row a panels}), 
  the abundance at the galaxy effective radius ({\em row b panels}),
  and at the optical radius  ({\em row c panels}) as a function of
  the stellar mass $M_{sp}$ ({\em left column panels}), as a function
  of $M_{sp}/R_{e,g}$  ({\em middle column panels}), and as a function of
   $M_{sp}/R_{e,g}^{2}$  ({\em right column panels}).
The grey points in each panel represent the data for individual MaNGA galaxes from our
sample, while the dark points are the binned mean values. 
}
\label{figure:de-oh}
\end{figure*}

Fig.~\ref{figure:v-oh} shows the central oxygen abundance (O/H)$_{0}$
(panel $a$) and the oxygen abundance at the optical radius
(O/H)$_{R_{25}}$ (panel $b$) as a function of the rotation
velocity $V_{rot}$. The points represent the individual galaxies; the
solid line is the broken (O/H)$_{0}$ -- $V_{rot}$ relation (best fit),
\begin{equation}
{\rm (O/H)}_{0}^{*} = 0.00212(\pm0.00021)\, V_{rot} + 8.246(\pm0.031) 
\label{equation:oho-v-l}
\end{equation}
for $90 < V_{rot} < 200$ km/s, and 
\begin{equation}
{\rm (O/H)}_{0}^{*} = 0.00029(\pm0.00010)\, V_{rot} + 8.587(\pm0.025) 
\label{equation:oho-v-h}
\end{equation}
for $V_{rot} > 200$ km/s.  The dashed line is the broken 
(O/H)$_{R_{25}}$ -- $V_{rot}$ relation (best fit),
\begin{equation}
{\rm (O/H)}_{R_{25}}^{*} = 0.00269(\pm0.00029)\, V_{rot} + 7.936(\pm0.042) 
\label{equation:oh25-v-l}
\end{equation}
for $90 < V_{rot} < 200$ km/s, and 
\begin{equation}
{\rm (O/H)}_{R_{25}}^{*} = 0.00048(\pm0.00026)\, V_{rot} + 8.351(\pm0.066) 
\label{equation:oh25-v-h}
\end{equation}
for $V_{rot} > 200$ km/s. The notation (O/H)$^{*}$ = 12 + log(O/H) 
is used for the sake of brevity. 

Fig.~\ref{figure:v-oh} shows that both the central oxygen abundance
(O/H)$_{0}$ and the oxygen abundance at the optical radius
(O/H)$_{R_{25}}$ correlate with the rotation velocity $V_{rot}$ in a
similar way, in the sense that the abundance rises with increasing
$V_{rot}$ and there is a break in the abundance growth rate at
$V_{rot}^{*} \sim 200$ km/s, i.e., the growth rate is lower for
galaxies with high rotation velocities.  It is difficult to establish
the exact value of the dividing rotation velocity $V_{rot}^{*}$
because of the scatter in the (O/H) -- $V_{rot}$ diagrams.  It should
be noted that the number of galaxies with $V_{rot} < 90$ km/s
is small in our sample. Therefore we do not discuss galaxies with low
rotation velocities in the current study. 

The mean value of the residuals of the (O/H)$_{0}$ -- $V_{rot}$
relation is 0.053 dex, and the mean scatter around the
(O/H)$_{R_{25}}$ -- $V_{rot}$ relation is 0.081 dex.  Thus the oxygen
abundance at any galactocentic distance can be estimated with a mean
error of around 0.08 dex from the rotation velocity through the
interpolation between the (O/H)$_{0}$ and (O/H)$_{R_{25}}$ values
obtained from the relations (O/H)$_{0} = f(V_{rot}$) and
(O/H)$_{R_{25}} = f(V_{rot}$), respectively.  

The increase of both the central oxygen abundance (O/H)$_{0}$ and the
oxygen abundance at the optical radius (O/H)$_{R_{25}}$ with
increasing rotation velocity $V_{rot}$ is compatible with the
so-called ``inside-out'' scenario for disc growth, where the formation
and evolution are faster in the inner part of the disc compared to the
outer disc. Within the simple model for the chemical evolution of
galaxies the local oxygen abundance is defined by the gas mass
fraction (or astration level) at a given galactocentric distance. It
should be noted that the optical radius moves outward during galaxy
evolution because the stellar surface mass density at each
galactocentric distance increases with time due to star formation.
Since the total (star + gas) surface mass density of the disc
decreases with galactocentric distance, a fixed value of the stellar
surface mass density at a larger galactocentric distance is reached at
a larger astration level and, consequently, at a higher oxygen
abundance. Thus, the more evolved (the more massive) galaxies have
higher astration levels (and, consequently, higher oxygen abundances)
both at the centre and at the optical radius.

The left column panels of Fig.~\ref{figure:v-x-oh} show the basic
parameters $X$ of a galaxy (where $X$ stands for $M_{sp}$ in 
panel $a1$, $L_{B}$ (panel $a2$),  $R_{25}$ (panel
$a3$), and  $M_{ma}$ in panel $a4$)
as a function of the rotation velocity $V_{rot}$.  The grey
points in each panel represent the data for individual MaNGA galaxies
from our sample; the dark points are mean values in bins of $V_{rot}$.
The middle column panels of Fig.~\ref{figure:v-x-oh} show the central
oxygen abundance (O/H)$_{0}$ as a function of the parameters $X$.  The
right column panels of Fig.~\ref{figure:v-x-oh} show the oxygen
abundance at the optical radius (O/H)$_{R_{25}}$ as a function of
parameters $X$.
The scatter in the central oxygen abundances and in the abundances
at the optical radius relative the mean  values in bins are reported
in Table \ref{table:ohscatter}.

\begin{table}
  \caption[]{\label{table:ohscatter}
  Scatter in the O/H values in the $X$ vs O/H diagrams where O/H is 
   the central oxygen abundance 12 + log(O/H)$_{0}$ (column 2),
  the oxygen abundances at the optical radius 12 + log(O/H)$_{R_{25}}$ (column 3),
  and at the galaxy effective radius  12 + log(O/H)$_{R_{e,g}}$ (column 4). 
  $X$ is the rotation velocity $V_{rot}$ (row 1),
  the stellar mass $M_{ma}$ (row  2),  the luminosity $L_{B}$ (row 3),
  the optical radius $R_{25}$ (row 4), the stellar mass $M_{sp}$ (row 5),
  and the parameters  $M_{sp}$/$R_{e,g}$ and   $M_{sp}$/$R_{e,g}^{2}$ (rows 6 and 7).
  The diagrams are shown in Fig.~\ref{figure:v-oh}, Fig.~\ref{figure:v-x-oh},
   and Fig.~\ref{figure:de-oh}.
   The scatter relative to the mean values in bins is given.
}
\begin{center}
\begin{tabular}{lcccccccc} \hline \hline
X-                                                       &
\multicolumn{3}{|c|}{scatter in 12+logO/H)$_{R_{x}}$}      \\           
                   &
(O/H)$_{0}$         &
(O/H)$_{R_{25}}$      & 
(O/H)$_{R_{e,g}}$     \\   \hline         
$V_{rot}$   (km/s)                             &   0.053     &  0.074   &         \\ 
log$M_{ma}$    ($M_{\sun}$)                     &  0.054      &  0.084   &         \\ 
log$L_{B}$    ($L_{\sun}$)                      &  0.069      &  0.101   &         \\ 
log$R_{25}$   (kpc)                            &  0.073      &  0.114   &         \\ 
log$M_{sp}$  ($M_{\sun}$)                       &  0.047      &  0.077   &  0.052  \\ 
log($M_{sp}/R_{e,g}$)   ($M_{\sun}$/kpc)         &  0.050      &  0.074   &  0.052  \\ 
log($M_{sp}/R_{e,g}^{2}$)   ($M_{\sun}$/kpc$^2$)  &  0.060      &  0.073   &  0.055  \\ 
                    \hline
\end{tabular}\\
\end{center}
\end{table}

Inspection of  Fig.~\ref{figure:v-oh} and Fig.~\ref{figure:v-x-oh}
shows that the variation of both (O/H)$_{0}$ and (O/H)$_{R_{25}}$ with
any macroscopic parameter $X$ ($X$ = $V_{rot}$, $M_{sp}$ or $M_{ma}$, $L_{B}$,
$R_{25}$) is similar in the sense that the abundance grows with the
increase of the value of the parameters $X$ and there is break in the
abundance growth rate at some value of $X^{*}$, i.e., the growth rate
becomes lower beyond some value of the parameter $X^{*}$.  Again, it
is difficult to establish the exact value of the $X^{*}$ because of
the scatter in the (O/H) -- $X$ diagrams.

Inspection of the Table \ref{table:ohscatter} shows that the scatter
in the central oxygen abundances is minimum in the (O/H)$_{0}$  --  $M_{sp}$
diagram being lower than that in the (O/H)$_{0}$  --  $V_{rot}$ diagram.
Is the correlation between the oxygen abundance and the stellar mass 
tighter than the correlation between the oxygen abundance and the rotation
velocity? From general considerations, this can be the case.  
If the intrinsic scatter in the baryonic Tully-Fisher relation is negligible
\citep{McGaugh2015} then the scatter in oxygen abundances in the
(O/H)$_{0}$  --  $M_{baryonic}$ and in the (O/H)$_{0}$  --  $V_{rot}$ diagram
for a sample of the purely rotating systems should be similar.
The stellar mass of the galaxy is defined by the baryonic mass 
of the galaxy and the astration level. The oxygen abundance in the 
galaxy is also defined by the astration level. 
Therefore the fluctuations in the astration level in the galaxies of a given 
baryonic mass (a given rotation velocity) result in a larger scatter
in the  (O/H)$_{0}$  --  $V_{rot}$ than in the (O/H)$_{0}$  --  $M_{sp}$ diagram.
On the other hand, the scatter in the oxygen abundances is a combination
of the intrinsic scatter and that from the random errors in the abundance
and stellar mass (rotation velocity) determinations. The difference between the
scatter in the (O/H)$_{0}$  --  $M_{sp}$ and the (O/H)$_{0}$  --  $M_{ma}$
diagrams is larger than the difference between the scatter in the
(O/H)$_{0}$  --  $M_{sp}$ and the (O/H)$_{0}$  --  $V_{rot}$ diagrams.
This may be considered as evidence in favour that the random errors
in the abundance and stellar mass (rotation velocity) determinations
make a significant contribution to the scatter in those diagrams.
Therefore it is difficult to conclude if the intrinsic scatter in the
oxygen abundance vs.\ stellar mass diagram is lower than that in
the  oxygen abundance vs.\ rotation velocity diagram.

\citet{DEugenio2018} have investigated the correlation of the
oxygen abundance with the stellar mass $M_{sp}$ and with two parameters
that are combinations of the value of the stellar mass and the
value of the galactic effective radius ($M_{sp}/R_{e,g}$ and
$M_{sp}/R_{e,g}^{2}$) for 68959 SDSS galaxies. They found that the scatter in
the O/H --  $M_{sp}/R_{e,g}$ diagram is lower than that in the 
O/H --  $M_{sp}$  and in the O/H --  $M_{sp}/R_{e,g}^{2}$ diagrams.
The oxygen abundances measured in the SDSS galaxies can be considered 
as central oxygen abundances.
The radial abundance gradients determined here provide a possibility
to examine the correlation of the oxygen abundance at any
fixed galactocentric distance with the parameters considered by \citet{DEugenio2018}.
We will consider the central oxygen abundance (O/H)$_{0}$, the oxygen
abundance at the galactic effective radius (O/H)$_{R_{e,g}}$, and the
abundance at the optical radius (O/H)$_{R_{25}}$.
Fig.~\ref{figure:de-oh} shows those values of the oxygen abundance as a function of 
stellar mass $M_{sp}$ and of the parameters $M_{sp}/R_{e,g}$ and  $M_{sp}/R_{e,g}^{2}$.
The grey points in each panel are the data for individual MaNGA galaxes from our
sample, while the dark points are the binned mean values. 
The scatter in the abundances in those diagrams relative to the mean
values in bins is reported in Table \ref{table:ohscatter}.

Inspection of Fig.~\ref{figure:de-oh} and Table \ref{table:ohscatter} shows
that the scatter in the central oxygen abundances (O/H)$_{0}$ for our sample of MaNGA
galaxies is minimum in the  (O/H)$_{0}$ --  $M_{sp}$ diagram,
the scatter in the oxygen abundances at the galaxy effective radius (O/H)$_{R_{e,g}}$ 
is similar (minimum) in the  (O/H)$_{R_{e,g}}$ --  $M_{sp}$ and in the (O/H)$_{R_{e,g}}$  --  $M_{sp}/R_{e,g}$
diagrams, and the scatter in the oxygen abundances at the optical radius (O/H)$_{R_{25}}$
is minimum in the (O/H)$_{R_{25}}$  --  $M_{sp}/R_{e,g}^{2}$ diagram.
Then it is difficult to conclude if our data are in agreement or in conflict
with the conclusion of  \citet{DEugenio2018}.

Thus, the variation of both (O/H)$_{0}$ and (O/H)$_{R_{25}}$ with
any macroscopic parameter $X$ ($X$ = $V_{rot}$, $M_{sp}$ or $M_{ma}$, $L_{B}$,
$R_{25}$) is similar in the sense that the abundance grows with the
increase of the value of the parameters $X$ and there is break in the
abundance growth rate at some value of $X^{*}$, i.e., the growth rate
becomes lower beyond some value of the parameter $X^{*}$.  It
is difficult to establish the exact value of the $X^{*}$ because of
the scatter in the (O/H) -- $X$ diagrams.

\subsection{The oxygen abundance gradient as a function of the 
macroscopic parameters}

\begin{figure*}
\resizebox{1.00\hsize}{!}{\includegraphics[angle=000]{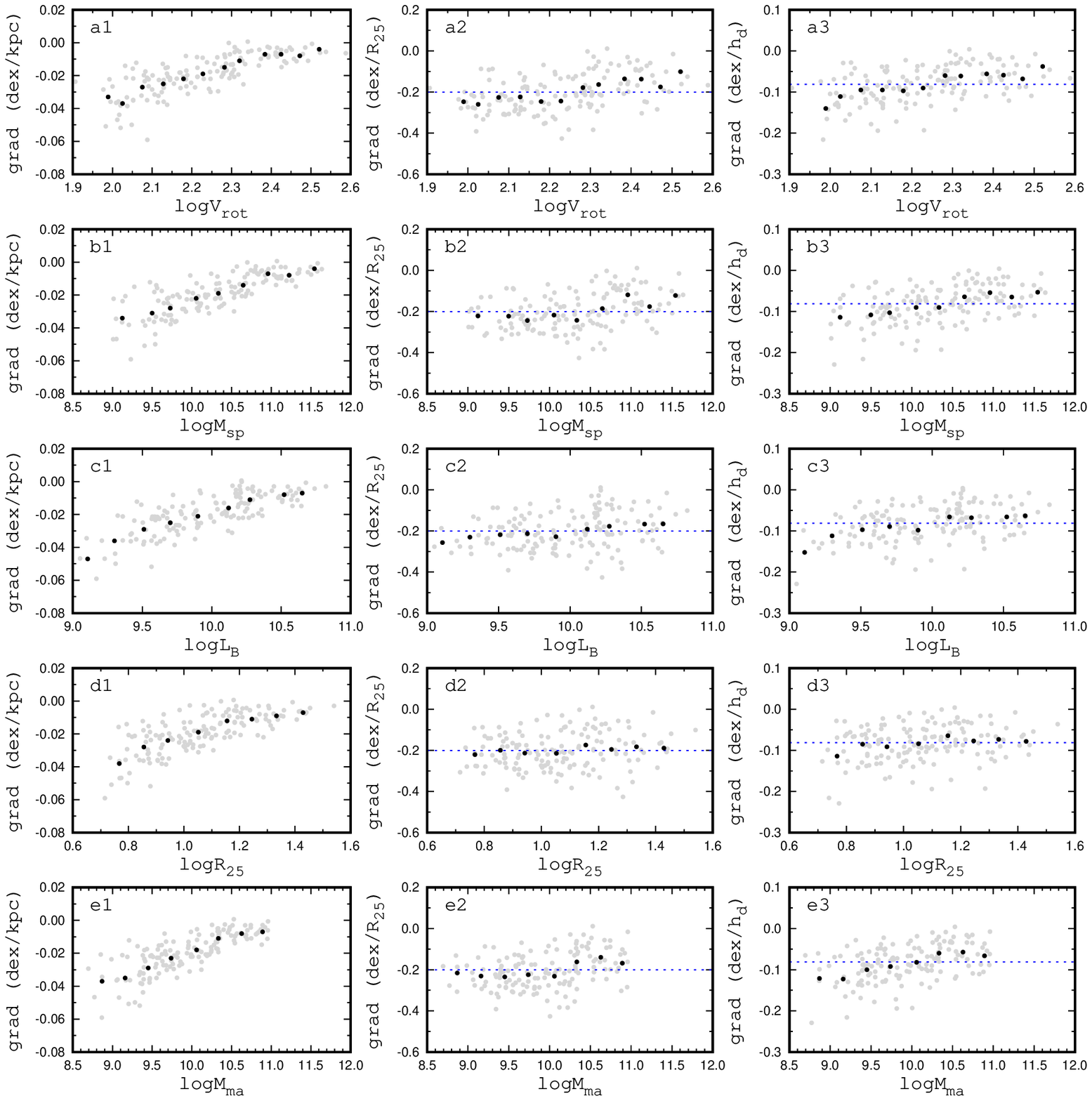}}
\caption{
The radial abundance gradient measured in dex/kpc ({\em left
column panels}), with respect to the optical radius $R_{25}$  ({\em
middle column panels}), and with respect to the disc scale length
$h_{d}$ ({\em right column panels}). The {\em row a panels} show the
radial abundance gradient as a function of the rotation velocity
$V_{rot}$, the {\em row b panels} as a function of the 
stellar mass $M_{sp}$, the {\em row c panels} as a function of galaxy
luminosity $L_{B}$, the {\em row d panels} as a function of
optical radius $R_{25}$, and  the {\em row e panels} as a function of the 
stellar mass $M_{ma}$,  The grey points at each panel are the data
for individual MaNGA galaxes from our sample, while the dark points
are the binned mean values.  The dashed line in the {\em middle} and
{\em right column panels} show the mean values of the gradient in units
of dex/$R_{25}$ and dex/$h_{d}$, respectively. 
}
\label{figure:v-x-gradoh}
\end{figure*}

\begin{figure}
\resizebox{1.00\hsize}{!}{\includegraphics[angle=000]{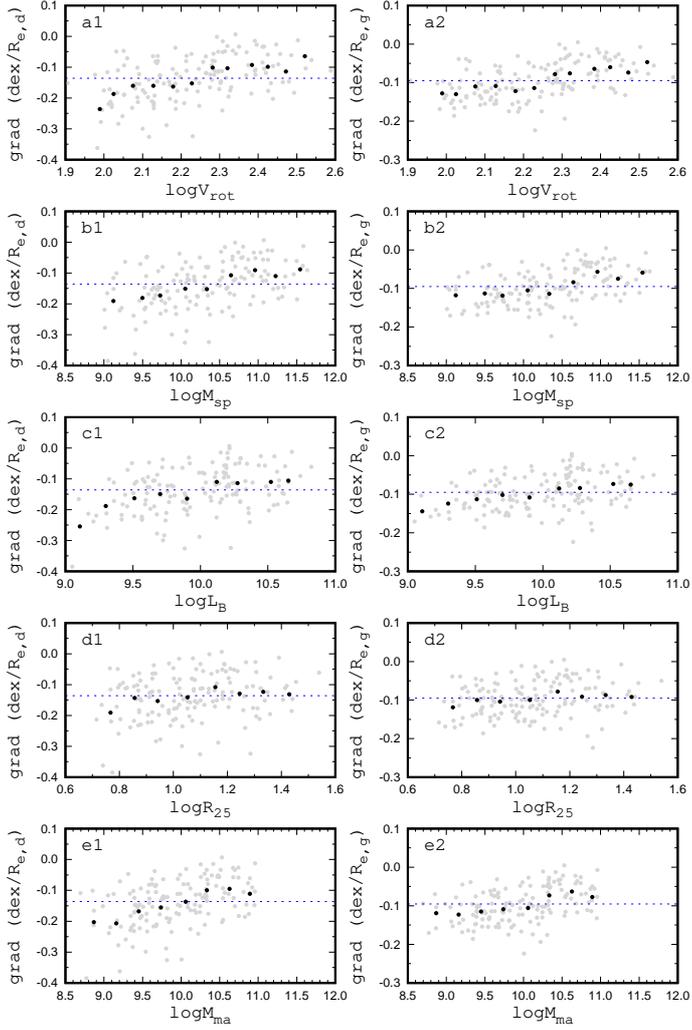}}
\caption{
The radial abundance gradient measured in dex/$R_{e,d}$ ({\em left
column panels}) and dex/$R_{e,g}$ ({\em right column panels}).
The {\em row a panels} show the
radial abundance gradient as a function of the rotation velocity
$V_{rot}$, the {\em row b panels} as a function of the 
stellar mass $M_{sp}$, the {\em row c panels} as a function of galaxy
luminosity $L_{B}$, the {\em row d panels} as a function of
optical radius $R_{25}$, and  the {\em row e panels} as a function of the 
stellar mass $M_{ma}$,  The grey points at each panel are the data
for individual MaNGA galaxes from our sample, while the dark points
are the binned mean values.  The dashed line in the {\em left} and
{\em right column panels} show the mean values of the gradient in units
of dex/$R_{e,d}$ and dex/$R_{e,g}$, respectively. 
}
\label{figure:gradre}
\end{figure}

\begin{figure}
\resizebox{1.00\hsize}{!}{\includegraphics[angle=000]{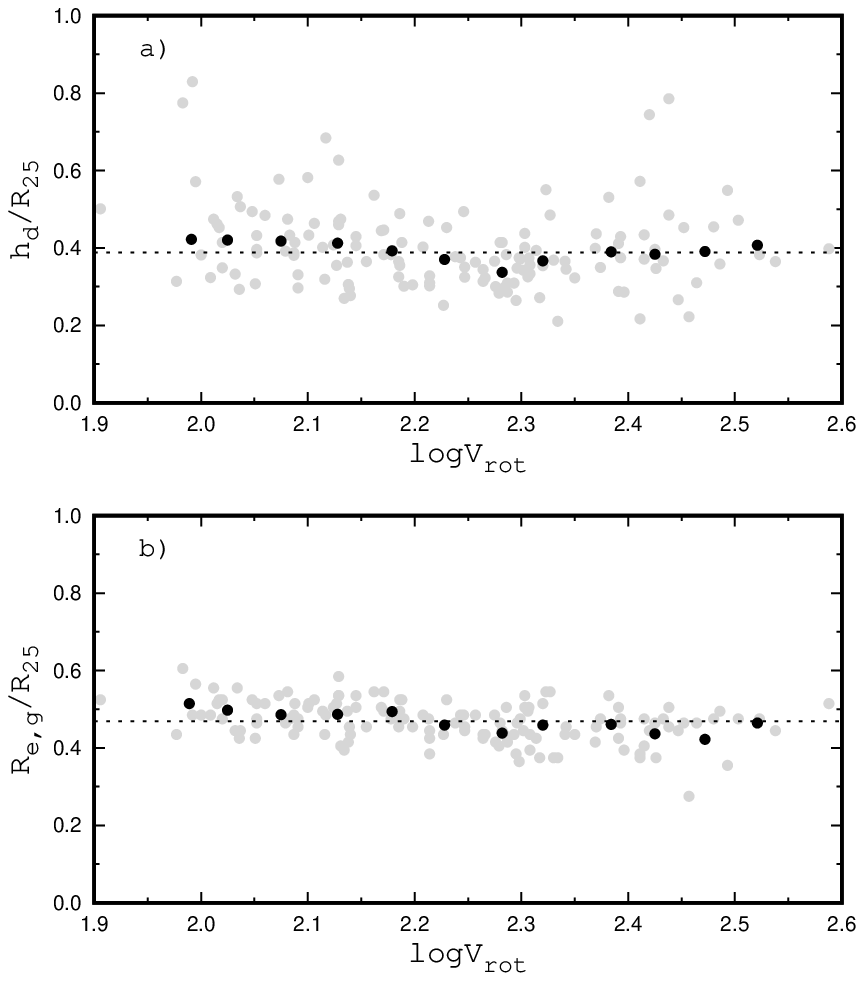}}
\caption{
{\em Panel} $a$ shows the ratio of the disc scale length to the optical radius
$h_{d}$/$R_{25}$ as a function of the value of the rotation
velocity $V_{rot}$.  The grey points represent the data for individual MaNGA
galaxies from our sample, and the dark points indicate the binned mean values
(omitting objects with $h_{d}/R_{25} > 0.7$).  The dotted line is the mean
$h_{d}$/$R_{25}$ value.
{\em Panel} $b$ shows the same but for the ratio of the effective
radius of the galaxy to the optical radius $R_{e,g}$/$R_{25}$. 
}
\label{figure:v-h}
\end{figure}

As noted above, the radial oxygen abundance distribution in a spiral galaxy is
traditionally described by a straight line of the form 
\begin{equation}
{\rm (O/H)}^{*} = {\rm (O/H)}_{0}^{*} + grad_{dex/R_{scale}} \, \times \, R/R_{scale}
\label{equation:grad}
\end{equation}
where (O/H)$^{*}$ $\equiv$ 12 + log(O/H)($R$) is the oxygen abundance
at the galactocentric distance $R$, (O/H)$_{0}^{*}$ $\equiv$ 12 +
log(O/H)$_{0}$ is the extrapolated central oxygen abundance,
$grad_{dex/R_{scale}}$ is the slope of the oxygen abundance gradient
expressed in terms of dex/$R_{scale}$, and $R_{scale}$ is the
adopted radial scale.

Different radial scalings are used to quantify the radial abundance
variation in the discs of spiral galaxies. An absolute physical radius
in kiloparsecs ($R_{scale} = 1$) and different dimensionless radii
normalized to the isophotal radius ($R_{scale} = R_{25}$) or to the
Holmberg radius ($R_{scale} = R_{26.5}$) of the galaxy, normalized to
the disc exponential scale length ($R_{scale} = h_{d}$), normalized to
the effective (or half-light) radius of the disc ($R_{scale} = R_{e,d}$)
or the effective radius of the galaxy ($R_{scale} = R_{e,g}$) were used to specify
the radial abundance distributions in galaxies \citep[][among many
others]{Smith1975,Alloin1979,VilaCostas1992,Oey1993,
Zaritsky1994,Ryder1995,Garnett1997,Garnett1998,Pilyugin2004,
Pilyugin2015,Sanchez2012b,Sanchez2014,Ho2015,Bresolin2015,Belfiore2017}. 
We consider here the values of the gradients for five
radial scales ($R_{scale} =1$, $h_{d}$, $R_{25}$, $R_{e,d}$, $R_{e,g}$),
i.e., the gradients are expressed in the terms of dex/kpc, dex/$h_{d}$,
dex/$R_{25}$, dex/$R_{e,d}$, and dex/$R_{e,g}$. 

The left column panels of Fig.~\ref{figure:v-x-gradoh} show the
oxygen abundance gradients in units of dex/kpc as a
function of rotation velocity $V_{rot}$ (panel $a1$),
spectroscopic stellar mass $M_{sp}$ (panel $b1$), luminosity in
the $B$-band $L_{B}$ (panel $c1$), optical radius $R_{25}$
(panel $d1$), and stellar mass $M_{ma}$ (panel $e1$)
for our sample of galaxies.  The grey points in
each panel stand for the individual galaxies, and the dark points are
the binned mean values.  The middle column panels of
Fig.~\ref{figure:v-x-gradoh} show the oxygen abundance gradients in
units of dex/$R_{25}$ as a function of $V_{rot}$, $M_{sp}$, $L_{B}$, 
$R_{25}$, and  $M_{ma}$. 
The dashed line is the mean value of the
gradient --0.200 dex/$R_{25}$ for our sample of galaxies.  The
right column panels of Fig.~\ref{figure:v-x-gradoh} show the
oxygen abundance gradients in units of dex/$h_{d}$ as a function
of $V_{rot}$, $M_{sp}$, $L_{B}$, $R_{25}$, and  $M_{ma}$. The dashed line
is the mean value of the gradient --0.081 dex/$h_{d}$ for our
sample of galaxies. The scatter in the gradients relative the mean
values in bins is reported in Table \ref{table:gradscatter}.

\begin{table}
  \caption[]{\label{table:gradscatter}
    Scatter in the radial oxygen abundance gradients in the gradient vs.\ $X$ diagrams
    where the radial gradient is given in dex/kpc (column 2),  dex/$h_{d}$ (column 3),
    dex/$R_{e,d}$ (column 4), dex/$R_{e,g}$ (column 5), and dex/$_{R_{25}}$ (column 6).
    $X$ is the rotation velocity $V_{rot}$ (row 1),
    the stellar masses $M_{sp}$ and $M_{ma}$ (rows 2 and 3),  the luminosity $L_{B}$ (row 4),
    and the optical radius $R_{25}$ (row 5).
    The diagrams are shown in Fig.~\ref{figure:v-x-gradoh} and  Fig.~\ref{figure:gradre}. 
    The scatter relative to the mean values in bins is listed. 
}
\begin{center}
\begin{tabular}{lccccc} \hline \hline
X                          &
\multicolumn{5}{|c|}{Scatter in the radial O/H gradient}    \\           
                   &
dex/kpc            &           
dex/$h_{d}$         &           
dex/$R_{e,d}$       &
dex/$R_{e,g}$       &
dex/$R_{25}$       \\   \hline         
log$V_{rot}$        &  0.0073    &  0.036   &  0.061  &  0.039 &  0.077    \\
log$M_{sp}$         &   0.0074    &  0.039   &  0.066  &  0.040  &  0.077   \\
log$M_{ma}$         &  0.0074    &   0.038  &  0.064  &  0.039 &  0.079    \\
log$L_{B}$          &  0.0074    &  0.038   & 0.065   & 0.042  &  0.085    \\
log$R_{25}$         &   0.0083    &  0.042   & 0.071   & 0.044  &  0.086    \\
                    \hline
\end{tabular}\\
\end{center}
\end{table}

The left column panels of Fig.~\ref{figure:gradre} show the radial abundance
 gradient in units of dex/$R_{e,d}$ as a
function of rotation velocity $V_{rot}$ (panel $a1$),
spectroscopic stellar mass $M_{sp}$ (panel $b1$), luminosity in
the $B$-band $L_{B}$ (panel $c1$), optical radius $R_{25}$
(panel $d1$), and stellar mass $M_{ma}$ (panel $e1$)
for our sample of galaxies.  The grey points in
each panel stand for the individual galaxies, and the dark points are
the binned mean values.  The dashed line is the mean value of the
gradient --0.136 dex/$R_{e,d}$ for our sample of galaxies.
The value of the disc effective radius $R_{e,d}$ is proportional to the disc
scale length, $R_{e,d}$  =   1.67835$h_{d}$, \citep{Sanchez2014}. 
The right column panels of Fig.~\ref{figure:gradre} show the
oxygen abundance gradients in units of dex/$R_{e,g}$ as a function
of $V_{rot}$, $M_{sp}$, $L_{B}$, $R_{25}$, and  $M_{ma}$, respectively. The dashed line
is the mean value of the gradient, $-0.095$ dex/$R_{e,g}$, for our
sample of galaxies.
The scatter in the gradients with respect to the mean
values in bins is listed in Table \ref{table:gradscatter}.

Examination of panel $a1$ of Fig.~\ref{figure:v-x-gradoh} shows
that the oxygen abundance gradients in units of dex/kpc flatten
with the increase of the rotation velocity and there is a break in
the rate of flattening at $V_{rot} \sim 200$ km/s in the sense
that the flattening rate with increasing $V_{rot}$ is lower for
galaxies with high rotation velocities.  The left column panels
of Fig.~\ref{figure:v-x-gradoh} show that the behaviour of the change
of the oxygen abundance gradients in units of dex/kpc is
similar for increasing $V_{rot}$, $M_{sp}$, $M_{ma}$, $L_{B}$, and $R_{25}$.

Inspection of panel $a2$ of Fig.~\ref{figure:v-x-gradoh} shows
that the abundance gradient expressed in dex/$R_{25}$ is rougly
constant for $V_{rot} \la 200$ km/s, shows a break at
$V_{rot} \sim 200$ km/s, and remains again roughly constant for
higher rotation velocities.  The middle column panels of
Fig.~\ref{figure:v-x-gradoh} show that the general picture of the
change of the oxygen abundance gradient in units of dex/$R_{25}$
kpc with any mascroscopic parameter $X$ ($X$ = $M_{sp}$, $M_{ma}$,
$L_{B}$, $R_{25}$) is similar to that for $V_{rot}$, in the sense
that the mean value of the oxygen abundance gradient for the
objects with $X$ $<$ $X^{*}$ differs from that for objects
with $X$ $>$ $X^{*}$, see Table \ref{table:grad2}.

\begin{table*}
  \caption[]{\label{table:grad2}
    The mean values of the radial abundance gradient for objects with
    $X < X^{*}$ and   $X > X^{*}$ where $X$ is the rotation velocity $V_{rot}$ (row 1),
    the stellar mass $M_{sp}$ and $M_{ma}$ (rows 2 and 3),  the luminosity $L_{B}$ (row 4),
    and the optical radius $R_{25}$ (row 5).
    The value of $X^{*}$ is reported in column 2, and the numbers of objects with
    $X < X^{*}$ and  $X > X^{*}$ are given in columns 3 and 4, respectively.
}
\begin{center}
\begin{tabular}{lcccccccccccc} \hline \hline
X                             &
X$^{*}$                        &
\multicolumn{2}{|c|}{n objects}       & 
\multicolumn{8}{|c|}{mean value of radial abundance gradient in dex/scale}   \\
                              &
                              &
                              &
                              &
\multicolumn{2}{|c|}{dex/$R_{25}$}  &
\multicolumn{2}{|c|}{dex/$h_{d}$}  &
\multicolumn{2}{|c|}{dex/$R_{e,d}$}  &
\multicolumn{2}{|c|}{dex/$R_{e,g}$}  \\           
                              &
                              &
X $<$ X$^{*}$                 &
X $>$ X$^{*}$                 &
X $<$ X$^{*}$                 &
X $>$ X$^{*}$                 &
X $<$ X$^{*}$                 &
X $>$ X$^{*}$                 &
X $<$ X$^{*}$                 &
X $>$ X$^{*}$                 &
X $<$ X$^{*}$                 &
X $>$ X$^{*}$                 \\   \hline         
log$V_{rot}$ (km/s)           &  2.3  & 92  & 47 & -0.226  & -0.149  &  -0.092  &  -0.059  & -0.155  &  -0.099  &  -0.109  &  -0.068    \\
log$M_{sp}$  ($M_{\sun}$)      &  10.5 & 76  & 59 & -0.232  & -0.158  &  -0.097  &  -0.060  & -0.164  &  -0.101  &  -0.113  &  -0.072    \\
log$M_{ma}$    ($M_{\sun}$)    &  10.3 & 91  & 45 & -0.231  & -0.148  &  -0.094  &  -0.058  & -0.159  &  -0.097  &  -0.112  &  -0.066    \\
log$L_{B}$    ($L_{\sun}$)     &  10.0 & 64  & 75 & -0.226  & -0.178  &  -0.099  &  -0.066  & -0.166  &  -0.110  &  -0.112  &  -0.081    \\
log$R_{25}$   (kpc)           &  1.1  & 82  & 57 & -0.213  & -0.182  &  -0.088  &  -0.070  & -0.148  &  -0.118  &  -0.103  &  -0.084    \\
                    \hline
\end{tabular}\\
\end{center}
\end{table*}

A comparison between the middle and right column panels of
Fig.~\ref{figure:v-x-gradoh} shows that the general picture of the
variation of the oxygen abundance gradient in units of
dex/$R_{25}$ kpc and in units of dex/$h_{d}$ is similar.
This is because there is no significant change in the
ratio of the disc scale length to the optical radius
$h_{d}$/$R_{25}$ as a function of rotation velocity $V_{rot}$ for
our sample of galaxies; see panel $a$ of Fig.~\ref{figure:v-h}.
A comparison between Fig.~\ref{figure:v-x-gradoh} and
Fig.~\ref{figure:gradre} shows that the qualitative trends in the
variation of the oxygen abundance gradient in units of
dex/$R_{25}$ kpc and in units of dex/$R_{e,g}$ is also similar.
Again, this is because of the change in the
ratio of the galactic effective radius to the optical radius
$R_{e,g}$/$R_{25}$ as a function of rotation velocity $V_{rot}$ for
our sample of galaxies is rather small as can be see in panel $b$ of Fig.~\ref{figure:v-h}.
The variation of the oxygen abundance gradient in units of
dex/$h_{d}$ kpc and in units of dex/$R_{e,d}$ is similar
because those gradients differ by a constant factor only. 
It was noted above
that the definition of the disc scale length $h_{d}$ may be debatable
for many galaxies since the surface brightness profile of the disc is
often broken.

Thus, the radial abundance gradient expressed
in dex/kpc flattens with the increase of the rotation velocity.
The slope of the relation is very low for galaxies with $V_{rot} \ga V_{rot}^{*}$.
The abundance gradient expressed in dex/$R_{25}$ is
rougly constant for galaxies with $V_{rot} \la  V_{rot}^{*}$, flattens
towards $V_{rot}^{*}$, and then again is roughly constant for galaxies
with $V_{rot} \ga V_{rot}^{*}$.  The change of the gradient expressed
in terms of dex/$h_{d}$, dex/$R_{e,d}$, and dex/$R_{e,g}$ 
with rotation velocity is similar to that for the gradient in
dex/$R_{25}$.  The qualitative trends in the relations between abundance gradients 
and other basic parameters (stellar mass, luminosity, and radius) are
are similar to that for the abundance gradient vs.\ rotation velocity relation.

\section{Discussion}

\begin{figure*}
\resizebox{1.00\hsize}{!}{\includegraphics[angle=000]{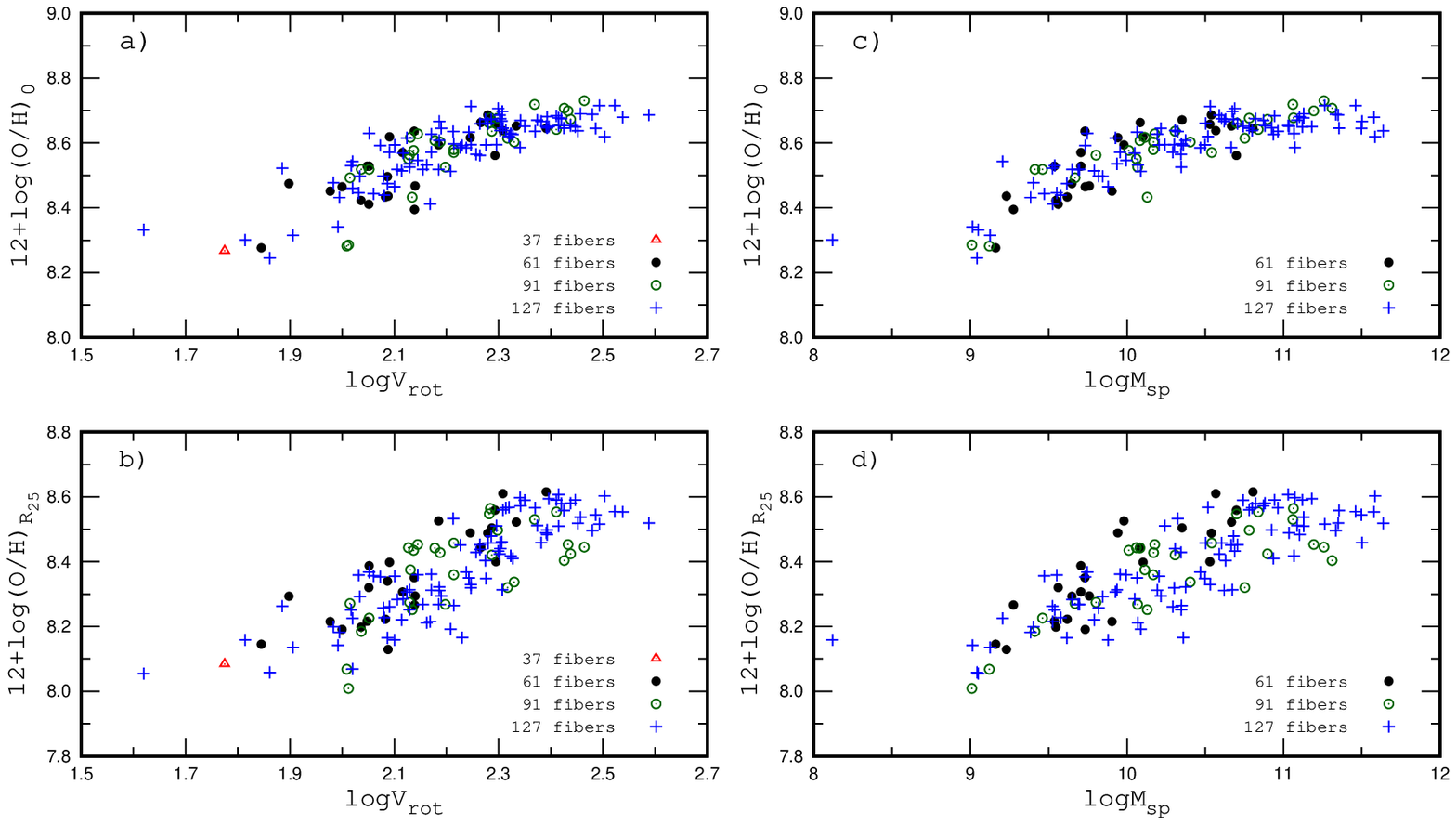}}
\caption{
{\em  Panel} $a$ shows the central oxygen abundance as a
function of rotation velocity for our sample of MaNGA galaxies.
Depending on their angular size, the number of fibers covering the
galaxies varied during the observations with the integral field units
employed by MaNGA.  The differing numbers of fibers are indicated by
different symbols.  {\em  Panel} $b$ shows the same as {\em panel}
$a$ but for the oxygen abundances at the isophotal radius.
{\em  Panel} $c$ shows the central oxygen abundance as a
function of stellar mass $M_{sp}$. 
Again the different numbers of fibers are indicated by
different symbols.  
{\em  Panel} $d$ shows the same as {\em panel}
$c$ but for the oxygen abundances at the isophotal radius.
The value of the $M_{sp}$ for the galaxy measured with 37 fibers
is not available.}
\label{figure:vr-oh-nfiber}
\end{figure*}

\begin{figure}
\resizebox{1.00\hsize}{!}{\includegraphics[angle=000]{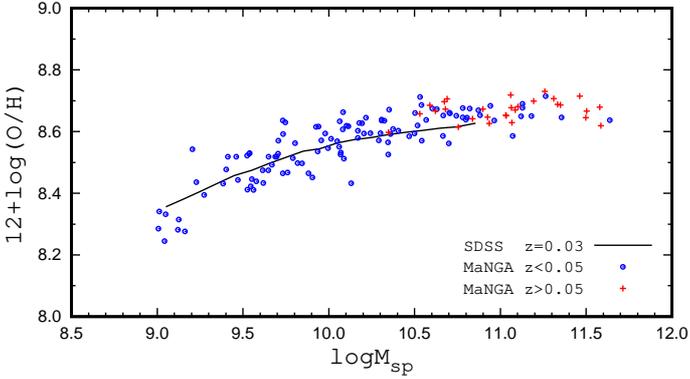}}
\caption{
The oxygen abundance as a function of the spectroscopic stellar mass.  The
oxygen abundances at the centres of individual MaNGA galaxies with
redshift $z < 0.05$ are shown by circles, and those for galaxies at 
$z > 0.05$ are shown by plus signs.  The solid line is the O/H --
$M_{sp}$ relation obtained for a large sample of SDSS galaxies with
redshift $z = 0.03$ in \citet{Pilyugin2017a}.
}
\label{figure:msp-oh-comparison}
\end{figure}

\begin{figure}
\resizebox{1.00\hsize}{!}{\includegraphics[angle=000]{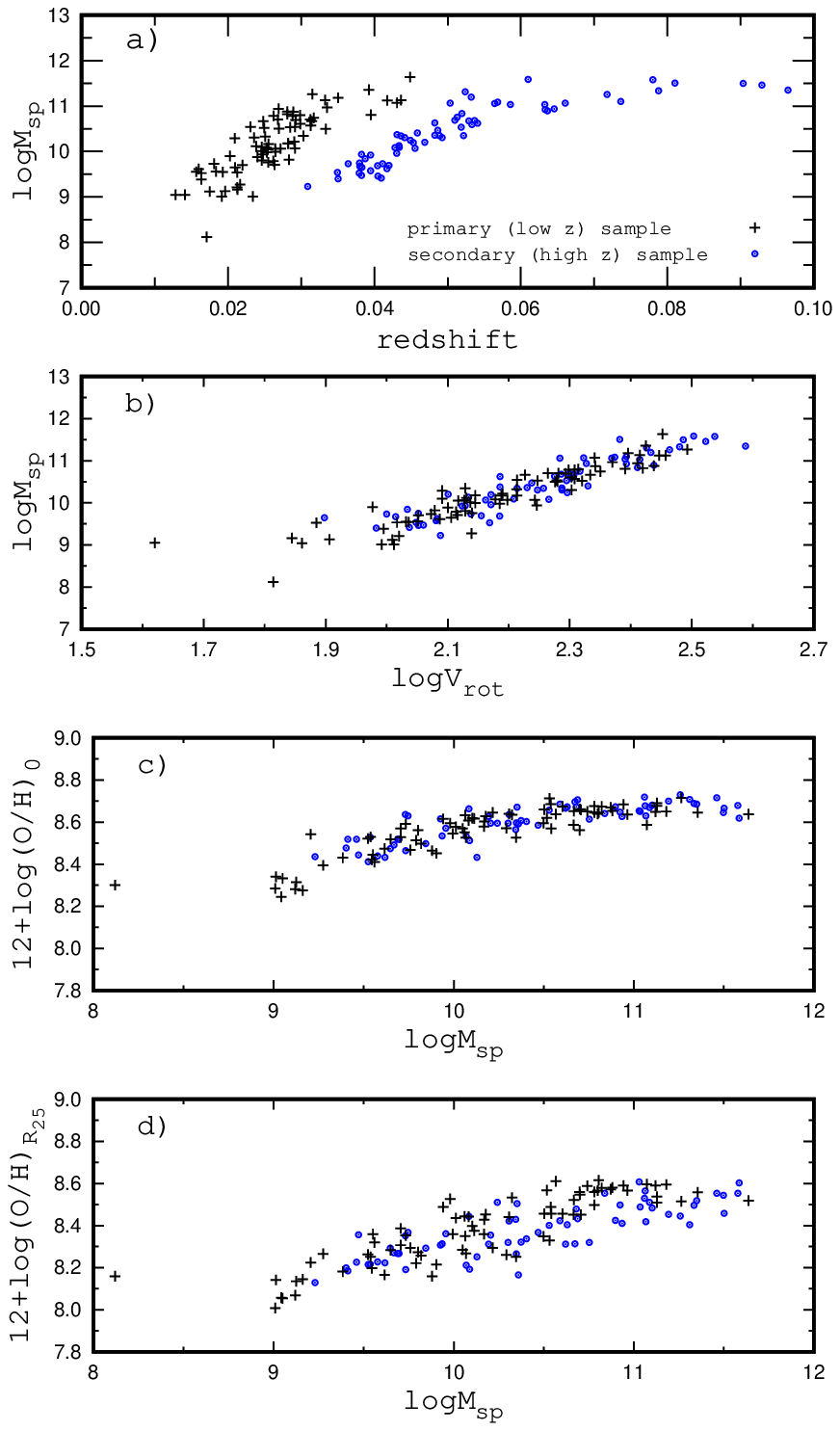}}
\caption{
Panel $a$ shows the stellar mass as a function of
redshift for our galaxy sample. The galaxies of
the primary (``low redshift'') and secondary (``high redshift'') subsamples
of the MaNGA targets \citep{Bundy2015,Wake2017} are shown by different symbols.
Panel $b$ shows the stellar mass Tully-Fisher relation for those galaxies. 
Panels $c$ and $d$ show the central oxygen abundance
(panel $c$) and the oxygen abundance at the optical radius (panel $d$) as a function
of stellar mass of the galaxy.
}
\label{figure:redshift-m}
\end{figure}

In this study, we are considering galaxies with different angular
sizes, implying that differing numbers of fibers are needed to cover
them when observing them with integral field units (IFUs) as employed
by MaNGA.  \citet{Belfiore2017} discussed that the abundance
distribution obtained in a MaNGA galaxy can depend on the conditions
of the observations, i.e., on the ratio of the galaxy effective radius
$R_{e,g}$ to the PSF or on the number of the
fibers with which a galaxy was covered.

Panel $a$ of Fig.~\ref{figure:vr-oh-nfiber} shows the (O/H)$_{0}$ -- $V_{rot}$
diagram (the same as panel $a$ of Fig.~\ref{figure:v-oh}) but
marking galaxies observed with different numbers of fibers by
different  symbols.  We see that galaxies observed with differing
numbers of fibers still follow the same trend in the (O/H)$_{0}$ --
$V_{rot}$ diagram.  Panel $b$ of
Fig.~\ref{figure:vr-oh-nfiber} shows the (O/H)$_{R_{25}}$ -- $V_{rot}$
diagram (the same as  panel $b$ of Fig.~\ref{figure:v-oh}) but
again the galaxies are marked with different symbols according to the
number of IFU fibers with which they were observed.  The panel
$b$ of Fig.~\ref{figure:vr-oh-nfiber} demonstrates that the
galaxies again follow a single trend in the (O/H)$_{R_{25}}$ --
$V_{rot}$ diagram. The panels $c$ and $d$ of
Fig.~\ref{figure:vr-oh-nfiber} show the (O/H)$_{R_{0}}$ -- $M_{sp}$
and the (O/H)$_{R_{25}}$ -- $M_{sp}$ diagrams.
The value of the $M_{sp}$ for the galaxy measured with 37 fibers
is not available.
As before, one can see that the galaxies observed with different numbers 
of IFU fibers  follow a single trend.
This suggests that the derived properties of our
galaxies do not depend on the number of fiber bundles.

\citet{Belfiore2017} found that the influence of the PSF on the
obtained value of the oxygen abundance is maximum at the centre of the
galaxy and can be as large as up to around 0.04 dex for a galaxy with
$R_{e,g}$/PSF = 1.5 and inclination angle $i$ = 60$\degr$.
The mean value of the $R_{e,g}$/PSF for our sample of galaxies is
around 2.75 (Fig.~\ref{figure:psf}). \citet{Belfiore2017} found that
the uncertainty in the central oxygen abundance due to the PSF for galaxies
with $R_{e,g}$/PSF = 3.0 is around around 0.02 dex or lower depending on
the value of the inclination angle. Thus, the lack of the dependence of
the obtained oxygen abundances on the number of the fiber bundles that we noted is in
line with the results of \citet{Belfiore2017}. 

The mass -- metallicity (or luminosity -- metallicity) relation at the
present epoch and its evolution with redshift has been considered
previously by numerous authors \citep[][among many
others]{Lequeux1979,Zaritsky1994, Garnett2002, Grebel2003,
Tremonti2004, Erb2006, Pilyugin2007, Cowie2008, Maiolino2008,Guseva2009, Thuan2010,
PilyuginThuan2011, RosalesOrtega2012, Pilyugin2013, Andrews2013,Zahid2013, Maier2014, 
Pilyugin2014b, Steidel2014, Izotov2015,Sanchez2017,Barrera2017}.
Different methods to estimate abundances
were used in different works, which led to considerable discrepancies
in the resulting oxygen abundances due to the differing abundance
scales in those papers.  Therefore, it is difficult to compare
quantitatively the mass (luminosity) -- metallicity relations derived
by other authors and the ones obtained here.  

In \citet{Pilyugin2017a}, mass -- metallicity relations for a large
sample of SDSS galaxies at different redshifts were determined,
measuring their oxygen abundances through the same calibration as used
here.  Fig.~\ref{figure:msp-oh-comparison} shows the comparison of the
mass -- metallicity diagram obtained in our current study for galaxies
from the MaNGA survey with the mass -- metallicity relation determined
for the SDSS galaxies of \citet{Pilyugin2017a}.  MaNGA galaxies with
redshift $z < 0.05$ are shown by circles, and galaxies with $z > 0.05$
are indicated by plus signs.  The solid line is the O/H -- $M_{sp}$
relation obtained for the large sample of SDSS galaxies with $z =
0.03$ in \citet{Pilyugin2017a}.  Fig.~\ref{figure:msp-oh-comparison}
shows that the locations of the MaNGA galaxies in the mass --
metallicity diagram agree well with the mass -- metallicity relation
determined for the other SDSS galaxies.
It should be noted that the abundances determined from the
SDSS spectra are also in good agreement with abundances determined from
the CALIFA spectra (figures 16 and 17 in \citet{Pilyugin2017b}).  

Panel $a$ of Fig.~\ref{figure:redshift-m} shows the stellar mass as a function of
redshift for our galaxy sample. One can see the sequences of
the primary (``low redshift'') and secondary (``high redshift'') subsamples
of the MaNGA targets \citep{Bundy2015,Wake2017}.  Fig.~\ref{figure:redshift-m}
shows that the galaxies with masses log$M_{sp}$ $\la$ 10 are at
redshifts  less than  $\sim$0.05 (i.e. spread over the small interval
of redshift) and only the most massive galaxies  (log$M_{sp}$ $\ga$ 11) are observed
at redshifts up to $\sim$0.1.
Panel $b$ of Fig.~\ref{figure:redshift-m} shows the stellar mass Tully-Fisher
relation for our sample of galaxies. 
Panels $c$ and $d$ of Fig.~\ref{figure:redshift-m} show the central oxygen abundance
(panel $c$) and the oxygen abundance at the optical radius (panel $d$) as a function
of stellar mass of the galaxies.
In each panel, the galaxies of the primary and secondary subsamles
are denoted by different symbols.
Examination of the panels $b$, $c$, and $d$ of  Fig.~\ref{figure:redshift-m} shows
that the regions of the locations of the galaxies of the primary and secondary subsamles
in different diagram are rather close to each other. A small shift 
of the locations of the massive galaxies of the secondary subsample towards the
lower abundances in comparison to the locations of the galaxies of the primary
subsample in the (O/H)$_{R_{25}}$ - $M_{sp}$ diagram 
cannot be excluded.  Since the shift is small (within around 0.05 dex)
and the number of galaxies in this region of the diagram is also small therefore
the validity of this shift is not beyond the question. If this shift is real then
this can be considered as evidence in favour of the evolutionary changes
at the outer parts of the galaxies of the primary (low redshift) subsample
as compared to the galaxies of the secondary (high redshift) subsample.
However, larger numbers of massive galaxies of both subsamples should be
considered in order to draw a solid conclusion.
The locations of the massive galaxies of the secondary subsample and of
the primary subsample in the (O/H)$_{0}$ - $M_{sp}$ diagram
are close to each other. It was noted  in
\citet{Pilyugin2017a} that the evolutionary change of the central oxygen
abundance in massive galaxies with redshifts from $z \sim 0.025$ to
$\sim 0.1$ is minor.

Abundance gradients in galactic discs were considered recently in
several studies.  Again it is difficult to compare quantitatively the
results from different works since different methods to estimate
abundances were used.  Thus we restricted ourselves to the comparison
of the general behaviour of the radial abundance gradient with stellar
mass obtained here and by other groups of investigators.

We found that the gradient expressed in units of dex/$R_{25}$ is
slightly shallower in massive galaxies although it remains roughly
constant both for the subsample of low-mass galaxies and for the
subsample of massive galaxies.  The behaviour of the gradient
expressed in units of dex/$h_{d}$, dex/$R_{e,d}$, and  dex/$R_{e,g}$
is similar to that for the
gradient expressed in dex/$R_{25}$.  The relation between the
gradient expressed in units of dex/$R_{e,d}$ and stellar mass was
recently considered by
\citet{Sanchez2012b,Sanchez2014,SanchezMenguiano2016,SanchezMenguiano2018},
and \citet{Belfiore2017}.
\citet{Sanchez2012b,Sanchez2014}, and \citet{SanchezMenguiano2016}
found that all the CALIFA galaxies without clear evidence of an
interaction present a common gradient in the oxygen abundance
expressed in terms of dex/$R_{e,d}$. The slope is independent of
morphology, the existence of a bar, the absolute magnitude, or the
mass.  This conclusion was confirmed by the investigation of gradients
based on MUSE observations of galaxies \citep{SanchezMenguiano2018}.
\citet{Ho2015} found that when the metallicity gradients are expressed
in dex/$R_{25}$ then there is no correlation between the
metallicity gradient and the stellar mass and luminosity.
\citet{Belfiore2017} determined the oxygen abundance gradients in a
sample of 550 nearby galaxies from the MaNGA survey.  They found that
the gradient in terms of dex/$R_{e,d}$ steepens with stellar
mass until $\sim 10^{10.5} {\rm M}_{\odot}$ and remains roughly
constant for higher masses.  It should be noted that the lack of a
correlation between the abundance gradient expressed in units of
dex/$h_{d}$ and the luminosity was already found in earlier
studies \citep{Zaritsky1994,Garnett1997}. 

In our current work, we found that the gradient expressed in
dex/kpc flattens with increasing stellar mass.  \citet{Ho2015}
and \citet{Belfiore2017} also examined the relation between the gradient
in units of dex/kpc and stellar mass.  \citet{Ho2015}
determined metallicity gradients in 49 local field star-forming
galaxies.  They found that when the metallicity gradients are
expressed in dex/kpc then galaxies with lower mass and
luminosity, on average, have steeper metallicity gradients.  In
contrast, \citet{Belfiore2017} found that the gradient in terms of
dex/kpc steepens with stellar mass until $\sim 10^{10.5} {\rm M}_{\odot}$,
while becoming flatter again afterwards. The change of
the abundance gradient expressed in units of dex/kpc with
stellar mass obtained here agrees with the result of \citet{Ho2015}. 

Thus, the oxygen abundance of a star-forming galaxy correlates tightly
with (and appears to be governed by) its rotational velocity. The
central oxygen abundance (O/H)$_{0}$ and the abundance at the optical
radius (O/H)$_{R_{25}}$ can be found from the rotation velocity with a
mean uncertainy of around 0.05 and 0.08 dex, respectively.
Consequently, the oxygen abundance at any galactocentric distance can
be found through interpolation between those values with a mean
uncertainty of 0.08 dex.
The observed scatter in the O/H -- $V_{rot}$ relation is a combination
of the intrinsic scatter and errors in the O/H and $V_{rot}$
measurements, and can be considered as an upper limit on the intrinsic
scatter. The abundance deviations of the majority of the target
galaxies from the (O/H)$_{0}$ -- $V_{rot}$ and the (O/H)$_{R_{25}}$ --
$V_{rot}$ relations are within $\sim$0.1 dex (see
Fig.~\ref{figure:v-oh}).  The uncertainty in the
central oxygen abundance and in the abundance at the optical radius
caused by the errors in the oxygen abundance determination in
individual regions (spaxels) and by the approximation of the radial
abundance distribution by a single straight line can be around $\sim$0.05
dex \citep{Pilyugin2016,Pilyugin2017b},
i.e., is comparable to the scatter in the (O/H)$_{0}$ --
$V_{rot}$ and/or (O/H)$_{R_{25}}$ -- $V_{rot}$ diagrams.

It is known that the star formation history of a galaxy is related to
its stellar mass, the downsizing effect
\citep{Gavazzi1993,Cowie1996,IbarraMedel2016}.  It has also been known
for a long time that the morphological Hubble type of a galaxy,
expressed in terms of the $T$-type, is an indicator of its star
formation history \citep[e.g.,][]{Sandage1986}.  The morphological
type can contribute to the relations between properties of a galaxy.
For example, \citet{MunozMateos2015} found that the galaxies of
different Hubble types populate different areas in the stellar mass --
$R_{25.5}$ diagram, defining clearly distinct sequences. A
well-defined transition takes place when moving from Sc to Scd
galaxies, in the sense that earlier-type galaxies are up to a factor
of $\sim 10$ more massive than later-type ones for the same size or
isophotal radius $R_{25.5}$.  Therefore, one may expect that the
morphological type of a galaxy should be considered as a basic
characteristic of a galaxy.  

However, \citet{Conselice2003} noted that spiral galaxy classification
systems are in many ways not very useful since the most popular
morphological classification systems have non-physical and often
purely descriptive classification criteria based on optical
morphologies that do not uniquely identify, or distinguish, the
morphological types of distant galaxies. The estimations of the
morphological type of some galaxies from our list can be found in the
HyperLeda database \citep{Paturel2003,Makarov2014}.  However, the
uncertainty of the numerical $T$-type can exceed 3.  Hence we do
not consider the morphological types of galaxies here.

\section{Summary}

The publicly available data obtained within the framework of the MaNGA
survey are the base of our current study. We measured the emission
lines in the spectrum of each spaxel from the MaNGA datacubes for a
sample of late-type galaxies.  The wavelengths of the H$\alpha$
line provide the observed velocity field. We derived rotation curves,
surface brightness profiles, and oxygen abundance distributions for
147 galaxies. These data coupled with the available estimations of the
stellar masses were used to examine the relations between abundance
characteristics (abundances at the centre and at the optical radius,
radial gradient) and  basic galaxy parameters (rotation velocity,
stellar mass, luminosity, radius).

We found that both the central oxygen abundance (O/H)$_{0}$ and the
abundance at the outer regions of a galaxy (O/H)$_{R_{25}}$ correlate
tightly with the rotational velocity. The mean value of the scatter
around the (O/H)$_{0}$ -- $V_{rot}$ relation is $\sim 0.053$ dex and
the scatter around the (O/H)$_{R_{25}}$ -- $V_{rot}$ relation is $\sim
0.081$ dex. Thus, the oxygen abundances at any galactocentric distance
can be found through interpolation between those values with a mean
uncertainty of around 0.08 dex.

The change of both (O/H)$_{0}$ and (O/H)$_{R_{25}}$ with $V_{rot}$ is
similar.  The abundance grows with increasing $V_{rot}$.  There is a
break in the abundance growth rate at $V_{rot} \sim 200$ km/s,
in the sense that the growth rate is lower for galaxies with high
rotation velocities.  Both the central oxygen abundance and the
abundance at the optical radius of a galaxy correlate with other basic
parameters (stellar mass, luminosity, radius) in a similar way as with
the rotational velocity. 

The oxygen abundance gradients in units of dex/kpc flatten with
increasing rotational velocity. Again there is a break in the
flattening rate at $V_{rot} \sim 200$ km/s in the sense that
the flattening rate is lower for galaxies with high rotation
velocities.  The behaviour of the variation of the oxygen abundance
gradients in units of dex/kpc as a function of $V_{rot}$,
stellar masses $M_{sp}$ and $M_{ma}$, $B$-band luminosities $L_{B}$, 
and $R_{25}$ is similar.

The gradient expressed in units of dex/$R_{25}$ is slightly
shallower in galaxies with high rotation velocities although remains
roughly constant both for the subsample of galaxies with low rotation
velocities ($V_{rot} \la 200$ km/s) and the subsample with high
rotation velocities ($V_{rot} \ga 200$ km/s).  The behaviour of
the gradient expressed in units of dex/$h_{d}$, dex/$R_{e,d}$, and dex/$R_{e,g}$
is similar to that for the gradient expressed in dex/$R_{25}$.

\section*{Acknowledgements}

We are grateful to the referee for his/her constructive comments. \\
L.S.P., E.K.G., and I.A.Z.\  acknowledge support within the framework
of Sonderforschungsbereich (SFB 881) on ``The Milky Way System''
(especially subproject A05), which is funded by the German Research
Foundation (DFG). \\ 
L.S.P.\ and I.A.Z.\ thank for hospitality of the
Astronomisches Rechen-Institut at Heidelberg University, where part of
this investigation was carried out. \\
I.A.Z.\ acknowledges the support of the Volkswagen Foundation 
under the Trilateral Partnerships grant No.\ 90411. \\
I.A.Z. acknowledges the support by the Ukrainian National Grid
project (especially subproject 400Kt) of the NAS of Ukraine. \\
J.M.V acknowledges financial support from projects AYA2017-79724-C4-4-P,
of the Spanish PNAYA, and Junta de Andalucia Excellence PEX2011-FQM705. \\
This work was partly funded by the subsidy allocated to Kazan Federal 
University for the state assignment in the sphere of scientific 
activities (L.S.P.).  \\ 
We acknowledge the usage of the HyperLeda database (http://leda.univ-lyon1.fr). \\
Funding for SDSS-III has been provided by the Alfred P. Sloan Foundation,
the Participating Institutions, the National Science Foundation,
and the U.S. Department of Energy Office of Science.
The SDSS-III web site is http://www.sdss3.org/. \\
SDSS-III is managed by the Astrophysical Research Consortium
for the Participating Institutions of the SDSS-III Collaboration
including the University of Arizona, the Brazilian Participation Group,
Brookhaven National Laboratory, Carnegie Mellon University,
University of Florida, the French Participation Group,
the German Participation Group, Harvard University,
the Instituto de Astrofisica de Canarias,
the Michigan State/Notre Dame/JINA Participation Group,
Johns Hopkins University, Lawrence Berkeley National Laboratory,
Max Planck Institute for Astrophysics,
Max Planck Institute for Extraterrestrial Physics,
New Mexico State University, New York University,
Ohio State University, Pennsylvania State University,
University of Portsmouth, Princeton University,
the Spanish Participation Group, University of Tokyo,
University of Utah, Vanderbilt University, University of Virginia,
University of Washington, and Yale University. \\ 
Funding for the Sloan Digital Sky Survey IV has been provided by the
Alfred P. Sloan Foundation, the U.S. Department of Energy Office of Science,
and the Participating Institutions. SDSS-IV acknowledges
support and resources from the Center for High-Performance Computing at
the University of Utah. The SDSS web site is www.sdss.org. \\
SDSS-IV is managed by the Astrophysical Research Consortium for the 
Participating Institutions of the SDSS Collaboration including the 
Brazilian Participation Group, the Carnegie Institution for Science, 
Carnegie Mellon University, the Chilean Participation Group,
the French Participation Group, Harvard-Smithsonian Center for Astrophysics, 
Instituto de Astrof\'isica de Canarias, The Johns Hopkins University, 
Kavli Institute for the Physics and Mathematics of the Universe (IPMU) / 
University of Tokyo, Lawrence Berkeley National Laboratory, 
Leibniz Institut f\"ur Astrophysik Potsdam (AIP),  
Max-Planck-Institut f\"ur Astronomie (MPIA Heidelberg), 
Max-Planck-Institut f\"ur Astrophysik (MPA Garching), 
Max-Planck-Institut f\"ur Extraterrestrische Physik (MPE), 
National Astronomical Observatories of China, New Mexico State University, 
New York University, University of Notre Dame, 
Observat\'ario Nacional / MCTI, The Ohio State University, 
Pennsylvania State University, Shanghai Astronomical Observatory, 
United Kingdom Participation Group,
Universidad Nacional Aut\'onoma de M\'exico, University of Arizona, 
University of Colorado Boulder, University of Oxford, University of Portsmouth, 
University of Utah, University of Virginia, University of Washington, University of Wisconsin, 
Vanderbilt University, and Yale University.

\end{document}